\documentclass[a4paper]{article}
\pdfoutput=1
\usepackage{jheppub}

\usepackage{amsmath}

\usepackage{amssymb}
\usepackage{graphicx,color}

\usepackage{amsmath,amsthm}
\usepackage{txfonts}

\usepackage{mathrsfs}

\usepackage[small,nohug,heads=littlevee]{diagrams}
\diagramstyle[labelstyle=\scriptstyle]

\usepackage{bm}

\newcommand{\Ref}[1]{(\ref{#1})}

\newtheorem{Theorem}{Theorem}[section]

\newtheorem{Lemma}[Theorem]{Lemma}

\newcommand{\Z}{\mathbb{Z}}
\newcommand{\R}{\mathbb{R}}
\newcommand{\C}{\mathbb{C}}

\newcommand{\half}{\frac{1}{2}}

\newcommand{\D}{\mathrm{d}}

\newcommand{\A}{{\mathscr{A}}}

\newcommand{\bA}{{\bar{A}}}

\newcommand{\Slc}{\mathrm{SL}(2,\mathbb{C})}
\newcommand{\PSlc}{\mathrm{PSL}(2,\mathbb{C})}
\newcommand{\slc}{\fs\fl_2\mathbb{C}}

\newcommand{\Su}{\mathrm{SU}(2)}

\def\be{\begin{eqnarray}}
\def\ee{\end{eqnarray}}

\newcommand{\cd}{\mathcal D}
\newcommand{\ce}{\mathcal E}
\newcommand{\cf}{\mathcal F}

\newcommand{\ch}{\mathcal H}

\newcommand{\ck}{\mathcal K}
\newcommand{\cl}{\mathcal L}
\newcommand{\cm}{\mathcal M}

\newcommand{\cp}{\mathcal P}

\newcommand{\cs}{\mathcal S}

\newcommand{\sa}{\mathscr{A}}
\newcommand{\sm}{\mathscr{M}}

  \newcommand{\Fa}{\mathfrak{A}}
  
  \newcommand{\Fc}{\mathfrak{C}}
  
\newcommand{\fe}{\mathfrak{e}}  
  
\newcommand{\fg}{\mathfrak{g}}  
  \newcommand{\Fh}{\mathfrak{H}}

\newcommand{\fl}{\mathfrak{l}}  \newcommand{\Fl}{\mathfrak{L}}
  \newcommand{\Fm}{\mathfrak{M}}
\newcommand{\fn}{\mathfrak{n}}  
\newcommand{\fo}{\mathfrak{o}}  
\newcommand{\fp}{\mathfrak{p}}

\newcommand{\fs}{\mathfrak{s}}

  \newcommand{\Fx}{\mathfrak{X}}

\renewcommand{\a}{\alpha}
\renewcommand{\b}{\beta}
\newcommand{\g}{\gamma}
\newcommand{\G}{\Gamma}
\renewcommand{\d}{\delta}
\newcommand{\eps}{\varepsilon}

\newcommand{\sig}{\sigma}
\newcommand{\Sig}{\Sigma}
\renewcommand{\l}{\lambda}
\renewcommand{\L }{\Lambda}
\renewcommand{\o}{\omega}
\renewcommand{\O}{\Omega}
\renewcommand{\t}{\tau}

\newcommand{\rmd}{\mathrm d}

\newcommand{\lt}{\left}
\newcommand{\rt}{\right}

\newcommand{\lag}{\left\langle}
\newcommand{\rag}{\right\rangle}

\newcommand{\tr}{\mathrm{tr}}

\newcommand{\Ar}{\text{a}}

\newcommand{\act}{\rhd}

\newcommand{\sgn}{\mathrm{sgn}}
\newcommand{\vth}{\vartheta}
\newcommand{\rmi}{\mathrm{i}}

\title{SL(2,C) Chern-Simons Theory, Flat Connections, and Four-dimensional Quantum Geometry}

\author[a,e]{Hal M. Haggard}  

\author[b,c]{\ Muxin Han}  

\author[d]{\ Wojciech Kaminski}  

\author[e]{\ Aldo Riello}

\affiliation[a]{Physics Program, Bard College, 30 Campus Rd, Annandale-on-Hudson, NY 12504, USA }

\affiliation[b]{Institut f\"ur Quantengravitation, Universit\"at Erlangen-N\"urnberg, Staudtstr. 7/B2, 91058 Erlangen, Germany}

\affiliation[c]{Department of Physics, Florida Atlantic University, FL 33431, USA}

\affiliation[d]{Faculty of Physics, University of Warsaw, Ho\.za 69, 00-681 Warszawa, Poland}

\affiliation[e]{Perimeter Institute for Theoretical Physics, 31 Caroline Street N, Waterloo, ON N2L 2Y5, Canada}

\emailAdd{hhaggard(AT)bard.edu}
\emailAdd{muxin.han(AT)gravity.fau.de} %
\emailAdd{Wojciech.Kaminski(AT)fuw.edu.pl}
\emailAdd{ariello(AT)perimeterinstitute.ca} %

\abstract{

A correspondence between three-dimensional flat connections and constant curvature four-dimensional simplices is used to give a novel quantization of geometry via complex $\Slc$ Chern-Simons theory. The resulting quantum geometrical states are hence represented by the 3d blocks of analytically continued Chern-Simons theory. In the semiclassical limit of this quantization the three-dimensional Chern-Simons action, remarkably, becomes the discrete Einstein-Hilbert action of a 4-simplex, featuring the appropriate boundary terms as well as the essential cosmological term proportional to the simplex's curved 4-volume. Both signs of the curvature and associated cosmological constant are present in the class of flat connections that give rise to this correspondence. We provide a Wilson graph operator that picks out this class of connections. We discuss how to promote these results to a model of Lorentzian covariant quantum gravity encompassing both signs of the cosmological constant. This paper presents the details for the results reported in \cite{HHKRshort}.\\
}


\arxivnumber{}

\begin{document}

\maketitle

\section{Introduction and Overview}\label{introduction}

Chern-Simons theory in 3-dimensions is the quintessential topological quantum field theory and has been studied extensively since the 1980's (see e.g. \cite{CSrev}). In addition to its importance in the formulation of topological quantum field theory \cite{TQFT}, Chern-Simons theory has applications in many branches of modern mathematics and physics. The celebrated work of Witten \cite{jones}, exposed the remarkable relation between Chern-Simons theory with compact gauge group and knot theory. This exchange has continued to the present day with, for example, Chern-Simons theory playing an important role in the formulation of the Volume Conjecture, which relates knot polynomials to the hyperbolic geometry of 3-manifolds \cite{VC,gukov,knots,analcs}. Many aspects of String theory, M-theory and Supersymmetric Gauge Theory also have close ties to Chern-Simons theory (e.g. \cite{CSstring,marino,CJ,ABJM,highspin,brane}). Most importantly for the present work, Chern-Simons theory has furnished exact solutions to quantum gravity in 3-dimensions \cite{3dgravity,carlip}, and provided interesting insights into Loop Quantum Gravity (LQG) in 4 dimensions, both in its covariant formulation and in black hole physics (e.g. \cite{smolin,wolfgang,QSFasymptotics,HHR,LQGBH}). Chern-Simons theory and its relation to four-dimensional quantum gravity with a cosmological constant (of either sign) will be the main focus of this paper. 

 Chern-Simons theories with a compact gauge group and their quantization have become well understood after the intensive investigations of the last 20 years. However, quantum Chern-Simons theory with complex gauge group $G_{\C}$, with  $G_{\C}$ the complexification of a compact Lie group $G$, is still a rather open subject. These Chern-Simons theories are noncompact and hence qualitatively different from those with compact group. In general, the Hilbert spaces associated to the complex case are infinite-dimensional \cite{witten,roche,knots,Dimofte}, while the Hilbert spaces in the compact cases are finite-dimensional. Recently, there has been substantial progress in understanding the complex gauge group case \cite{gukov,analcs,Dimofte,knots,DGLZ}. This is an active area of research.

\begin{figure}[h]
\begin{center}
\includegraphics[width=4.5cm]{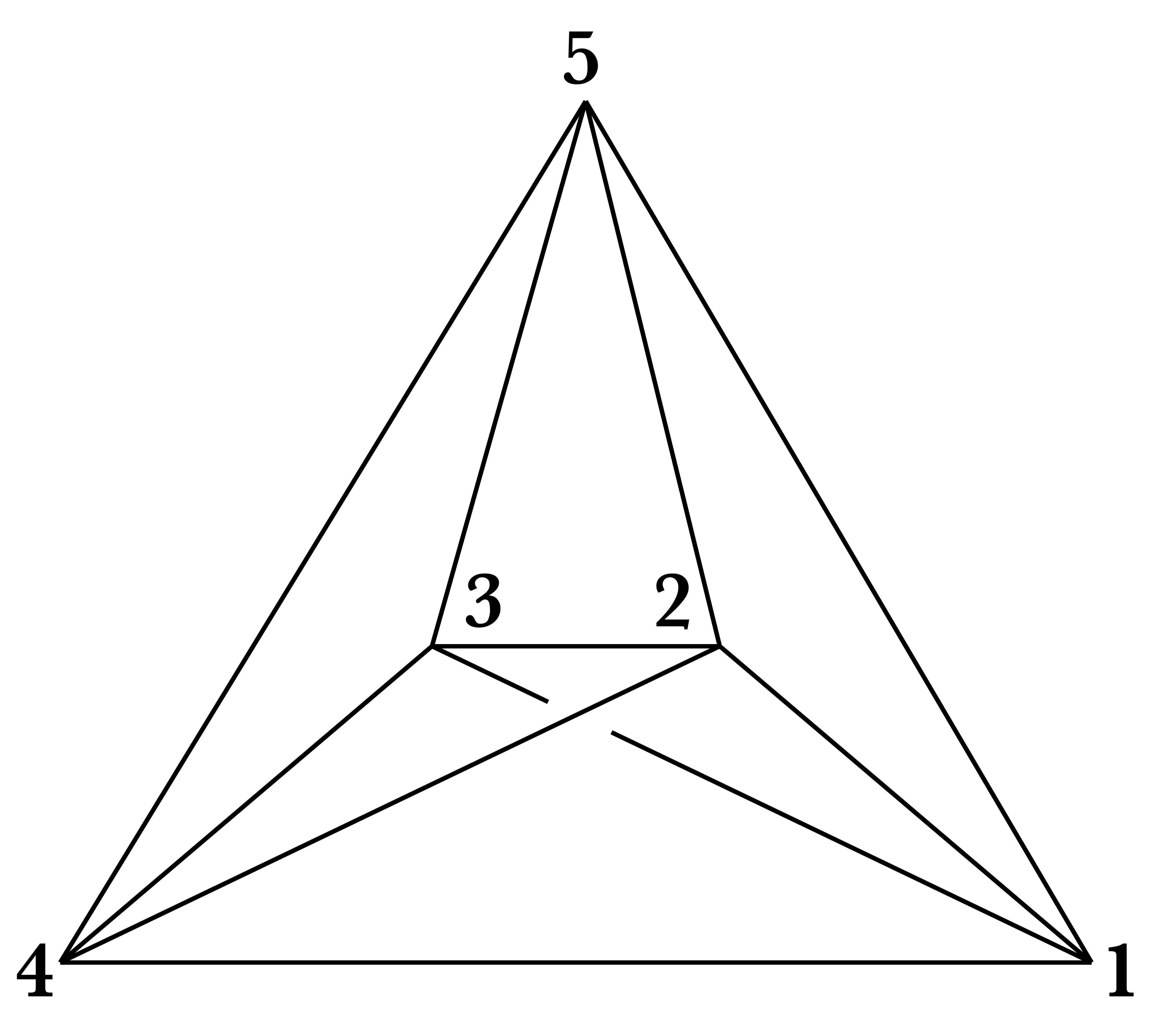}
\caption{The $\G_5$ graph can be drawn with five 4-valent vertices, ten edges $\ell_{ab}$, and the curve $\ell_{24}$ over-crossing $\ell_{13}$. It can also be drawn with all vertices being 3-valent by expanding each 4-valent vertex into two connected 3-valent vertices, which results in 10 vertices and 15 edges. Both ways of drawing $\G_5$ lead to the same 3-manifold $S^3\setminus\G_5$. }
\label{gamma5}
\end{center}
\end{figure}

This paper focuses on Chern-Simons theory with a complex $\Slc$ gauge group on a compact oriented 3-manifold $\sm_3$. The action for this theory is 
\be
{CS}\lt[\sm_3\,|\,A,\bar{A}\rt]=\frac{t}{8\pi}\int_{\sm_3}\tr\lt(A\wedge \rmd A+\frac{2}{3}A\wedge A\wedge A\rt)+\frac{\bar{t}}{8\pi}\int_{\sm_3}\tr\lt(\bA\wedge \rmd \bA+\frac{2}{3}\bA\wedge \bA\wedge \bA\rt),
\ee
and might include boundary terms when $\sm_3$ has a boundary. Here $t=k+is$ is the Chern-Simons coupling with $k,s\in\R$, and $\bar{t}$ is taken to be the complex conjugate of $t$. The connection 1-form is $A=A^j\t_j$, where $j \in \{1,2,3\}$, $\t_j=-\frac{i}{2}\sig_j$ are generators that take values in the complex Lie algebra $\slc$, and $\sig_{j}$ are the Pauli matrices. We will focus on a certain class of 3-manifolds $\sm_3$, the simplest example of which is the graph complement 3-manifold $\sm_3 =  S^3\setminus\G_5$, where $\G_5$ is the graph with five 4-valent vertices and the single essential crossing depicted in Fig. \ref{gamma5}. For a graph embedded in $S^3$, the graph complement manifold is obtained by removing the graph as well as the interior of its tubular neighborhood from $S^3$. The boundary of $S^3\setminus\G_5$ is a genus-6 closed 2-surface, which we denote $\Sig_6$.

Chern-Simons theory with graph defects has been considered in \cite{knotgraph} in the case of a compact gauge group; and the volume conjecture has been generalized to quantum spin-networks with knotted graphs in \cite{roland,satoshi}. From the mathematical point of view, the space of knotted graphs may be more interesting than the space of knots---due to the fact that the space of trivalent knotted graphs is finitely generated. This means that there is a finite (and small) set of trivalent knotted graphs that can generate all trivalent knotted graphs via just a few algebraic operations, while the space of knots is a proper subset of the space of trivalent graphs \cite{T}. A recent study of trivalent knotted graphs, from the perspective of perturbative Vassiliev-Kontsevich invariants, specifies these algebraic operations, \cite{dancso}. 

Classically, the equations of motion for $\Slc$ Chern-Simons theory are 
\be
F=\rmd A+A\wedge A=0,\ \quad \text{and} \quad \ \bar{F}=\rmd \bar{A}+\bar{A}\wedge \bar{A}=0,
\ee
that is, the connections $A$ and $\bar{A}$ are flat on the 3-manifold $\sm_3$. The moduli space of flat connections $\cm_{\mathrm{flat}}(\sm_3,\Slc)$ is the space of solutions. When $\sm_3$ has boundary a closed 2-surface $\Sig_g=\partial\sm_3$, of genus-$g$, the space of boundary values of $A\in \cm_{\mathrm{flat}}(\sm_3,\Slc)$ is a subvariety inside $\cm_{\mathrm{flat}}(\Sig_g,\Slc)$, which is the moduli space of $\Slc$ flat connections on the two-dimensional manifold $\Sig_g$. In general, $\cm_{\mathrm{flat}}(\Sig_g,\Slc)$, known as the Hitchin moduli space, is a hyper-K\"ahler variety of $\dim_\C=6g-6$, which has 3 distinct complex structures $I,J,$ and $K$ \cite{hitchin}.\footnote{The complex structure $I$ is induced from that of $\Sig_g$, $J$ is from the complex structure of the complex group $\Slc$, and $K$ is obtained through $K=IJ$.}  The three corresponding K\"ahler forms are denoted $\o_I,\o_J,$ and $\o_K$. When we think of $\cm_{\mathrm{flat}}(\Sig_g,\Slc)$ as the phase space of $\Slc$ Chern-Simons theory, the holomorphic Chern-Simons (Atiyah-Bott-Goldman) symplectic structure $\o_{CS}$ is given by
\be
\o_{CS}=\frac{t}{4\pi}\int_{\Sig_{g}}\tr\lt[\delta_1 A\wedge \delta_2 A\rt]=\frac{t}{\pi}\lt[\o_I-i\o_K\rt],
\ee
which comes from the holomorphic part of ${CS}\lt[\sm_3\,|\,A,\bar{A}\rt]$. The space of flat connections on $\sm_3$ can be embedded as a subvariety $\cl_{\mathbf{A}}$ of complex dimension $\dim_\C=3g-3$ in $\cm_{\mathrm{flat}}(\Sig_g,\Slc)$ by considering the boundary values of these flat connections,
\be
\cl_{\mathbf{A}}\simeq \cm_{\mathrm{flat}}(\sm_3,\Slc).
\ee
The subvariety $\cl_{\mathbf{A}}$ is holomorphic with respect to the complex structure $J$, and is Lagrangian with respect to $I$ and $K$, i.e. $\o_I$ and $\o_K$, and hence $\o_{CS}$, vanish on $\cl_{\mathbf{A}}$ \cite{DGV,DGG}. 

The fact that $\cl_{\mathbf{A}}$ is Lagrangian has a clear physical meaning as well. Consider an analogy with particle mechanics, which can be seen as a field theory over the time axis. The boundary values of a physical trajectory are the phase space points at the initial and final times $t_0$ and $t$. Introduce a boundary phase space, which is just the Cartesian product of two copies of the phase space one at each of these times. This doubled phase space has a symplectic form $\Omega = d p \wedge dq- d p_0 \wedge d q_0$.  The sign on the second term indicates that the initial space is to the past. The statement that the dynamics is a canonical transformation, i.e. that $dp \wedge dq$ is invariant under time evolution, is precisely the statement that the space of orbits of the equations of motion corresponds to a Lagrangian manifold of the doubled boundary phase space. That is, $\Omega|_{\cl_{\mathbf{D}}} =0$, where $\cl_{\mathbf{D}}$ is the subset of points of the boundary phase space connected by a dynamical orbit. This mechanical analogy was  introduced by Tulcyjew precisely with the generalization to field theory in mind \cite{Tulcyjew:1979}.  The connections of $\cm_{\mathrm{flat}}(\sm_3,\Slc)$ provide dynamical interpolations of the boundary data. So, not only is  $\cm_{\mathrm{flat}}(\Sig_g,\Slc)$ of larger dimension, e.g. there are non-contractible loops in $\Sig_g$ that are contractible in $\sm_3$, but $\cm_{\mathrm{flat}}(\sm_3,\Slc)$ is exactly half-dimensional and is Lagrangian. 

The complex Fenchel-Nielsen (FN) coordinates $x_m,y_m\in\C,\ {m=1\cdots3g-3}$ can be used to locally parametrize the connections of $\cm_{\mathrm{flat}}(\Sig_g,\Slc)$ \cite{FN,kabaya}, using a trinion (or pants) decomposition of the closed 2-surface $\Sig_g$. Here the complex FN ``length variable'' $x_m$ is the eigenvalue of the holonomy along a closed curve $c_m$ transverse to a tube of the trinion decomposition. The complex FN ``twist variable'' $y_m$ is the conjugate variable such that $\o_{CS}$ is written as
\be
\o_{CS}=\lt(-\frac{t}{2\pi}\rt)\sum_{m=1}^{3g-3}\frac{\rmd y_m}{y_m}\wedge \frac{\rmd x_m}{x_m}.
\ee
The explicit relation between $y_m$ and holonomies is given in e.g. \cite{kabaya,DGV}, and is briefly reviewed in Section \ref{quantum}. In terms of $\{x_m,y_m\}_{m=1}^{3g-3}$, the holomorphic Lagrangian subvariety $\cl_{\mathbf{A}}\simeq \cm_{\mathrm{flat}}(\sm_3,\Slc)$ can be expressed locally as a set of holomorphic polynomial equations 
\be
\mathbf{A}_m(x,y)=0,\ \ \ \ m=1,\cdots,3g-3.
\ee
When $\sm_3$ is the complement of a knot, so that $\partial\sm_3= T^2$, we have $\cm_{\mathrm{flat}}( T^2,\Slc)\simeq \C^*\times \C^*/\Z_2$, and $\cl_{\mathbf{A}}$ is the zero-locus of a single holomorphic polynomial $\mathbf{A}(x,y)$, known as the A-polynomial \cite{gukov,Apoly}. This provides an interesting and quite different perspective on the quantum gravity quantizations discussed below.

\begin{figure}[h]
\begin{center}
\includegraphics[width=.75\textwidth]{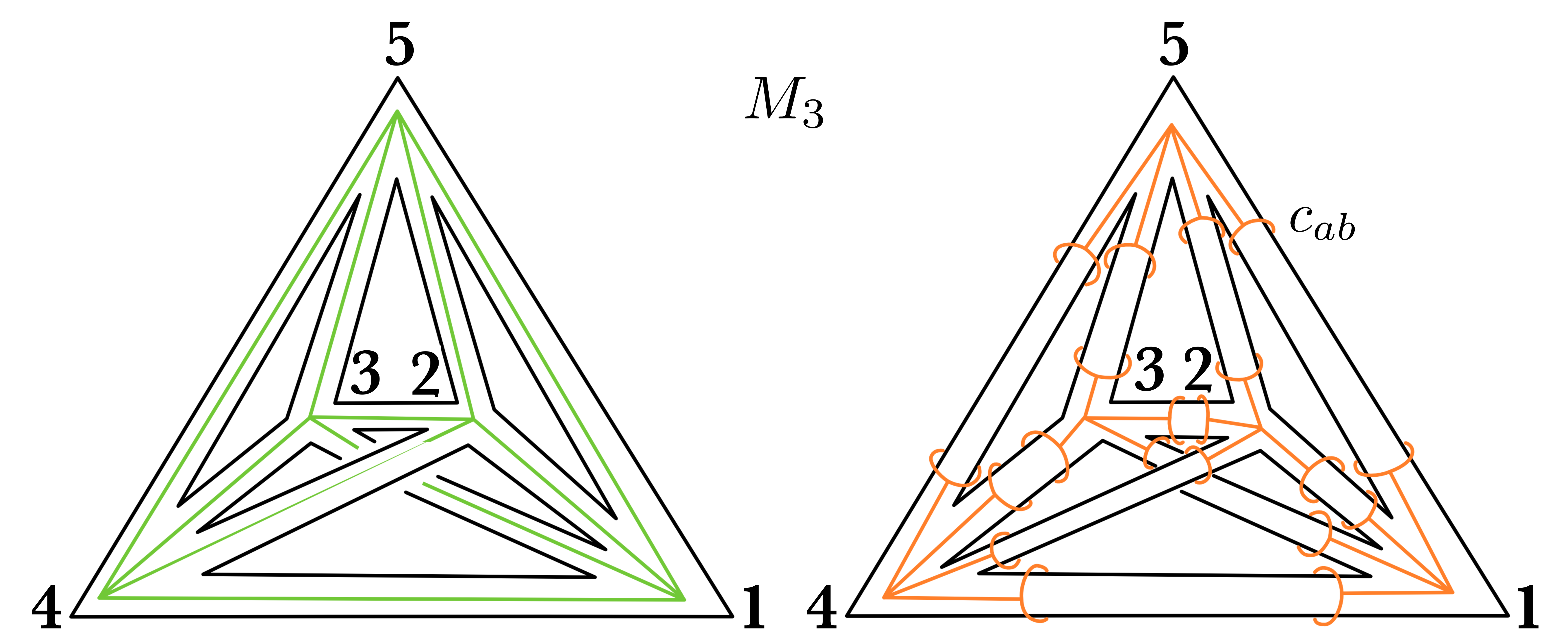}
\caption{The graph complement 3-manifold $M_3$ after removing the thickened $\G_5$-graph from $S^3$. The 2d boundary $ \Sig_{6}=\partial M_3$ of the graph complement $M_3$ is a genus-6 closed 2-surface. Left: The longitudinal paths (in green) used below to calculate holonomies $G_{ab}$. Right: A set of meridian closed curves $c_{ab}$ (in orange) defined on $\Sig_{6}$ such that $\Sig_{6}\setminus\{c_{ab}\}$ is a set of 4-holed spheres. These curves are used to calculate the holonomies $H_{b}(a)$ below.  The vertices of the graph are labeled by $a,b=1,\dots,5$.} 
\label{ribbon}
\end{center}
\end{figure}

In this paper we use these tools to present a novel model of four-dimensional Lorentzian quantum gravity based on a discretized path integral over gravitational holonomy variables where the cosmological constant emerges as a consequence of our quantization procedure via complex Chern-Simons theory. In particular, both signs of $\L$ are treated on an equal footing in the model.

Our discretization of the path integral decomposes spacetime into a simplicial complex, with each simplex of constant curvature $\L$. We choose to work with parallel transports along closed paths (holonomies) as they are the most natural gravitational observables. They also fit nicely with the use of constant curvature simplices, as we have shown previously \cite{HHKR, HHR}, whose geometry can be completely encoded in a finite number of these holonomies. From the Chern-Simons perspective, these holonomies arise as the non-contractible cycles of the particular graph complement manifold $M_3 =  S^3\setminus\G_5$. 

Because the tools used in this paper are drawn from different areas and many will be unfamiliar to sections of our intended audience, we give a detailed overview of the paper in the next few subsections. A brief outline of the paper appears in Section \ref{structure} and pointers to the detailed arguments of the main body of the paper can be found there. 

\subsection{Classical Correspondence}

Let us focus on the 3-manifold $S^3\setminus\G_5$ whose boundary is a genus-6 closed surface $\Sig_6$ (see Fig. \ref{ribbon}). We are interested in a subspace of $\cl_{\mathbf{A}}= \cm_{\mathrm{flat}}(S^3\setminus\G_5,\Slc)$ in which the $\Slc$ flat connections can be interpreted in terms of simplicial 4-geometries. More precisely, a flat connection in this particular subspace will determine the geometry of a convex 4-simplex in 4-dimensional Lorentzian constant curvature spacetime (de-Sitter or Anti-de-Sitter).\footnote{Four-dimensional simplices can be used as the elementary building blocks of the simplicial decomposition of a 4-dimensional manifold. This is analogous to tetrahedral decompositions in 3 dimensions and triangulations in 2 dimensions. See Figures \ref{gamma5} and  \ref{4simplex} for two different projections of the 4-simplex that give some insight into its combinatorial structure.} Fix the Lorentzian signature to $(-,+,+,+)$. We will find that all non-degenerate convex constant curvature 4-simplices with both $\L>0$ and $\L<0$ can be described by a class of $\Slc$ flat connections on the graph complement 3-manifold $S^3\setminus\G_5$. In brief: 
\be
\text{A class of $(A,\bA)$ in }\cm_{\mathrm{flat}}\lt(S^3\setminus\G_5,\,\Slc\rt)\ =\ \text{constant curvature 4-simplex geometries}.\label{relation4simplex}
\ee

The particular subspace of flat connections that determines 4-simplex geometries is specified by certain boundary conditions imposed on their boundary values on $\Sig_6$. These boundary conditions are introduced in Section \ref{b.c.}, and can be summarized in the following way: $\Sig_6$ can be decomposed into five 4-holed spheres $\cs_{a=1,\cdots,5}$ by cutting through the 10 meridian closed curves on the right in Fig. \ref{ribbon}. The boundary conditions require that the boundary value of $A\in \cm_{\mathrm{flat}}(S^3\setminus\G_5,\Slc)$ reduces to an SU(2) flat connection, up to gauge transformations, when it is restricted to each of the 4-holed spheres $\cs_a$. {This does not imply that $A$ is an SU(2) flat connection on all of $\Sig_6$, since the different 4-holed spheres may correspond to different SU(2) subgroups in $\Slc$.}

These boundary conditions are motivated by a geometrical interpretation of the SU(2) flat connections on a 4-holed sphere $\cs_a$. Each of these connections determines uniquely a convex tetrahedron in constant curvature 3d space (spherical or hyperbolic). This  holds for a dense subset of $\cm_{\mathrm{flat}}(\cs_a,\Su)$, and only excludes the flat connections corresponding to degenerate geometries. If we consider PSU(2) flat connections instead of SU(2), the correspondence becomes 1-to-1 (see Theorem \ref{2d/3d}). This interpretation of SU(2) flat connections on a 4-holed sphere was introduced in \cite{HHR} and is reviewed in Section \ref{b.c.} (see \cite{HHKR} for a thorough exploration).  

A flat connection $A\in \cm_{\mathrm{flat}}(S^3\setminus\G_5,\Slc)$ on the $\G_5$ graph complement manifold that satisfies the above boundary conditions on the full complement goes further and determines uniquely a convex 4-simplex geometry in 4-dimensional Lorentzian spacetime with constant curvature $\L$ (see Theorem \ref{3d/4d} and the analysis of \cite{HHR}). The closed boundary of the 4-simplex determined by $A$ is formed by 5 constant curvature tetrahedra, which are congruent to the tetrahedral geometries determined by the boundary data of $A$ on the 4-holed spheres $\cs_a$. Again the statement holds up to those flat connections that correspond to degenerate 4-simplex geometries. If we consider $\PSlc$ flat connections instead of $\Slc$, the correspondence once again becomes 1-to-1. In the following, we will refer to flat connections satisfying the boundary conditions that put them into correspondence with a 4-simplex geometry as simplicial flat connections. 

A simple intuition lies behind the above correspondence between flat connections on a 3-manifold and the geometry of a 4-manifold. The 1-skeleton of a 4-simplex gives a triangulation of the 3-sphere, thought of as the boundary of the 4-simplex. The $\G_5$ graph can be viewed as a ``dual'' graph of the 4-simplex skeleton, in the sense that the fundamental group of $S^3\setminus\G_5$ is isomorphic to the fundamental group of the 4-simplex skeleton $\pi_1(\mathrm{simplex})$. The isomorphism is unique under a few natural assumptions (see Lemma \ref{topounique}). On the one hand, an $\Slc$ flat connection on $S^3\setminus\G_5$ is a representation of the fundamental group $\pi_1(S^3\setminus\G_5)$ up to conjugation. On the other, if the 4-simplex is embedded in a geometrical 4d spacetime $(\Fm_4,g_{\a\b})$, the spin connection on $\Fm_4$ gives a representation up to conjugation of $\pi_1(\mathrm{simplex})$ using holonomies. The isomorphism between $\pi_1(S^3\setminus\G_5)$ and $\pi_1(\mathrm{simplex})$ identifies the flat connection on $S^3\setminus\G_5$ and the spin connection on the 4-simplex. More precisely, it identifies the holonomies of the flat connection along the loops in $\pi_1(S^3\setminus\G_5)$ and the holonomies of the spin connection along the closed paths of $\pi_1(\mathrm{simplex})$. In terms of a commutative diagram,
\be
&\pi_1(S^3\setminus\G_5)\ \ \ \ \ \ \ \ \ \ \ \ \ \ \ \ \ \stackrel{X}{\longleftarrow}\ \ \ \ \ \ \ \ \ \ \ \ \ \ \ \ \ \pi_1(\text{simplex})&\nonumber\\
&&\nonumber\\
&\o_{\mathrm{flat}}\searrow\ \ \ \ \ \ \ \ \ \ \ \ \ \ \ \ \ \ \ \ \ \ \ \ \ \ \swarrow\o_{\mathrm{spin}}&\nonumber\\
&&\nonumber\\
&\lag\ \{\tilde{H}_{ab}\in\Slc\}_{a<b}\ \big|\ \text{algebraic relations Eqs}.\Ref{vertex1} - \Ref{X}\ \rag\big/\mathrm{conjugation},&
\ee 
where $X$ denotes the isomorphism between $\pi_1(S^3\setminus\G_5)$ and $\pi_1(\mathrm{simplex})$ and $\o_{\mathrm{flat}}$ and $\o_{\mathrm{spin}}$ denote the representations by the flat connection on $S^3\setminus\G_5$ and the spin connection on $\Fm_4$, respectively. In this way, the $\Slc$ flat connections on $S^3\setminus\G_5$ relate to the spin connections on a spacetime $(\Fm_4,g_{\a\b})$. If we take $(\Fm_4,g_{\a\b})$ to be a Lorentzian spacetime with constant curvature $\L$, and all 10 triangles of the 4-simplex flatly embedded in $(\Fm_4,g_{\a\b})$ (i.e. with vanishing extrinsic curvature), the holonomy of the spin connection along a closed path in $\pi_1(\mathrm{simplex})$ enclosing a single triangle determines the area of the triangle, as well as the embedding property of the triangle, i.e. the 2 normal directions of the triangle embedded in $\Fm_4$. The above relation between $\o_{\mathrm{flat}}$ and $\o_{\mathrm{spin}}$, as well as the geometrical properties of the spin connections, result in the correspondence between the $\Slc$ flat connections on $S^3\setminus\G_5$ and the 4d geometry of constant curvature 4-simplices.

Each geometrical flat connection $A\in\cm_{\mathrm{flat}}(S^3\setminus\G_5,\Slc)$ is naturally accompanied by an $\tilde{A}\in\cm_{\mathrm{flat}}(S^3\setminus\G_5,\Slc)$, which is the complex conjugate of $A$ with respect to the complex structure $J$ induced from the complex group $\Slc$. The pair $A$ and $\tilde{A}$ determine the same 4-simplex geometry but result in 2 opposite 4d orientations for the 4-simplex. We call $(A,\tilde{A})$ a ``parity pair,'' because complex conjugation using $J$ naturally relates to a parity inversion in 4d spacetime \cite{HHR}. This complex conjugation leaves the SU(2) flat connections invariant, so $A$ and $\tilde{A}$ induce the same SU(2) flat connections on the 4-holed spheres $\cs_{a=1,\cdots,5}$. This is consistent with the fact that the 4-simplex geometries determined by $A$ and $\tilde{A}$ are the same, and give the same set of geometrical tetrahedra on the boundary.

Consider $\cm_{\mathrm{flat}}(S^3\setminus\G_5,\Slc)\simeq\cl_{\mathbf{A}}$ as a holomorphic Lagrangian subvariety in $\cm_{\mathrm{flat}}(\Sig_6,\Slc)$. Given $A\in\cm_{\mathrm{flat}}(S^3\setminus\G_5,\Slc)$ corresponding to a constant curvature 4-simplex, the complex Fenchel-Nielsen (FN) variables of $A$ have direct interpretations in terms of the 4-simplex geometry (see Section \ref{quantum}). The 10 length variables $x_{ab}$ for the closed curves $c_{ab}$ in Fig. \ref{ribbon} relate respectively to the 10 areas $\Ar_{ab}$ of the triangles $\Delta_{ab}$ in the 4-simplex. The 10 conjugate twist variables $y_{ab}$ relate respectively to the 10 hyperdihedral angles $\Theta_{ab}$ of the 4-simplex. Each hyperdihedral angle $\Theta_{ab}$ between a pair of boundary tetrahedra is hinged by the triangle $\Delta_{ab}$ shared by the tetrahedra. Interestingly the canonical conjugacy of $\Ar_{ab}$ and $\Theta_{ab}$ that follows from the correspondence between flat connections and their geometrical counterparts, relates to the canonical structure induced by the 4-dimensional Einstein-Hilbert action in General Relativity (GR), see \cite{bianca} for a derivation in the GR case. This further motivates the relation between the flat connections on 3-manifolds and (simplicial) gravity on 4-dimensional manifolds. 

The phase space of flat connections has complex dimension $\dim_{\C}[ \cm_{\mathrm{flat}}(\Sig_6,\Slc)]=30$. In addition to the 20 coordinates $\{x_{ab},y_{ab}\}_{a<b}$, there are 5 pairs of variables $\{x_a,y_a\}_{a=1}^5$ that parametrize the SU(2) flat connections on $\cs_{a=1\cdots,5}$. Geometrically they correspond to the shapes of the 5 constant curvature tetrahedra on the boundary of the 4-simplex.

\subsection{Quantum Correspondence}

Our correspondence between $\Slc$ flat connections on $S^3\setminus\G_5$ and the constant curvature geometry of 4-simplices inspires a new understanding of 4-dimensional quantum simplicial geometry in terms of the quantization of flat connections on a 3-manifold. For any 3-manifold $\sm_3$ with boundary $\Sig_g$, the quantization of $\cm_{\mathrm{flat}}(\Sig_g,\Slc)$ with the symplectic structure $\o_{CS}$ results in an operator algebra for the canonically conjugate variables, e.g. the operators representing the complex FN variables $\hat{x}_m$ and $\hat{y}_m$ satisfy $\hat{x}_m\hat{y}_m=e^{-\frac{2\pi i\hbar}{t}}\hat{y}_m\hat{x}_m$ $(\hbar\in\R)$ and $\hat{x}_m\hat{y}_n=\hat{y}_n\hat{x}_m$ for $n\neq m$. The states are represented as the wave functions $Z(u)$, where $u$ is the logarithmic coordinate $u_m=\ln x_m$. The reader is referred to, e.g \cite{teschner,Qteich}, for details of quantizing $\cm_{\mathrm{flat}}(\Sig_g,\Slc)$. The quantization of the holomorphic Lagrangian subvariety $\cm_{\mathrm{flat}}(\sm_3,\Slc)\simeq\cl_{\mathbf{A}}$ gives a set of operator constraint equations:
\be
\hat{\mathbf{A}}_m(\hat{x},\hat{y},\hbar)\,Z(u)=0,\ \ \ \ m=1,\cdots,3g-3.
\ee
The solutions $Z(u)$ of the above operator constraint equations are the physical states of $\Slc$ Chern-Simons theory on $\sm_3$. A basis of solutions $Z^{(\a)}_{CS}(u)$ can be found using semiclassical, WKB methods \cite{analcs,knots,DGLZ}:
\be
Z^{(\a)}_{CS}(u)=\exp\lt[\frac{i}{\hbar}\int_{\Fc\subset \cl_{\mathbf{A}}}^{u,v{(\a)}}\vth +\cdots\rt]\label{expZCS}.
\ee
The leading term is completely determined by the classical phase space and a Lagrangian subvariety within it. Here the Liouville 1-form $\vth$ satisfies $\rmd \vth =\o_{CS}$ and is integrated along a contour $\Fc$ in the Lagrangian subvariety $\cl_{\mathbf{A}}$. The logarithmic coordinates $u_m$ and $v_m$ are related to $x_m,y_m$ by $x_{m}=e^{u_m}$ and $y_{m}=e^{-\frac{2\pi}{t} v_{m}}$. The label $\a$ indexes the branches of $\cl_{\mathbf{A}}$. On each of these branches the defining equation of the subvariety $\mathbf{A}_m({x},{y})=0$ can be solved to give a unique set of $v_m$ as functions of the $u_m$. The end point of the contour $\Fc$, which labels $Z^{(\a)}_{CS}(u)$, is a flat connection determined by $u,v{(\a)}$ in $\cm_{\mathrm{flat}}(\sm_3,\Slc)\simeq\cl_{\mathbf{A}}$ (or more precisely, in the cover space of $\cm_{\mathrm{flat}}(\sm_3,\Slc)\simeq\cl_{\mathbf{A}}$). Thus each $Z^{(\a)}_{CS}(u)$  is associated to a unique flat connection $A\in\cm_{\mathrm{flat}}(\sm_3,\Slc)$. The starting point of $\Fc$ is conventional and corresponds to a choice of overall phase for $Z^{(\a)}_{CS}(u)$. The ellipsis ``$\cdots$'' in Eq.~\eqref{expZCS} stands for the quantum corrections, which in principle can be obtained recursively from the operator constraint equations. The semiclassical wave function $Z^{(\a)}_{CS}(u)$, often called an \emph{holomorphic 3d block}, can also be formulated nonperturbatively as a ``state-integral model,'' see \cite{knots,andersen}. 

The holomorphic 3d block $Z^{(\a)}_{CS}(u)$ can also be defined by a functional integral of the holomorphic part of ${CS}\lt[\sm_3\,|\,A,\bar{A}\rt]$ over a certain integration cycle, known as a Lefschetz thimble \cite{analcs}. The Lefschetz thimble is an integration cycle that only contains a single critical point of the action; this provides another way to understand the association between $Z^{(\a)}_{CS}(u)$ and a single flat connection on $\sm_3$.

The holomorphic 3d block $Z^{(\a)}_{CS}(u)$ plays a central role in the quantum part of this work. We again specialize to the 3-manifold $S^3\setminus\G_5$ with boundary $\Sig_6$ and impose boundary conditions on $\Sig_6$ to pick out the flat connections in $\cm_{\mathrm{flat}}(S^3\setminus\G_5,\Slc)$ corresponding to constant curvature 4-simplices. Given such an $A\in\cm_{\mathrm{flat}}(S^3\setminus\G_5,\Slc)$, as well as its parity partner $\tilde{A}$, we can construct an holomorphic 3d block $Z^{(\a)}_{CS}(u)$ associated with $A$ and using $\tilde{A}$ as a reference. We simply let $A$ be the end point of the contour $\Fc$ and use $\tilde{A}$ as its initial point. Our classical correspondence between flat connections on $S^3\setminus\G_5$ and constant curvature 4-simplex geometries suggests that the so constructed $Z^{(\a)}_{CS}(u)$ is a wave function for the \emph{quantum 4d geometry} of a constant curvature 4-simplex. Schematically,
\be
Z^{(\a)}_{CS}(u)\ \text{with boundary conditions}\ =\ \text{quantum constant curvature 4-simplex geometry}.\label{Qrelation4simplex}
\ee
This quantum correspondence indicates that the asymptotic expansion of $Z^{(\a)}_{CS}(u)$ in Eq.\Ref{expZCS} should have the  classical action for the simplicial 4d geometry as its leading term. In particular, due to the relation between the symplectic structures of flat connections and 4d simplicial gravity, it is natural to expect that the leading term should give the action of 4d gravity in the simplicial context. 

This expectation is confirmed by the analysis in Section \ref{3dblock}. We show that the leading asymptotic behavior of $Z^{(\a)}_{CS}(u)$ is a simplicial discretization of the four-dimensional Einstein-Hilbert action on a constant curvature 4-simplex
\be
S^\L_{Regge}=\sum_{a<b}{\bf a}_{ab}\Theta_{ab}-\L \mathrm{Vol}_4^\L,\label{Rgaction}
\ee
we call this the curved Regge action, and it is expressed here up to an integration constant and a term depending on the lift to the logarithmic variables $(u,v)$. The coefficient $\L$ is the cosmological constant and can also be identified as the constant curvature of the 4-simplex, while $\mathrm{Vol}_4^\L$ is its 4-volume. We refer the reader to, e.g. \cite{curvedRegge,Regge,foxon,FFLR}, for the derivation of the curved 4d Regge action through a discretization of the Einstein-Hilbert action (see also \cite{HHR} for a summary). 

Because $Z^{(\a)}_{CS}(u)$ is holomorphic, its leading asymptotic behavior is not necessarily an oscillatory phase. In studying the full $\Slc$ Chern-Simons action ${CS}\lt[\sm_3\,|\,A,\bar{A}\rt]$, including both holomorphic and anti-holomorphic parts, we are interested in the 3d block $Z^{(\a)}_{CS}(u)Z^{(\overline{\a})}_{CS}(\bar{u})$, where $Z^{(\overline{\a})}_{CS}(\bar{u})$ is associated to $\bar{A}$. For a flat connection with corresponding 4-simplex geometry, the leading asymptotic behavior of $Z^{(\a)}_{CS}(u)Z^{(\overline{\a})}_{CS}(\bar{u})$ is an oscillatory phase:
\be
\label{simplexlevel}
Z_{CS}^{(\a)}\lt(u\rt)Z_{CS}^{(\overline{\a})}\lt(\bar{u}\rt)=\exp\lt[\frac{i}{\hbar}2\mathrm{Re}\lt(\frac{\L t}{12\pi i}\rt)S^\L_{Regge}+\frac{i}{\hbar}2\mathrm{Re}\lt(\frac{\L t}{6}\rt)\sum_{a<b} \mathbf{N}_{ab}\mathbf{a}_{ab}+i\mathrm{C}_{\mathrm{int}}+\cdots\rt].
\ee
This is shown in section \ref{quantum}. Thus we see that $Z^{(\a)}_{CS}(u)Z^{(\overline{\a})}_{CS}(\bar{u})$ is an analog of the functional integral quantization of the Einstein-Hilbert action in the simplicial context,
\be
Z_{\text{EH}}(\Fm_4) = \exp\lt[\frac{i}{2 \ell_P^2} \int_{\Fm_4} R-2\L+\text{``Quantum Corrections''}\rt].
\ee
With this analogy in mind, we identify the gravitational constant $G_{\mathrm{N}}$ in terms of Chern-Simons coupling $t$ and cosmological constant $\L$ as 
\be
\label{GNewton}
G_{\mathrm{N}}=\lt|\frac{3}{2 \mathrm{Im}(t)\L}\rt|.
\ee
The quantity $\mathrm{C}_{\mathrm{int}}\in\R$ in \Ref{simplexlevel} is an (integration) constant that is independent of the 4-simplex geometry. The additional term $\frac{i}{\hbar}2\mathrm{Re}\lt(\frac{\L t}{6}\rt)\sum_{a<b} {N}_{ab}\mathbf{a}_{ab}$ $({N}_{ab}\in\Z)$ in the leading asymptotics comes from the choice of lift of the FN variables $x_m$ and $y_m$ to the logarithmic variables $u_m$ and $v_m$. This term disappears trivially when $t\in i\R$. However, for general complex $t$, the additional term can also be made to disappear by imposing a quantization condition on the triangle areas ${\bf a}_{ab}$:
\be
2\mathrm{Re}\lt(\frac{\L t}{6}\rt)\sum_{a<b} {N}_{ab} \mathbf{a}_{ab}\in 2\pi\hbar\Z. \label{Qcondition0}
\ee
Indeed, this quantization condition is natural: when the boundary conditions on $A\in\cm_{\mathrm{flat}}(S^3\setminus\G_5,\Slc)$ are imposed using a Wilson graph operator, the quantization condition is automatically satisfied (see Section \ref{wilson}). The quantization condition is also consistent with the discrete area spectrum in Loop Quantum Gravity (LQG) \cite{RS,Hal}.  

The bulk of this paper is devoted to the flat-connection-to-geometry correspondences at the single 4-simplex level because this is the most crucial step in building models for more general situations. The analysis is generalized, in Section \ref{beyond}, to a simplicial complex with an arbitrary number of 4-simplices.  In the resulting simplicial geometry, the 4-simplices are of constant curvature $\L$, while the large simplicial geometry built by many 4-simplices can approximate an arbitrary smooth geometry on a 4-manifold. However, a generic $A\in\cm_{\mathrm{flat}}(\sm_3,\Slc)$ that corresponds to a 4d simplicial geometry may result in a non-uniform 4d orientation throughout the simplicial complex, that is, different 4-simplices may obtain different 4d orientations. For an orientable simplicial complex  $\ck_4$, we find the class of flat connections on $\sm_3$ that not only determine all possible (nondegenerate) 4-dimensional simplicial geometries, but also induce consistent global 4d orientations. Each flat connection $A$ in the class is accompanied by its \emph{global parity} partner $\tilde{A}$. We construct the Chern-Simons 3d block $Z_{CS}^{(\a)}\lt(\sm_3\big|\,u\rt)Z_{CS}^{(\overline{\a})}\lt(\sm_3\big|\,\bar{u}\rt)$ associated with $A$ (and reference $\tilde{A}$) in the same way as above. The asymptotic expansion in $\hbar$ of the resulting 3d block generalizes Eq. \Ref{simplexlevel} to the level of a simplicial complex:
\be
Z_{CS}^{(\a)}\lt(\sm_3\big|\,u\rt)Z_{CS}^{(\overline{\a})}\lt(\sm_3\big|\,\bar{u}\rt) =\exp\lt[\frac{i}{\hbar}2\mathrm{Re}\lt(\frac{\L t}{12\pi i}\rt)\mathbf{S}_{Regge}^\L+\frac{i}{\hbar}2\mathrm{Re}\lt(\frac{\L t}{6}\rt)\sum_{\Delta} \mathbf{N}_{\Delta}\mathbf{a}_{\Delta}+i\mathbf{C}_{\mathrm{int}}+\cdots\rt], \label{complexlevel}
\ee
where $\mathbf{S}_{Regge}^\L$ is the 4-dimensional Lorentzian Regge action on the entire simplicial complex $\ck_4$: 
\be
\mathbf{S}_{Regge}^\L=\sum_{\Delta\ \text{internal}} \mathbf{a}({\Delta}){\eps}({\Delta})+\sum_{\Delta\ \text{boundary}} \mathbf{a}({\Delta})\Theta({\Delta})-\L \sum_\sig \mathrm{Vol}^\L_4(\sig).
\ee
Here $\Delta$ denotes a triangle in $\ck_4$ and $\sig$ denotes a 4-simplex. If we denote the hyperdihedral boost angle of $\Delta$ in the 4-simplex $\sig$ by $\Theta_\Delta(\sig)$ (the same as $\Theta_{ab}$ above), then ${\eps}({\Delta})$ is the Lorentzian deficit angle defined by ${\eps}({\Delta}):=\sum_{\sig,\Delta\subset\sig}\Theta_\Delta(\sig)$ for $\Delta$ an internal triangle, and ${\Theta}({\Delta})$ is the Lorentzian boundary hyperdihedral angle defined by ${\Theta}({\Delta}):=\sum_{\sig,\Delta\subset\sig}\Theta_\Delta(\sig)$ for $\Delta$ a boundary triangle. In Eq. \Ref{complexlevel} the additional term $\frac{i}{\hbar}2\mathrm{Re}\lt(\frac{\L t}{12}\rt)\sum_{\Delta}\mathbf{N}_{\Delta}\mathbf{a}_{\Delta}$ ($\mathbf{N}_{\Delta}\in\Z$) again disappears when $t\in i\R$, or when the quantization condition Eq. \Ref{Qcondition0} is satisfied, for general $t$.

This asymptotic expansion in $\hbar$ suggests that the Chern-Simons 3d block $Z_{CS}^{(\a)}\lt(u\rt)Z_{CS}^{(\overline{\a})}\lt(\bar{u}\rt)$, which associates with a flat connection on $\sm_3$ a corresponding 4d simplicial geometry on $\ck_4$, is a wave function for 4-dimensional simplicial quantum gravity; its subleading terms in $\hbar$ should give the quantum corrections to the classical Einstein-Hilbert action.

\subsection{Wilson Graph Operator and Loop Quantum Gravity}

The analysis in the present paper is a continuation of the work done in \cite{HHR}, where a class of Wilson graph operators are studied in $\Slc$ Chern-Simons theory on $S^3$. The Wilson graph operators are defined by a $\G_5$ graph embedded in $S^3$ colored by certain principle unitary irreducible representations of $\Slc$. The definition is summarized in Section \ref{wilson}. In \cite{HHR}, we have studied the Chern-Simons expectation value $\sa$ of the Wilson graph operators on $S^3$, and in particular the asymptotic behavior of $\sa$ in the ``double-scaling limit'', that is, when both the Chern-Simons coupling $t$ and the Wilson-graph representation labels are scaled to infinity, but their ratio is kept fixed. In this double-scaling limit, the Chern-Simons expectation value $\sa$ of the Wilson graph operator again yields the 4d Regge action $S_{Regge}^\L$ of a constant curvature 4-simplex as its leading asymptotics,
\be
\A = e^{\frac{i}{\ell_P^2}S_{Regge}^\L+\ \cdots}+e^{-\frac{i}{\ell_P^2}S_{Regge}^\L+\ \cdots}
\ee
up to a possible overall phase factor. Here the ellipsis ``$\cdots$" represent the subleading terms in the double-scaling limit. These asymptotics and their relation with simplicial gravity suggest that $\sa$ can be viewed as a 4d gravity analog of the quantum $6j$-symbol in the Turaev-Viro model of 3d quantum gravity \cite{TV,q6j}.\footnote{The double-scaling limit of quantum $6j$-symbol gives the 3d Regge action on a constant curvature tetrahedron \cite{q6j}. }

The Chern-Simons expectation value $\sa$ has a close relationship with Loop Quantum Gravity (LQG). LQG is an attempt to make a background independent, non-perturbative, quantization of 4-dimensional gravity; for reviews, see \cite{book,rev,sfrevs}. The central objects in the covariant dynamics of LQG, which adapts the idea of path integral quantization to the framework of LQG, are the \emph{spinfoam amplitudes}. A spinfoam amplitude is defined on a 4-dimensional simplicial complex $\ck_4$ and encodes the transition amplitude for a given boundary quantum 3-geometry. In LQG, the quantum 3-geometries are described by \emph{spin-network states}. A spinfoam amplitude sums over the history of spin-networks, and suggests a foam-like quantum spacetime structure. An important building block for a general spinfoam amplitude is the Engle-Pereira-Rovelli-Livine (EPRL) partial 4-simplex amplitude $\sa_{EPRL}$ associated to a 4-simplex $\sig$ in $\ck_4$ \cite{EPRL}.\footnote{It is also called the EPRL/FK amplitude, including Freidel and Krasnov, when referring to the version for Euclidean gravity \cite{FK}.} The Chern-Simons expectation value $\sa$ of the Wilson graph is a deformation of the EPRL 4-simplex amplitude, in the sense that $\sa$ approaches $\sa_{EPRL}$ asymptotically in the decoupling limit, that is, when the Chern-Simons coupling $t$ is scaled to infinity keeping the Wilson graph representation labels fixed (see \cite{HHR} or Section \ref{LQG}). This deformation is largely motivated by two streams of research: (1) studies of the relation between LQG and Topological Quantum Field Theory \cite{smolin,kodama,wolfgang} and (2) the quantum group deformation of spinfoam amplitudes that  include a cosmological constant \cite{QSF,QSFasymptotics}. We have the following commutative diagram for the relations among $\sa$, $\sa_{EPRL}$, and the Regge action $S_{Regge}^\L$ (or $S_{Regge}$) with (or without) a cosmological constant term:
\be
\A \ \ \ \ \ \ \ \ \ \ \ &\ \ \  \ \stackrel{\text{double-scaling limit}}{\longrightarrow}\ \ \ \ &\ e^{\frac{i}{\ell_P^2}S_{Regge}^\L}+e^{-\frac{i}{\ell_P^2}S_{Regge}^\L}\nonumber\\
&& \nonumber\\
\Big\downarrow \stackrel{\text{decoupling}}\ &&\ \ \ \ \ \Big\downarrow \stackrel{\L\to 0}\nonumber\\
&& \nonumber\\
\A_{EPRL} \ \ \ \ \ \ \ \  &\ \ \ \ \stackrel{\text{large-j limit}}{\longrightarrow}\ \ \ \ &\ e^{\frac{i}{\ell_P^2}S_{Regge}}+e^{-\frac{i}{\ell_P^2}S_{Regge}}
\ee
The asymptotic behavior on the bottom line has been established for the EPRL 4-simplex amplitude $\sa_{EPRL}$ in \cite{semiclassical,HZ}. The action $S_{Regge}$ that results from the asymptotic analysis of $\sa_{EPRL}$ is the Regge action without cosmological constant for a \emph{flat} 4-simplex, while the action $S^\L_{Regge}$ that comes out of the Chern-Simons expectation value $\sa$ is the Regge action with cosmological constant $\L$ for a \emph{constant curvature} 4-simplex, that of Eq. \Ref{Rgaction}. In this sense $\sa$ is a deformation of the spinfoam amplitude $\sa_{EPRL}$ that includes the cosmological constant in the framework of LQG.

The 4-dimensional Lorentzian Regge action $S^\L_{Regge}$ appears in both the leading asymptotics of the Chern-Simons expectation value $\sa$ of the Wilson graph operator and in the Chern-Simons 3d block $Z_{CS}^{(\a)}\lt(u\rt)Z_{CS}^{(\overline{\a})}\lt(\bar{u}\rt)$.
This is not a coincidence (see Section \ref{wilson}). Firstly it turns out that the double-scaling limit of Chern-Simons theory on $S^3$ with a Wilson graph insertion is the same as the semiclassical limit $\hbar\to0$ of Chern-Simons theory on the graph complement, keeping the boundary data fixed. Secondly the Chern-Simons expectation value $\sa$ can be understood as an inner product 
\be
\sa=\lag N(\G_5)\, \big|\, S^3\setminus\G_5\rag,
\ee
where $|N(\G_5)\rangle$ is the Chern-Simons state on the tubular neighborhood of $\G_5$ excited by the Wilson graph operator, and $| S^3\setminus\G_5\rangle$ is the Chern-Simons ground state on $S^3\setminus\G_5$. In the double-scaling limit, the Wilson graph operators in \cite{HHR} that define $\sa$ impose the right boundary conditions on the boundary $\Sig_6$ of $S^3\setminus\G_5$ (including the quantization condition Eq. \Ref{Qcondition0}). Right in the sense that these boundary conditions pick out the parity pair of flat connections $A$ \& $\tilde{A}$ on $S^3\setminus\G_5$ and determine a constant curvature 4-simplex geometry. In other words, the state $|N(\G_5)\rangle$ is a ``semiclassical state'' peaked at the right phase space point in $\cm_{\mathrm{flat}}(\Sig_6,\Slc)$. The state $| S^3\setminus\G_5\rangle$ is a linear combination of Chern-Simons 3d blocks $Z_{CS}^{(\a)}\lt(u\rt)Z_{CS}^{(\overline{\a})}\lt(\bar{u}\rt)$ on $S^3\setminus\G_5$. The peakedness of $|N(\G_5)\rangle$ selects the pair of 3d blocks that associate to $A $ and $\tilde{A}$ respectively, and which have respectively $e^{\frac{i}{\ell_P^2}S_{Regge}^\L}$ and $e^{-\frac{i}{\ell_P^2}S_{Regge}^\L}$ in their leading asymptotics. 

Separate study of the Chern-Simons 3d block and the Wilson graph operator clarify the different roles they play in the asymptotics of $\sa$. The Regge-action asymptotic behavior of $\sa$ crucially depends on the peakedness of $|N(\G_5)\rangle$ created by the Wilson graph operator. However, different Wilson graph operators can produce the same peakedness in the phase space,\footnote{For instance, for an harmonic oscillator, different squeezed coherent state can have the same peakedness.} and thus lead to the same asymptotics of $\sa$. The close relationship with the EPRL 4-simplex amplitude motivates us to study the particular type of Wilson graph operators in \cite{HHR}. In principle other types of Wilson graph operators could work equally well, as long as they produce the same peaking.\footnote{The different types of Wilson graphs having the same peakedness may relate to the spinfoam amplitudes defined in \cite{DingYou}. } However, independent of the choice of Wilson graph, the essential ingredient leading to the Regge-action asymptotics of $\sa$ is the Chern-Simons 3d block on $S^3\setminus\G_5$ with the right boundary conditions imposed. This means that the Chern-Simons 3d block $Z_{CS}^{(\a)}\lt(u\rt)Z_{CS}^{(\overline{\a})}\lt(\bar{u}\rt)$ studied here plays an important role in the covariant formulation of LQG. Both the classical and the quantum correspondences between flat connections on 3-manifolds and simplicial geometries on 4-manifolds studied here may be viewed as a re-formulation of covariant LQG that emphasizes its relationship with $\Slc$ Chern-Simons theory.

In the quantum case, this correspondence suggests that the Chern-Simons 3d block $Z_{CS}^{(\a)}\lt(u\rt)Z_{CS}^{(\overline{\a})}\lt(\bar{u}\rt)$ is the wave function of simplicial quantum gravity in 4 dimensions. Given its relation with LQG, this 3d block may be understood as the physical wave function for LQG in 4 dimensions, at least for simplicial geometries. In future research it will be interesting to find the behavior of $Z_{CS}^{(\a)}\lt(u\rt)Z_{CS}^{(\overline{\a})}\lt(\bar{u}\rt)$ under refinement of the simplicial complex $\ck_4$. This should shed light on the continuum limit in covariant LQG.

The physical wave function of LQG describes quantum transitions in a 4-dimensional region that go between boundary quantum 3d geometries. In this logic, the boundary data of $Z_{CS}^{(\a)}\lt(u\rt)Z_{CS}^{(\overline{\a})}\lt(\bar{u}\rt)$, namely the flat connections on the 2d boundary of the graph complement 3-manifold, should describe the quantum 3d geometry in LQG. Indeed, as discussed in Section \ref{LQG}, the boundary data of $Z_{CS}^{(\a)}\lt(u\rt)Z_{CS}^{(\overline{\a})}\lt(\bar{u}\rt)$ relate naturally to spin-network states, which quantize 3d geometry in the kinematical framework of LQG.

\subsection{Structure of the Paper}
\label{structure}

The structure of the paper is as follows: Section \ref{classical} explains the classical correspondence between the $\Slc$ flat connections on $S^3\setminus\G_5$ specified by certain boundary conditions and constant curvature 4-simplex geometries in 4 dimensions. Section \ref{quantum} discusses the correspondence between quantum $\Slc$ Chern-Simons theory on $S^3\setminus\G_5$ and quantum 4-simplex geometry. After a brief review of quantum Chern-Simons theory, Fenchel-Nielsen coordinates, and the holomorphic 3d block in Sections \ref{FNCoords} and \ref{reviewCS}, we analyze, in Section \ref{3dblock}, the asymptotic expansion of the Chern-Simons 3d block. The leading order asymptotics of this block gives the 4-dimensional Regge action on a constant curvature 4-simplex and includes a cosmological constant. Section \ref{wilson} discusses the relation with \cite{HHR}, in which Wilson graph operators were used to impose the correct boundary conditions. Section \ref{LQG} treats the relationship between $\Slc$ Chern-Simons theory and Loop Quantum Gravity in 4 dimensions. In Section \ref{beyond}, the correspondence is generalized from a single 4-simplex to a 4d simplicial complex. A particular class of Chern-Simons 3d block is defined and the asymptotics of this block gives the 4d Regge action on a full simplicial complex. The two Appendices go deeper into the mathematical structure of (\ref{Ktheory}) $K_2$-Lagrangian subvarieties, alluded to in Section \ref{quantum}, and ( \ref{coadj}) of the coadjoint orbit quantization used in Section \ref{wilson}.

\section{From Flat Connections on a 3-Manifold to 4d Simplicial Geometry}\label{classical}

In this section we explain the classical correspondence between $\Slc$ flat connections on $\mathrm S^3 \setminus \G_5$ subject to a certain set of boundary conditions and constant curvature 4-simplices in four-dimensions. In order to explain the boundary conditions that allow us to achieve this correspondence, we begin in subsection \ref{conn} by explaining the Wirtinger algorithm for generating the fundamental group of a graph complement manifold. The desired boundary conditions on the flat connection are most easily expressed in terms of the generators of this fundamental group and are made explicit in subsection \ref{b.c.}. This section concludes by connecting these boundary conditions to our previous  work on constant curvature tetrahedra \cite{HHKR} and hence establishes that these boundary conditions allow the reconstruction of geometrical constant curvature tetrahedra around each of the vertices of the graph. In subsection \ref{3/4} these tetrahedral pieces are assembled into the full geometry of a constant curvature 4-simplex. This section also provides a commutative diagrams that helps explicate how such a correspondence is possible in abstract terms. Finally subsection \ref{PP} explains some discrete symmetries of the reconstructed geometries that will be useful in what follows. 

\subsection{Flat Connections on a Graph-Complement 3-Manifold}\label{conn}

Consider the embedding of the pentagon graph $\G_5$ in a 3-sphere $\mathrm S^3$, Figure \ref{gamma5}, and let $N(\G_5)$ be (the interior of) its tubular neighborhood. Define the 3-manifold $M_3 : = \mathrm S^3 \setminus N(\G_5)$, which has boundary $\partial M_3 =  \overline{\partial N(\G_5)}$. With a slight abuse of notation we will often write 
\be
M_3 = \mathrm S^3\setminus \G_5.
\ee

The moduli space of flat $\slc$ connections on $M_3$ is defined as
\be
\cm_\text{flat}\lt( M_3,\Slc\rt) = \mathrm{Hom}\lt(\pi_1(M_3),\Slc\rt)\big/\text{conjugation},
\ee
i.e. as the space of representations $\rho$ of the fundamental group of $M_3$ in the group $\Slc$, up to conjugation. 
As defined above, the moduli space of flat connections is often badly behaved, e.g. it is non-Hausdorff. It is customary---and enough for our purposes---to make a further restriction to the so-called `character variety', which is an algebraic variety. For details see \cite{charactervariety}. 

The fundamental group $\pi_1(M)$ of a graph complement $M$ is easily characterized via a generalized Wirtinger presentation \cite{topo}. This construction proceeds in four steps: (\textit{i}) Project the graph onto a plane; (\textit{ii}) Take a point $\ast$ not lying in this plane as base point; (\textit{iii}) Take as generators of $\pi_1(M)$ the independent loops starting and ending at $\ast$. These go around each edge once and cross the plane of the projected graph twice; (\textit{iv}) Every crossing of the initial graph breaks the original undergoing edge into two pieces in the planar projection---the associated loops  $\fl^{(1)}$ and $\fl^{(2)}$ should be considered as independent generators. 

The generators obtained in this manner are required to satisfy the following two sets of relations:

\begin{itemize}
\item When $n$ edges meet at a vertex (all oriented ingoing for the moment), we require
\be
 \fl_n \cdots \fl_2\;\fl_1 = \fe , \qquad \qquad \quad \raisebox{-1.2cm}{\includegraphics[width=2cm]{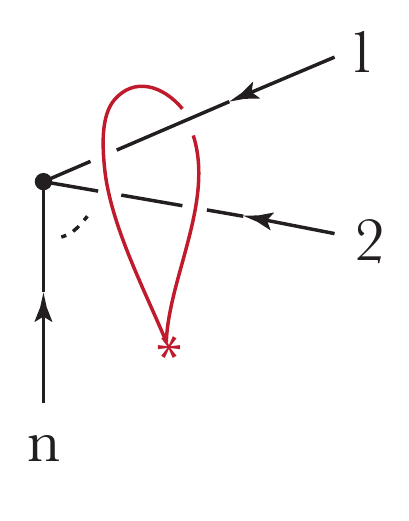}},  \label{rule1}
 \label{Wirt_vertex}
\ee
where $\fe$ denotes the identity in $\pi_1(M)$, and we have supposed them to be numbered from 1 to $n$ in a clockwise fashion on the projection plane. To change the $i$-th edge from ingoing to outgoing, substitute $\fl_i$ with $\fl_i^{-1}$;
\item Upon projection onto the plane, an edge with generator $\tilde \fl$ over-crosses another edge, the latter gets associated with two independent generators $\fl^{(1)}$ and $\fl^{(2)}$ as in point (iv) above, see figure below. These three generators $\tilde\fl$, $\fl^{(1)},$ and $\fl^{(2)}$ are required to satisfy
\be
\fl^{(1)} = \tilde\fl \; \fl^{(2)} \tilde \fl^{-1}, \qquad \qquad \raisebox{-1.2cm}{\includegraphics[width=2.2cm]{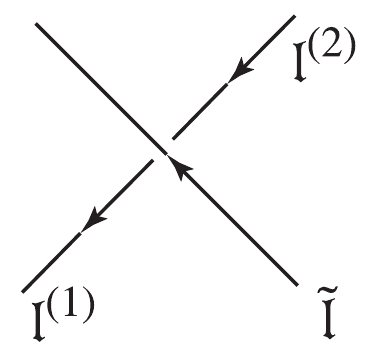}}.\label{rule2}
\label{Wirt_crossing}
\ee
\end{itemize}

Using this algorithm, $\pi_1(M_3=\mathrm S^3\setminus\G_5)$ can be computed in a straightforward manner; this is the task we take up now. To fix notation, label the vertices of $\G_5$ as in Figure \ref{gamma5} with an index $a\in\{1,\dots,5\}$, and call its (unoriented) edges $\ell_{ab}=\ell_{ba}.$ The generators of $\pi_1(M_3)$ are then the loops $\fl_{ab}$ associated to every edge $\ell_{ab}$ of $\G_5$ except $\ell_{13}$, which is broken by a crossing, and hence is associated to two distinct generators $\fl_{13}^{(1)}$ and $\fl_{13}^{(2)}$. A representation $\rho\in \mathrm{Hom}\lt( \pi_1(M_3), \Slc \rt)$ maps each of these generators to an element of $\Slc$, i.e. $\rho\lt(\fl_{ab}\rt) = \tilde H_{ab}$ for every $(ab)\neq(13)$ and $\rho\lt(\fl_{13}^{(i)}\rt) = \tilde H_{13}^{(i)}$, for $i\in\{1,2\}$. The requirements of Eqs. \eqref{Wirt_vertex} and \eqref{Wirt_crossing}, when expressed in terms of these group elements (holonomies) are:
\begin{subequations}
\be
\text{vertex 1}:&&\tilde{H}_{14}\tilde{H}_{13}^{(1)}\tilde{H}_{12}\tilde{H}_{15}=1,\label{vertex1}\\
\text{vertex 2}:&&\tilde{H}_{12}^{-1}\tilde{H}_{24}^{}\tilde{H}_{23}\tilde{H}_{25}=1,\\
\text{vertex 3}:&&\tilde{H}_{23}^{-1}(\tilde{H}_{13}^{(2)})^{-1}\tilde{H}_{34}\tilde{H}_{35}=1,\\
\text{vertex 4}:&&\tilde{H}_{34}^{-1}\tilde{H}_{24}^{-1}\tilde{H}_{14}^{-1}\tilde{H}_{45}=1,\\
\text{vertex 5}:&&\tilde{H}_{25}^{-1}\tilde{H}_{35}^{-1}\tilde{H}_{45}^{-1}\tilde{H}_{15}^{-1}=1,\label{vertex5}
\ee
\label{vertex1-5}
\end{subequations}

\vspace{-3.4em}\be
\text{crossing}:&&\tilde{H}_{13}^{(1)}=\tilde{H}_{24}\tilde{H}_{13}^{(2)}\tilde{H}_{24}^{-1}.\label{X} \qquad
\ee 
Notice that all the above holonomies, collectively referred to as $\lt\{\tilde H_{ab}\rt\}$, have the same base-point $\ast\in\mathrm S^3\setminus\G_5$.

The moduli space $\mathcal M_\text{flat} \lt( M_3, \Slc \rt)$ is defined as the group elements $\lt\{\tilde H_{ab}\rt\}$ modulo simultaneous conjugation by a $g\in\Slc$, i.e. $\lt\{\tilde H_{ab}\rt\}\sim\lt\{g\tilde H_{ab} g^{-1}\rt \}$.

\subsection{The Boundary Conditions and their Geometrical Interpretation as Curved Tetrahedra}\label{b.c.}

For our geometrical purposes, we are not interested in a generic connection in $\mathcal M_\text{flat} \lt( M_3, \Slc \rt)$. Rather, we want to restrict to connections satisfying a certain type of boundary conditions on the graph complement manifold's boundary 
\be
\Sigma_6 := \partial M_3 =  \overline{\partial N( \G_5)},
\ee
which is a closed 2-surface of genus 6. The restriction of a connection $A\in\mathcal M_\text{flat} \lt( M_3, \Slc \rt)$ to the boundary surface $\Sigma_6$ gives an element of $\mathcal M_\text{flat} \lt( \Sigma_6, \Slc \rt)$. In this sense one can write
\be
\mathcal M_\text{flat} \lt( M_3, \Slc \rt)  \subset   \mathcal M_\text{flat} \lt( \Sigma_6, \Slc \rt).
\ee

On $\Sigma_6$, we specify 10 meridian curves $\{c_{ab}\}$ each cutting one edge of $\G_5$ transversally. Hence,
\be
\Sigma_6 \setminus \{c_{ab}\} = \bigcup_{a=1,\dots,5} \mathcal S_a
\ee
where $\mathcal S_a \cong \mathrm S^2 \setminus\{4\text{pts}\}$ is a four-punctured sphere associated to the $a$-th vertex of $\G_5$. A representation $\sigma\in\mathrm{Hom}\lt(\pi_1(\Sigma_6), \Slc \rt)$ when restricted to $\mathcal S_a$ gives a representation $\sigma|_{\mathcal S_a}\in\mathrm{Hom}\lt(\pi_1(\mathcal S_a), \Slc \rt)$ (defined up to global $\Slc$ conjugation). We think of these punctured spheres as (the boundaries of) tetrahedra whose `quanta of area' are `concentrated' at the punctures in the form of defects. We want each of these tetrahedra to define a three-dimensional space-like frame in (A)dS.

With this geometrical picture in mind we define the following boundary conditions: a representation $\sigma\in\mathrm{Hom} \lt( \pi_1(\Sigma_6), \Slc \rt)$ is said to satisfy  \textit{geometric boundary conditions} if there exists five elements $g_a \in \Slc$, such that
\be
g_a \lt(\sigma|_{\mathcal S_a}\rt) g_a^{-1} \in \mathrm{Hom}\lt( \pi_1(\mathcal S_a), \Su\rt).
\ee
In words, an $\Slc$ representation of the fundamental group of $\Sigma_6$ is said to satisfy the geometric boundary conditions if on each four-punctured sphere $\mathcal S_a$ it restricts to an $\Su$ representation up to conjugation by an element $g_a\in\Slc$:
\be
\forall a \;\exists g_a \in\Slc \quad \text{such that} \quad g_a \tilde H_{ab} g_a^{-1} =: H_{b}(a) \in \mathrm{SU}(2) \quad \forall b, \; b\neq a.
\label{defHols}
\ee
We call the gauge associated to such a set of $\{g_a\}$, the `time gauge'.

An immediate consequence of the geometric boundary conditions is that Eqs. \eqref{vertex1-5} can be written after conjugation by $g_a\in\Slc$ as equations in $\Su$:
\begin{subequations}
\be
\text{vertex 1}:&&{H}_{4}(1){H}_{3}(1){H}_{2}(1){H}_{5}(1)=1,\label{vertex1}\\
\text{vertex 2}:&&{H}_{1}^{-1}(2){H}_{4}^{}(2){H}_{3}(2){H}_{5}(2)=1,\\
\text{vertex 3}:&&{H}_{2}^{-1}(3){H}_{1}^{-1}(3){H}_{4}(3){H}_{5}(3)=1,\\
\text{vertex 4}:&&{H}_{3}^{-1}(4){H}_{2}^{-1}(4){H}_{1}^{-1}(4){H}_{5}(4)=1,\\
\text{vertex 5}:&&{H}_{1}^{-1}(5){H}_{2}^{-1}(5){H}_{3}^{-1}(5){H}_{4}^{-1}(5)=1,\label{vertex5}
\ee
\label{vertex1-5su2}
\end{subequations}
where again the argument of the parentheses indicates the vertex where the holonomy is based, see Eq. \eqref{defHols}. 
We will refer to these equations as the `closure equations'.

The missing information, with respect to Eqs. \eqref{vertex1-5} and \eqref{X}, can be encoded in terms of a $G_{ab}\in\Slc$ defined by
\be
G_{ba} := g_b^{-1} g_a \quad \text{for all $(ab)$, except} \quad G_{13} := g_1^{-1}\left[ g_2 H_{4}(2) g_2^{-1} \rt] g_3 .
\ee
This information can be interpreted as a set of `parallel transport equations' encoding the relation $\tilde H_{ab} = \tilde H_{ba}$ through
\be
G_{ab} H_b(a) G_{ba} = H_{a}(b),
\ee
and as a set of `bulk equations' encoding the position of the crossing
\begin{subequations}
\be
G_{ac}G_{cb}G_{ba} && = {1} \quad (abc)\in\{125,235,345,124,234\},\\
G_{13}G_{32}G_{21} && = H_{4}(2).
\ee
\end{subequations}
Connections satisfying the geometric boundary conditions are denoted 
\be
\mathcal M^\text{BC}_\text{flat}(\Sigma_6, \Slc ) \subset \mathcal M_\text{flat}(\Sigma_6, \Slc ).
\ee

In section \ref{3dblock}, we will come back to these boundary conditions and express them in terms of a set of preferred coordinates, the complex Fenchel-Nielsen coordinates. These are Darboux coordinates on $\mathcal M_\text{flat}(\Sigma_6, \Slc )$ with respect to the canonical Atiyah-Bott-Goldman symplectic structure induced by the Chern-Simons theory.

As anticipated above, there is a precise correspondence between $\Su$ flat connections on a four-holed sphere and tetrahedral geometries flatly embedded in $\mathrm S^3$ and $\mathrm H^3$. This result was proved and discussed in detail in \cite{HHKR,HHR}, and hence in this paper we will limit ourselves to a brief account of this geometry before connecting it with the boundary conditions just discussed.

\begin{Theorem}\label{2d/3d}
There is a bijection between flat connections in $\cm_{\mathrm{flat}}\lt(\cs_a,\mathrm{PSU}(2)\rt)$ and the convex constant curvature tetrahedron geometries in 3d, excepting degenerate geometries. Non-degenerate tetrahedral geometries are dense in $\cm_{\mathrm{flat}}\lt(\cs_a,\mathrm{PSU(2)}\rt)$.

The correspondence applies to both spherical and hyperbolic tetrahedra. Both positive and negative constant curvature geometries are included in $\cm_{\mathrm{flat}}\lt(\cs_a,\mathrm{PSU(2)}\rt)$.
\end{Theorem}

The theorem is primarily built on two observations: (\textit{i}) the fundamental group of the four-holed two-sphere is isomorphic to that of a tetrahedron's one-skeleton, and both are defined by a closure constraint; and (\textit{ii}) in the flat case, a tetrahedron's geometry can be fully reconstructed from four vectors that add up to zero, once these vectors' directions are interpreted as the tetrahedron's face normals and their magnitudes as the respective face areas. Observations (\textit{i}) and (\textit{ii}) are related: the spin-connection holonomy around the boundary of a surface that is flatly-embedded in a homogeneous space contains information about both the area and the orientation of the surface. This means that the curved-space closure constraint, e.g. any of the Eqs. \eqref{vertex1-5su2}, could be a sound generalization of (\textit{ii}).

Observation (\textit{ii}) is a special case of a more general classic result due to H. Minkowski \cite{Minkowski}, known as Minkowski's theorem. This theorem states that an $N$-tuple of vectors that sum to zero corresponds to the set of face vectors of a unique convex polyhedron with $N$ faces. The convexity hypothesis, which primarily guarantees the uniqueness in the case of flat tetrahedra, is particularly crucial in the curved-space generalization of Minkowski's theorem \cite{HHKR}. 

Before proceeding, let us further define what we mean by a flatly embedded simplicial geometry. Take the case of a constant curvature tetrahedron flatly embedded in a unit $\mathrm S^3$.  The zero-simplices (vertices) are 4 points on $\mathrm S^3$. The one-simplices are the \textit{shortest} geodesic arcs connecting 2 zero-simplices. These are given by arcs along great circles in $\mathrm S^3$. Notice that the restriction to the shorter geodesic arc is because we are considering only convex simplices, which will turn out to be crucial for the uniqueness part of the reconstruction theorem. Finally for faces, a triple of vertices identifies uniquely a great 2-sphere in $\mathrm S^3$ and the face is just the convex hull of the three vertices in this two-sphere. As portions of a great two-sphere these surfaces are flatly embedded in $\mathrm S^3$. In particular this means that vectors normal to the surface remain so under parallel transport. Finally, the tetrahedron itself is the convex hull defined by its four faces.

The simplest way to visualize this construction is to consider the unit three-sphere as embedded in one more dimension. Then, the edges of the tetrahedron are defined by intersection of the three-sphere with the unique plane passing through the origin and two of the tetrahedron's vertices. Similarly, the tetrahedron's faces are given by the intersection of the three-sphere and the unique hyperplane passing through the origin and three of the tetrahedron's vertices. This construction makes it obvious how to generalize the definitions to the hyperpolic, higher dimensional, and Lorentzian cases.

Now we would like to relate this geometry to the output of the boundary conditions from above. The idea is to find a relation between the holonomies $U_{ab}$ of the spacetime spin connection $\omega_\mathrm{spin}$ around the faces of a flatly embedded, constant curvature tetrahedron and the $H_b(a)$'s of Eq. \eqref{vertex1-5su2}. To completely define the $U_{ab}$ label the vertices of a four-simplex flatly embedded in (A)dS by $a,b,\dots\in\{1,\dots,5\}$, and the tetrahedron opposite to vertex $a$ by the same label. The triangle shared by tetrahedra $a$ and $b$ is thus labeled by $(ab)$. Call its boundary $\Delta_{ab}$. Make a partial gauge fixing at the base point $O$, such that the tetrad components $e_0^\a=\delta^a_0$ are given by the unit time-like normal to $\Delta_{ab}$, i.e. fix to \emph{time gauge}. Then it is not hard to show that
\be
U_{ab} \equiv U^0_{\partial \Delta_{ab}}(\omega_\mathrm{spin})  = \exp\lt[\frac{\L}{3}\ \mathrm{a}_{ab}\ \hat{\fn}_{ab}\cdot{\vec{\t}}\;\rt]\in\mathrm{SU(2)},
\ee
where $\mathrm{a}_{ab}$ is the area of the triangle $(ab)$, $\hat{\fn}_{ab}$ is its spacelike normal (expressed within the frame $e_{\a}^I$ at $O$), $\vec{\t}$ is a basis for the Lie algebra $\mathfrak{su}(2)$, and $\Lambda\gtrless0$ is the cosmological constant associated with (A)dS. 

A mapping between the $U_{ab}$ and the $H_b(a)$ can now be made explicit. To do this we construct an isomorphism between the fundamental group of the 4-holed sphere $\mathcal S_a$ and that of the $a$-th tetrahedron's 1-skeleton $\tau_a$. It is important to notice that there is no canonical isomorphism. We will come back to this point when dealing with the reconstruction of the 4-simplex geometry. For the moment, we limit our study to a single tetrahedron, say $a=5$, and hence drop the relative label. This allows us to label triangles by their opposite vertex within the tetrahedron.

\begin{figure}[t]
\begin{center}
\includegraphics[width=12cm]{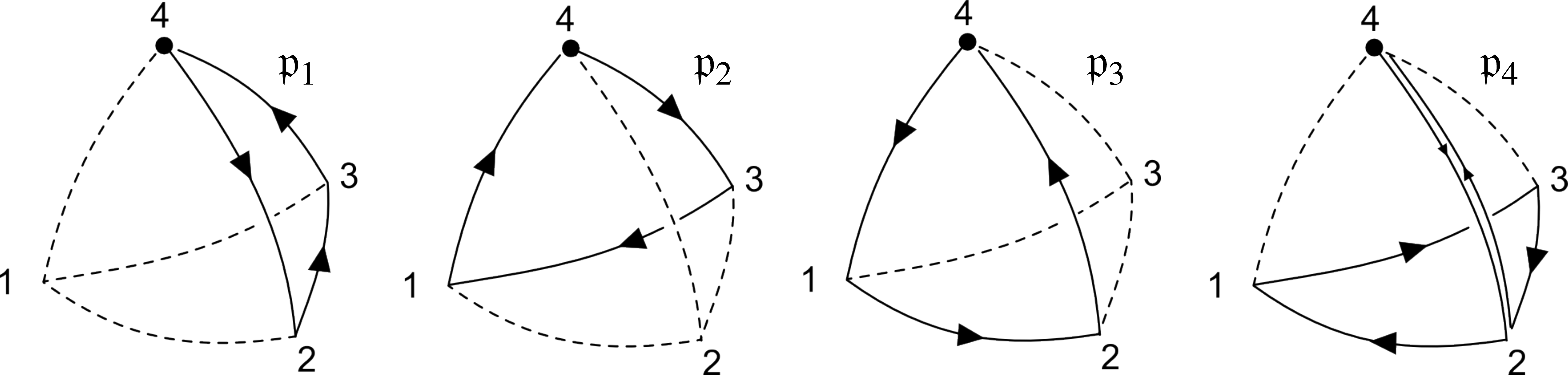}
\caption{The paths $\fp_1,\dots, \fp_4$ used to generate the fundamental group on a constant curvature tetrahedron. The edge $(2,4)$, which is arbitrarily singled out to base the path $\fp_4$ at vertex 4, we call the ``special edge''.} 
\label{paths}
\end{center}
\end{figure}

Denote the to-be-constructed isomorphism between the fundamental groups by $\mathcal{I}_5$, so that $\mathcal I_5 :  \pi_1(\tau_5)  \rightarrow  \pi_1(\mathcal S_5).$ To specify $\mathcal I_5$, consider the basis of $ \pi_1(\tau_5)$ formed by the set of four paths $\{\fp_a\}$ depicted in Figure \ref{paths}. The paths $\fp_a$ go around each face of the tetrahedron once. Hence, we require these paths to be in 1-to-1 correspondence with the $\fl_a$ introduced at the beginning of this section, which circumnavigate each puncture of $\mathcal S_5$ once:
\be
\mathcal{I}_5(\fp_a) = \fl_a.
\ee
This is possible thanks to the fact that both sets of paths satisfy the same defining constraints
\be
\fp_4\fp_3\fp_2\fp_1=\fe \qquad \text{and} \qquad \fl_4\fl_3\fl_2\fl_1=\fe.
\ee

Following \cite{HHKR,HHR}, we call the paths $\{\fp_a\}$, `simple paths'. The name comes from the fact that this is arguably the shortest set of paths satisfying the defining constraint (up to relabeling of the vertices). It is also clear that edge (42) is singled out by this choice of paths (see \cite{HHR} for the explicit role it plays in the reconstruction theorem). We call it the `special edge'. 

Notice that the simple paths are such that the faces are traversed counterclockwise (as seen from the outside of the tetrahedron). Adopting a right-handed convention, this means that the normals to the triangles have to be understood as outgoing.

This isomorphism allows us to interpret the holonomies of a flat connection $\omega_\text{flat}$  on $\mathcal S_5$ as the parallel transports of a spin-connection $\omega_\text{spin}$ on $\tau_5$: 
\be
&\pi_1(\cs_5)\ \ \ \ \ \ \ \ \ \ \ \ \ \ \ \ \ \stackrel{\mathcal I_5}{\longleftarrow}\ \ \ \ \ \ \ \ \ \ \ \ \ \ \ \ \ \pi_1(\tau_5)&\nonumber\\
&&\nonumber\\
&\o_{\mathrm{flat}}\searrow\ \ \ \ \ \ \ \ \ \ \ \ \ \ \ \ \ \ \ \ \ \ \ \ \ \ \swarrow\o_{\mathrm{spin}}&\nonumber\\
&&\nonumber\\
&\lag\,U_1,\cdots,U_4\in\mathrm{SU(2)}^{\otimes 4}\,\big|\, U_4\cdots U_1=1\,\rag\big/\mathrm{conjugation}&\label{Smap}
\ee 
at least provided we find a canonical lift of $H_a\in\mathrm{PSU}(2)$ to $\Su$. The prescription for the canonical lift is actually provided by the convexity condition, as we now explain.

An element $H\in\mathrm{PSU}(2)$ is given by the equivalence class formed by the following two elements of $\Su$:
\be
\exp \lt[ a \, \hat n \cdot \vec\tau \,   \rt] \; \sim \; - \exp \lt[ a \, \hat n \cdot \vec\tau \,   \rt]  \,=\, \exp \lt[ (2\pi-a) \, (-\hat n) \cdot \vec\tau \,   \rt] 
\ee
for some $a\in[0,2\pi]$ and $\hat n \in \mathrm S^2$. This correspondence suggests that we interpret
\be
\hat n \text{, or }-\hat n, \text{ as  } \text{sgn}(\Lambda)\hat{\fn} 
\quad \text{and} \quad 
a \text{, or }(2\pi-a) \text{ respecitvely, as  } \frac{|\Lambda|}{3}\mathrm{a}.
\ee

Using outward normal conventions set by the simple paths, and the tetrahedron's convexity, one sees that the triple products $\hat\fn_a \cdot (\hat\fn_b \times \hat\fn_c)$, with the labels $\{a,b,c\}$ properly ordered, must all be positive (e.g. at vertex 4, $\hat\fn_1 \cdot (\hat\fn_2 \times \hat\fn_3)>0$).\footnote{Notice that parallel transport of one of the vectors might be needed to make sense of these vector products. This happens when one has to compare the normal relative to face 4 to the others. However, since this is parallel transport of a 3-vector, it makes use of the vector representation of the $H_a$'s, and hence is immune to the ambiguity we are trying to solve here. See \cite{HHR}  for details.}  It is hence clear that the convexity conditions fully determine the lift from $\mathrm{PSU}(2)$ to $\mathrm{SU}(2)$, at least up to a global sign, equal to $\text{sgn}(\Lambda)$. Perhaps surprisingly, this final sign can also be determined from the $\hat n_a$ that we have just calculated. 
To do so, use the $\hat n_a$ to calculate the scalar products $\cos\theta_{ab} \equiv \hat n_a \cdot \hat n_b$.\footnote{Again, in some cases a parallel transport of the normals is needed before taking the scalar product. In this case, using the flat-embedding condition, it is not hard to convince oneself that the only dihedral angle needing a `twisted' formula is $\cos\theta_{24} = \hat n_4 \cdot (\mathbf{H}_3\hat n_2)$. Here, $\mathbf{H}_3\in\mathbf{SO}(3)$ is the vector representation of $H_3$. See \cite{HHR}.} Notice that these quantities are insensitive to the global sign ambiguity associated with $\text{sgn}(\Lambda)$ itself. These scalar products are nothing but the (external) dihedral angles of the tetrahedron. It is a classical result in discrete geometry, that the Gram matrix 
\be
(\text{Gram})_{ab} = - \cos\theta_{ab}
\ee
contains all the information needed to reconstruct the tetrahedron's geometry. In particular
\be
\text{sgn}(\text{det}(\text{Gram})) = \text{sgn}(\Lambda).
\ee

To conclude the proof of the reconstruction theorem, one only needs to prove the consistency of the geometry reconstructed from the Gram matrix and the areas implicitly contained in the original group elements. This can be done for example via a counting argument. Again, for all the details of the proof see \cite{HHR}.

For future reference, we note here the formula interpreting the transverse holonomies $H_{b}(a)$ as the spin-connection holonomies around the face $(ab)$ of the four-simplex:
\be
H_{b}(a) = \exp\left[  \frac{\Lambda}{4} \mathrm{a}_{ab} \hat{\fn}_{ab} \cdot \vec \tau \, \right].
\ee

\subsection{Flat Connections on 3-Manifold and Curved 4-Simplex Geometries}\label{3/4}

This subsection discusses the reconstruction of a full 4-simplex geometry from the flat connections on the graph complement manifold subject to the boundary conditions of subsection \ref{b.c.}. There is little conceptual novelty with respect to the reconstruction of the tetrahedron, although some intriguing subtleties arise, and this subsection can safely be skipped on a first reading after taking a look at Theorem \ref{3d/4d} below. 

\begin{figure}[h]
\begin{center}
\includegraphics[width=4cm]{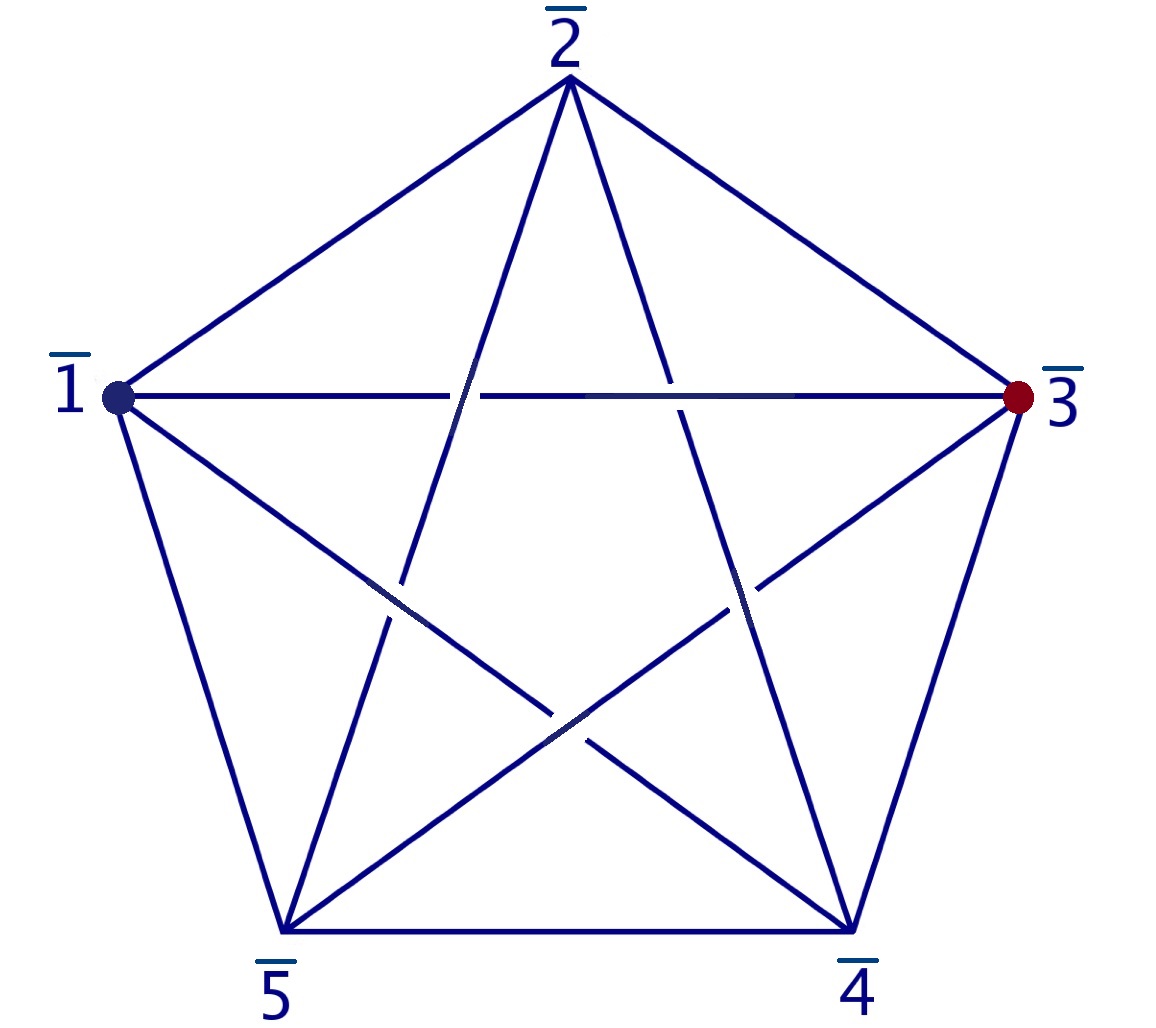}
\caption{An abstract 4-simplex, whose vertices are labeled by $\bar{1},\cdots,\bar{5}$. We denote $\tau_a$ the tetrahedron that does not have the vertex $\bar{a}$. The symbol $\Delta_{ab}$ (resp. $\Delta_{ba}$) denotes the triangle belonging to $\tau_a$ (resp. $\tau_b$) shared by $\tau_a$ and $\tau_b$. The edges are denoted by $(\bar{a},\bar{b})$ oriented from $\bar{b}$ to $\bar{a}$.} 
\label{4simplex}
\end{center}
\end{figure}

Analogously to the discussion surrounding the commuting diagram \Ref{Smap}, we consider the fundamental group for the 1-skeleton of an abstract 4-simplex, see Figure \ref{4simplex}, which we denote by $\pi_1(\sigma_4)$, with $\sigma_4$ denoting the 1-skeleton of the 4-simplex. Closed paths $\fp_{ab}$ along the 1-skeleton and circling each triangle $\Delta_{ab}$ specify a set of generators. A convenient choice of paths, either $\fp_{ab}$ or $\fp_{ab}^{-1}$, is specified by the sets of simple paths for all 5 tetrahedra. All the paths $\fp_{ab}$ can be based at the same point, which we choose to be vertex $1$ of the 4-simplex. 

Explicitly, we choose the paths as follows:
Tetrahedron $\tau_2$ has special edge $(31)$, and its closure relation is\footnote{Note that all the paths $\fp_{21}^{-1},\fp_{24},\fp_{23},$ and $\fp_{25}$ are closed paths circling around a single triangle in a counter-clockwise fashion when viewed from the outside of the tetrahedron. The same holds for Eqs. \Ref{vertex3p} - \Ref{vertex1p}.  } 
\be
\fp_{21}^{-1}\fp_{24}\fp_{23}\fp_{25}=\fe.\label{vertex2p}
\ee 
Tetrahedron $\tau_3$ has special edge $(51)$, and its closure relation is
\be
\fp_{32}^{-1}\fp_{31}^{-1}\fp_{34}\fp_{35}=\fe.\label{vertex3p}
\ee
Tetrahedron $\tau_4$ has special edge $(31)$, and its closure relation is
\be
\fp_{43}^{-1}\fp_{42}^{-1}\fp_{41}^{-1}\fp_{45}=\fe.\label{vertex4p}
\ee
Tetrahedron $\tau_5$ has special edge $(31)$, and its closure relation is
\be
\fp_{52}^{-1}\fp_{53}^{-1}\fp_{54}^{-1}\fp_{51}^{-1}=\fe.\label{vertex5p}
\ee
Tetrahedron $\tau_1$ is the `special tetrahedron,' which is non-adjacent to the base vertex $1$. All the paths associated to $\tau_1$ travel from $1$ to $3$ along $(31)$, then circle around the relevant triangle of $\tau_1$ as in Figure \ref{paths}, and finally go back from ${3}$ to ${1}$ along $(13)$. When we draw the paths on $\tau_1$ starting and ending at ${3}$, the special edge is $({53})$. The closure relation is then 
\be
\fp_{14}\fp_{13}\fp_{12}\fp_{15}=\fe.\label{vertex1p}
\ee

The above list specifies all the (closed) paths $\fp_{ab}$. One can check the following properties: $\fp_{ab}=\fp_{ba}$ for $(a,b)\neq (1,3)$, and for $\fp_{13}$ this relation becomes $[53][31] \fp_{13}[13][35]=[51] \fp_{31} [15]$, where $[ab]$ indicates the path along the edge $ab$. Equivalently, 
\be
\fp_{13}=\fp_{24}\fp_{31} \fp_{24}^{-1}\label{Xp}
\ee
where $\fp_{24}=[13][35][51]$. 

The fundamental group $\pi_1(\sigma_4)$ is generated by the closed paths $\fp_{ab}$ subjected to the set of closure relations Eqs. \Ref{vertex2p} - \Ref{vertex1p} together with the relation \Ref{Xp}. A quick comparison shows that $\pi_1(\sigma_4)$ is isomorphic to $\pi_1(S^3\setminus \G_5)$. In fact, the relations above for the $\fp_{ab}$'s generating $\pi_1(\sigma_4)$ are identical to the relations associated to the generators $\fl_{ab}$ of $\pi_1(S^3\setminus \G_5)$ (see Section \ref{conn}). The isomorphism maps the generators of $\pi_1(S^3\setminus\G_5)$ to the generators of $\pi_1(\mathrm{simplex})$, which delivers the flat connection on $S^3\setminus\G_5$ to the spin connection as a representation of $\pi_1(\mathrm{simplex})$.

If we require that the isomorphism $\mathcal I:\pi_1(\sigma_4) \to \pi_1(S^3\setminus\G_5)$ maps the counterclockwise simple paths (the $\fp_{ab}$) to the loop generators in $\pi_1(S^3\setminus\G_5)$ oriented in a right-handed manner (the $\fl_{ab}$) according to the orientation of the edges $\ell_{ab}\subset \G_5$, then the isomorphism $\mathcal I$ is unique in the following sense:

\begin{Lemma}\label{topounique}

A map $\iota: a\mapsto\tau_a$ identifying a vertex in $\G_5$ with a tetrahedron on the boundary of the 4-simplex, induces an identification between the edges $\ell_{ab}$ of $\G_5$ and the triangles $\Delta_{ab}=\tau_a\cap\tau_b$ of the 4-simplex. Given an isomorphism $\mathcal I: \pi_1(\sigma_4)\to\pi_1(S^3\setminus\G_5)$ such that $\mathcal I(\fp_{ab})=\fl'_{ab}$ is a loop generator in $\pi_1(S^3\setminus\G_5)$ transverse to the edge $\ell_{ab}$ near the vertex $a$, requiring that $\fl'_{ab}$ cycles $\ell_{ab}$ in a right-handed manner according to the orientation of $\ell_{ab}$,\footnote{The orientation condition for $\fl'_{ab}$ corresponds to the counter-clockwise choice for the paths $\fp_{ab}$ or $\fp_{ab}^{-1}$ in Eqs. \Ref{vertex2p} - \Ref{vertex1p}.} the isomorphism $\mathcal I$ is unique. Hence $\mathcal I(\fp_{ab})=\fl'_{ab}=\fl_{ab}$ is the generator for the presentation in Section \ref{conn} associated to the projection of $\G_5$ on a plane, Figure \ref{ribbon}.

\end{Lemma}

\textbf{Proof:} The set of loops $\mathcal I(\fp_{ab})=\fl'_{ab}$, whose common base point could be anywhere in $S^3\setminus\G_5$, can be understood as the generators of a generalized Wirtinger presentation of $\pi_1(S^3\setminus\G_5)$ from a certain projection of $\G_5$ on a plane, which could be different from that of Figure \ref{ribbon}. However, $\fp_{ab}=\fp_{ba}$ implies $\fl'_{ab}=\fl'_{ba}$ for $(a,b)\neq (1,3)$ because of the isomorphism $\mathcal I$. This means that in this projection of $\G_5$, the loops $\fl'_{ab}$ for $(a,b)\neq (1,3)$ can be continuously deformed along the whole edge $\ell_{ab}$ without meeting a crossing. Therefore the crossing only occurs between $\ell_{13}$ and $\ell_{24}$. Then this new projection is either ($a$) as in Figure \ref{ribbon}, with $\ell_{24}$ over-crossing $\ell_{13}$, or (b) as it would appear if Figure \ref{ribbon} was viewed from the back, i.e. with $\ell_{24}$ under-crossing $\ell_{13}$. Without loss of generality, we assume the base point of $\fl'_{ab}$ is in front of the projected graph in both cases (a) and (b). Furthermore the relations Eqs. \Ref{vertex2p} - \Ref{vertex1p} imply the same relations for $\fl'_{ab}$ up to cyclic permutation. These relations for $\fl'_{ab}$ imply that in the case (a), each loop $\fl'_{ab}$ circles $\ell_{ab}$ in a right-handed manner (as in Eq. \Ref{rule1}) with respect to the orientation of $\ell_{ab}$, while in case (b) each loop $\fl'_{ab}$ circles $\ell_{ab}$ in a left-handed manner. Both (a) and (b) imply $\fl'_{13}=\fl'_{24} \fl'_{31} {\fl'}_{24}^{-1}$. However, (b) is ruled out by the requirement that $\fl'_{ab}$ cycles $\ell_{ab}$ in a right-handed manner. Therefore we conclude that the case (a) is singled out, and $\fl'_{ab}=\fl_{ab}$. $\Box$\\

The identification map $\iota: a\mapsto\tau_a$ produces the numbering of the tetrahedra (or vertices) of an abstract 4-simplex using the numbering of the $\G_5$ vertices and the convention that $\tau_a$ labels the tetrahedron \emph{not} containing vertex $\bar{a}$, as in Figure \ref{4simplex}. Given such an identification, we have the following diagram if the 4-simplex is embedded in a geometrical space with spin connection $\o_{\mathrm{spin}}$:
\be
&\pi_1(S^3\setminus\G_5)\ \ \ \ \ \ \ \ \ \ \ \ \ \ \ \ \ \stackrel{\mathcal I}{\longleftarrow}\ \ \ \ \ \ \ \ \ \ \ \ \ \ \ \ \ \pi_1(\sigma_4)&\nonumber\\
&&\nonumber\\
&\o_{\mathrm{flat}}\searrow\ \ \ \ \ \ \ \ \ \ \ \ \ \ \ \ \ \ \ \ \ \ \ \ \ \ \swarrow\o_{\mathrm{spin}}&\nonumber\\
&&\nonumber\\
&\lag\ \{\tilde{H}_{ab}\}\ \big|\ \text{Eqs}.\Ref{vertex1} - \Ref{X}\ \rag\big/\mathrm{conjugation}&\label{Xmap}
\ee 
where the isomorphism $\mathcal I$ is unique in the sense of the previous Lemma. The isomorphism $\mathcal I$ determines by restriction the isomorphisms $\mathcal I_a$ associated to each of the five tetrahedra. This means that the isomorphisms $\mathcal I_a$ in the diagram \Ref{Smap} are unique if embedded in a 4-simplex context. 

The connection $\o_{\mathrm{spin}}$ associates to the set of paths $\fp_{ab}$ the holonomies of an $\Slc$ spin connection:
\be
\o_\text{spin}(\fp_{ab}) = U_{ab}.
\ee
On the other hand, the flat connection representation on $\mathrm S^3\setminus\G_5$ discussed in Section \ref{conn}, gives
\be
\o_\text{flat}(\fl_{ab}) = \tilde H_{ab}.
\ee
The above diagram shows that $\pm \o_{\mathrm{spin}} = \o_{\mathrm{flat}}\circ \mathcal I$ and hence
\be
\pm U_{ab} = \tilde H_{ab}.
\ee
This relation allows us to interpret the holonomies of a flat connection $\tilde{H}_{ab}$ as the holonomies of a spin connection along the paths $\fp_{ab}$ around the 1-skeleton of an embedded 4-simplex. The $\pm$ sign comes from the fact that Theorem \ref{2d/3d} holds for PSU(2) flat connections, and $H_{ab}$ is identified with the spin connection $U_{ab}$ up to a sign, as discussed in Section \ref{b.c.}.

Here we are relating the flat connection $A$ on $\mathrm S^3\setminus\G_5$ to the geometry of a 4-simplex embedded in a constant curvature (Lorentzian) spacetime, whose boundary tetrahedra are constant curvature spacelike tetrahedra with flatly embedded surfaces. The flat connection $A$ on $\mathrm S^3\setminus\G_5$ is taken to satisfy the boundary conditions of Section \ref{b.c.}, which give us $H_b(a)=g_a^{-1}\tilde{H}_{ab}g_a\in \mathrm{SU(2)}$ . In turn, the reconstruction theorem of that section grants us that the equation $\overleftarrow\prod_b H_b(a)=1$ associates to $\tau_a$ the geometry of a non-degenerate convex spacelike tetrahedron with constant curvature $\L_a$.\footnote{We only consider the boundary data corresponding to nondegenerate tetrahedral geometries. These data are dense in the space of all boundary data.} Hence the interpretation of the $H_b(a)$ in terms of face vectors $\mathrm a_{ab} \hat\fn_{ab}$ is
\be
H_b(a)=\exp\lt[\frac{\L_a}{3}\mathrm{a}_{ab}\hat{\fn}_{ab}\cdot\vec{\t}\rt],
\label{HAr}
\ee
where $\L_a = \pm_a |\Lambda|$. For future convenience we introduce
\be
\nu_a = \text{sgn}\,\L_a \quad \text{and} \quad \nu = \text{sgn}\, \L.
\ee
The parameter $\L$ is a constant for all $\tau_a$ and its sign will be determined shortly. This constant introduces a length unit. Once again $\Ar_{ab}$ are the areas of the convex constant curvature tetrahedron. Note that at this stage we do not know whether the boundary data induce a sign $\nu_a$ that is constant throughout the 4-simplex. However, we will prove that this follows from the requirement that the boundary data are given by the boundary value of $A$. 

If we let $\eps_{ab}\hat{\fn}_{ab}$ be the outward-pointing normal to $\tau_a$ and choose the time-like normal of tetrahedron $a$ to be gauge fixed to $(1,0,0,0)^T$, then $\hat{\fn}_{ab}=\nu_a\hat{n}_{ab}$ is the spatial normal 3-vector to the triangle $\Delta_{ab}$, parallel transported to the base point of $\fp_{ab}$, i.e. to the vertex ${1}$ of the 4-simplex. In fact, a parallel transport is only needed when $\Delta_{ab}$ is not adjacent to ${1}$ (depending on the pattern of $\fp_{ab}$).

Up to this point we have studied only the geometry of the individual tetrahedra that make up a 4-simplex. We turn now to assembling the full geometry of the 4-simplex from these pieces, and show how this can be achieved using the holonomies $H_b(a)$ and $G_{ab}$ alone. 

The group elements $\pm g_a\in\Slc$ that allow one to put each of the simplex's five tetrahedra into the time-gauge also specify the Lorentz frame of the four surfaces contained in each tetrahedron. As argued at the end of the last subsection $\tilde{H}_{ab}=g_a H_b(a) g^{-1}_a$ can be interpreted as
\be
\tilde{H}_{ab}=\exp\lt[\frac{\L_a}{3}\ \mathrm{a}_{ab} \ce_{ab}({1})\rt]
\label{tildeHAr}
\ee
where $\ce_{ab}(\bar{1})$ is the surface area bivector located at $\bar{1}$: 
\be
\ce_{ab}=\lt[\eps^{\a\b}e_\a e_\b\rt]_+({1})\ \text{of $\Delta_{ab}$},
\ee 
here the $+$ subscript indicates the self-dual part of the bivector viewed as an $\slc$ Lie algebra element. The $\slc$ algebra is viewed as a 6-dimensional real Lie algebra with generators $\vec{J} := \vec{\tau}$ and $\vec{K} := -i \tau$; the duality is $\star \vec{J} = - \vec{K}$ and $\star \vec{K} = \vec{J}$. 
Note that $\ce_{ab}({1})$ is related to $\hat{\fn}_{ab}\cdot\vec{\t}$ by 
\be
\ce_{ab}({1})=-g_a(\hat{\fn}_{ab}\cdot\vec{\t})g_a^{-1}.
\ee
The set of $\ce_{ab}({1})$ is defined up to a simultaneous adjoint action of $\Slc$, which is a local Lorentz transformation in the base frame at ${1}$.

Much like what happens for the $\hat{\fn}_{ab}$, a parallel transport (which depends on the specific pattern of the $\fp_{ab}$) relates $\ce_{ab}({1})$ to the actual bivector on $\Delta_{ab}$, whenever $\Delta_{ab}$ is not adjacent to ${1}$. For the $\Delta_{ab}$'s adjacent to ${3}$, their bivectors are given by $\ce_{ab}({3})=\O[{3}{1}]\ce_{ab}({1})\O[{3}{1}]^{-1}$ where $\O[\bar{a},\bar{b}]\in\mathrm{SL}(2,\C)$ is the holonomy of the spin connection $\o_{\mathrm{spin}}$ along the edge $(ab)$.

Finally note that the tetrahedral reconstructions do not automatically guarantee  that the areas of the triangles $\Delta_{ab}$ as seen from tetrahedra $a$ and $b$ coincide. This is because of the ambiguity between $\Ar_{ab}$ and $2\pi-\Ar_{ab}$ mentioned above. This potential ambiguity does not arise as shown in the main reconstruction result of \cite{HHKR}:

\begin{Theorem}\label{3d/4d}

The flat connections $A$ drawn from a dense subset of the space $\cm_{\mathrm{flat}}^{BC}(S^3\setminus\G_5,\Slc)$, i.e. such that their restriction to the boundary $\Fa\in\cm_{\mathrm{flat}}(\Sig_6,\Slc)$ satisfy the boundary conditions corresponding to 5 non-degenerate convex constant curvature tetrahedra, each determine a unique non-degenerate convex Lorentzian 4-simplex geometry with constant curvature $\L$, whose boundary geometry is consistent with the tetrahedral geometries determined by $\Fa$. 

\end{Theorem}

The proof of the theorem (see \cite{HHKR}) is analogous to that of the three-dimensional case, and also employs the reconstruction of the 4-simplex's Gram matrix
\be
\text{Gram}_4 \equiv \cosh \Theta_{ab},
\ee
where $\Theta_{ab}$ are the boost dihedral angles of the four-simplex. This matrix contains all the information needed to reconstruct the 4-simplex geometry, and again this includes the sign of the reconstructed simplex's curvature. The Gram matrix is calculated via the equation
\be
\label{GdihedralRel}
\cosh\Theta_{ab} = - u_I (\hat G_{ab})^I_{\phantom{I}J} u^J,
\ee
where $u^I=(1,0,0,0)^T$, and $\hat G_{ab}\in\mathrm{SO}^+(1,3)$ is the vectorial representation of $G_{ab}\in\Slc$.
The non-degeneracy condition corresponds to the requirement that the connection does \emph{not} produce $G_{ab}$ such that $ u_I (\hat G_{ab})^I_{\phantom{I}J} u^J = 1$.

 Notice that the theorem implies in particular that all five of the boundary tetrahedra share the same sign of the curvature, hence
\be
\nu_a = \nu = \text{sgn}\,\L
\ee
is a \textit{global} sign. The theorem also allows one to reconstruct the meaning of the rotation part of $G_{ab}$. This is associated to the plane of the triangle $\Delta_{ab}$, and corresponds to the relative rotation by an angle $\theta_{ab}$ between the frames of $\Delta_{ab}$ as seen from tetrahedra $a$ and $b$.

\subsection{Parity Pairs}\label{PP}

In this final subsection we summarize the flat connection-geometry correspondence and indicate relations between discrete symmetries of the reconstructed geometry and properties of the flat connections. 

The boundary conditions discussed in Section \ref{b.c.} require that the flat connections in $\cm_{\mathrm{flat}}^\text{BC}\lt(S^3\setminus\G_5,\ \Slc\rt)$ restrict to flat connections on the boundary of the graph complement manifold $\Sig_6 = \partial M_3$, i.e. to connections in $\cm_{\mathrm{flat}}^\text{BC}\lt(\Sig_6,\ \Slc\rt)$. In turn these boundary connections reduce to SU(2) flat connections on each of the 4-holed spheres around the vertices of $\G_5$. The boundary data on $\Sig_6$ are completely determined by specifying at the same time: 

\begin{itemize}

\item[$i$)] the conjugacy class of the holonomies around a path $\ell_{ab}$ transverse to each edge of $\G_5$. This is equivalent to specifying (one of) the eigenvalues $x_{ab}$ of these holonomies. In particular, the boundary conditions impose that $x_{ab}\in \mathrm{U}(1)$, instead of being a general complex number;

\item[$ii$)] the eigenvalues $x_a$ and $x_a''$ of the products of two pairs of holonomies computed along the path pairs $(c_{ab},c_{ab'})$ and $(c_{ab},c_{ab''})$, which encircle three different edges adjacent to the same vertex $a$, see Figure \ref{ribbon}. Name the paths associated to the above compositions $c_a$ and $c_a''$. Again, $x_a$ and $x_a''$ must be complex numbers of unit norm, i.e. $x_a, x_a''\in \mathrm U(1)$. In the next section, we will discuss why it is far more convenient to substitute $x_a''$ with a coordinate $y_a$, which turns out to be canonically conjugated (in the sense of symplectic geometry)  to $x_a$. In terms of these variables, known as the Fenchel-Nielsen length and twist respectively, the boundary conditions reduce again to $x_a,y_a\in\mathrm U(1)$.

\end{itemize}

The boundary data $\{x_{ab}; x_a , y_a\}\subset \mathrm U(1)$ fully specify the $\Su$ flat connections on the five 4-punctured spheres $\{\mathcal S_a\}_{a=1}^{5}$. The geometrical reconstruction theorems discussed above imply that these same data encode completely the geometry of five geometrical constant-curvature tetrahedra. These tetrahedra are characterized by the fact that the value of their faces' areas are shared by couples of tetrahedra. This is because, geometrically, the $\{x_{ab}\}_b$ encode the areas of the faces of tetrahedron $a$. On the other hand the $\{x_a,y_a\}$ fix the remaning two degrees of freedom (a tetrahedron is determined by 6 independent numbers; think of the edge lengths). At this stage nothing is enforcing the fact that the {\em shapes} of the equi-area faces of two different tetrahedra are the same, more on this below. 

Note that at fixed areas, the space of tetrahedra parametrized by $(x_a,y_a)$ turns out to carry a natural symplectic structure \cite{HHR}, such that the logarithms of these variables are conjugated. We will come back to this fact in the next section.

Denote a given value of the boundary data $\{x_{ab}; x_a , y_a\}$ by $\lt\{\mathring{x}_{ab};\mathring{x}_a,\mathring{y}_a\rt\}$. The following questions and their answers turn out to be interesting and useful in later analysis: Does a flat connection $A \in \cm^\text{BC}_{\mathrm{flat}}\lt(S^3\setminus\G_5,\ \Slc\rt)$ that has boundary value consistent with a given set of the boundary data $\lt\{\mathring{x}_{ab};\mathring{x}_a,\mathring{y}_a\rt\}$ always exist? Provided such a consistent flat connection exists, is it uniquely determined by the $\lt\{\mathring{x}_{ab};\mathring{x}_a,\mathring{y}_a\rt\}$? 

Both of the above questions have negative answers. Let us explain why: A generic flat connection in $\cm^\text{BC}_{\mathrm{flat}}\lt(\mathrm  S^3\setminus\G_5,\ \Slc\rt)$ satisfies the hypothesis of Theorem \ref{3d/4d} and hence corresponds to a geometric 4-simplex. However, as we discussed above, within the boundary data $\lt\{\mathring{x}_{ab};\mathring{x}_a,\mathring{y}_a\rt\}$ there is nothing that guarantees the correspondence of the shapes of the triangular faces. Hence, not every set of boundary conditions $\lt\{\mathring{x}_{ab};\mathring{x}_a,\mathring{y}_a\rt\}$ is the  boundary of a flat connection in $\cm^\text{BC}_{\mathrm{flat}}\lt(\mathrm  S^3\setminus\G_5,\ \Slc\rt)$.

Turning to uniqueness, consider a set of boundary data $\lt\{\mathring{x}_{ab};\mathring{x}_a,\mathring{y}_a\rt\}$, and a flat connection $A\in\cm^\text{BC}_{\mathrm{flat}}\lt(\mathrm S^3\setminus\G_5,\ \Slc\rt)$ consistent with them. Theorem \ref{3d/4d} states that  $A$ corresponds uniquely to a geometric 4-simplex $\sig_4$. However, as the next theorem shows, it is easy to produce from $A$ another flat connection $\tilde A\in\cm^\text{BC}_{\mathrm{flat}}\lt(\mathrm  S^3\setminus\G_5,\ \Slc\rt)$ whose boundary value is also consistent with $\lt\{\mathring{x}_{ab};\mathring{x}_a,\mathring{y}_a\rt\}$. Notice that this does not mean that $A$ and $\tilde A$ must have the same boundary values when restricted to $\Sigma_6$, since the data  $\lt\{\mathring{x}_{ab};\mathring{x}_a,\mathring{y}_a\rt\}$ do not contain information about the \textit{longitudinal} holonomies $G_{ab}$. In fact, it turns out that $A$ and $\tilde A$  correspond to different constant curvature 4-simplices $\sig_4$ and $\tilde{\sig}_4$ related by a parity inversion, and $\tilde G_{ab} = {G_{ab}^{-1}}^\dagger$. In analogy with the previous discussion, we can introduce the variables $y_{ab}$, conjugated to the $x_{ab}$, which supply a complete set of coordinates on $\cm_{\mathrm{flat}}\lt(\Sig_6,\ \Slc\rt)$. In these coordinates, the parity pair is described by $\lt\{\mathring{x}_{ab}, \mathring{y}_{ab};\mathring{x}_a,\mathring{y}_a\rt\}$ and $\lt\{\mathring{x}_{ab},\widetilde{\mathring{y}_{ab}};\mathring{x}_a,\mathring{y}_a\rt\}$, where $\widetilde{\mathring{y}_{ab}} = 1/\overline{\mathring{y}_{ab}}$, with the bar indicating complex conjugation.

\begin{Theorem}\label{parity}

Given a set of boundary data $\lt[\mathring{x}_{ab};\mathring{x}_a,\mathring{y}_a\rt]$ corresponding geometrically to 5 constant curvature tetrahedra forming the boundary of a constant curvature 4-simplex, there exists exactly 2 flat connections $A,\tilde{A}\in \cm_{\mathrm{flat}}\lt(S^3\setminus\G_5,\ \Slc\rt)$ on the graph complement 3-manifold, whose boundary values are consistent with $\lt[\mathring{x}_{ab};\mathring{x}_a,\mathring{y}_a\rt]$. The connections $A$ and $\tilde{A}$ correspond to the constant curvature 4-simplices $\sig$ and $\tilde{\sig}$ which have the same intrinsic geometry but different 4d orientations. The pair $A$ \& $\tilde{A}$ are called a ``{parity pair}''.

\end{Theorem}

The proof can be found in \cite{HHR}. The existence of the parity pair $A$ \& $\tilde{A}$ is natural, because these connections are complex conjugated to one another with respect to the complex structure on $\cm_{\mathrm{flat}}(S^3\setminus\G_5,\Slc)$, which, in turn, is induced from the complex group $\Slc$.\footnote{Namely $A=A^j\t_j$ and $\tilde{A}=\bar{A}^j\t_j$.} So the boundary values of $A$ \& $\tilde{A}$ give the same SU(2) flat connection on each 4-holed sphere $\cs_a$; this implies that they give the same data $\lt[\mathring{x}_{ab};\mathring{x}_a,\mathring{y}_a\rt]$.

\section{Complex Chern--Simons Theory: from Quantization of a 3d Flat Connection to 4d Quantum Gravity}\label{quantum}

In the previous sections we established a correspondence between a class of $\Slc$ flat connections on $M_3 = S^3\setminus \G_5$ and homogeneously-curved 4d simplicial geometries. 
Since a natural way of quantizing flat connections on $M_3$ exists, and is given by Chern--Simons theory, such a correspondence provides us a natural way to quantize 4d simplicial geometry. 
Somewhat surprisingly, the resulting quantum states are related to discrete general relativity, in the sense of Regge.
More precisely, one finds that the physical Chern--Simons states induced by the boundary conditions discussed in previous sections reproduce semiclassically the Hamilton--Jacobi functional of 4d Regge gravity with a cosmological constant, as discretized on homogeneosuly curved 4-simplices. 

In this section, we are going to prove the previous claims by means of a WKB approximation of the 3d holomorphic blocks of $\Slc$ Chern--Simons theory on $M_3$, with the appropriate boundary conditions imposed. The main technical tool to this end will be the Schl\"afli identities. Let us, however, proceed in order. 

The $\Slc$ Chern--Simons theory on $M_3=S^3\setminus \G_5$ is defined by the following action \cite{witten}:
\be
CS\lt[M_3\,\big|\,A,\bA\rt]&=&\frac{t}{8\pi}\int_{M_3}\tr\lt(A\wedge \rmd A+\frac{2}{3}A\wedge A\wedge A\rt)+\frac{\bar{t}}{8\pi}\int_{M_3}\tr\lt(\bA\wedge \rmd \bA+\frac{2}{3}\bA\wedge \bA\wedge \bA\rt)\nonumber\\
&&+\ \frac{t}{8\pi}\int_{\partial M_3}\tr\lt(A_1\wedge A_2\rt)+\frac{\bar{t}}{8\pi}\int_{\partial M_3}\tr\lt(\bA_1\wedge \bA_2\rt),\label{CSstart}
\ee
where $A$ and $\bar A$ are the holomorphic and antiholomorphic parts of the $\Slc$ connection, respectively, where holomorphicity is defined with respect to the natural complex structure of $\Slc$.
We will assume the Chern--Simons couplings 
\be
t=k+is \quad \text{and} \quad \bar{t}=k-is
\ee
are complex conjugates of one another, that is we will assume $k,s\in\R$. 
Notice that if $k\in\Z$, then $\exp\lt[i\, CS\rt]$ is invariant under large gauge transformation. Nonetheless, in most of the following discussion, we will avoid this requirement, and keep $k$ an arbitrary real number \cite{analcs}. 

The boundary terms of equation \eqref{CSstart} are crucial for imposing the correct boundary conditions. Or, in other words, they are crucial for the path integral on $M_3$ to be a well-defined ``wave-functional'' on {\it half} of the Atiyah--Bott--Goldman phase-space defined on $\mathcal M_\text{flat}(M_3,\Slc)$.
For this, coordinates $(s^1,s^2)$ have been chosen on $\Sigma_6 = \partial M_3$, such that $s_1$ is the meridian direction of $\G_5$. 
Thus, in the boundary action, the $A_i$ with $i=1,2$, are the components of the $\Slc$ connection $A$ along the directions defined by $s^i$. 
The sign chosen in front of the boundary terms implies that it is the values of $(A_1,\bar A_1)$, i.e. the meridian part of the connection form, which set the boundary conditions for the path integral \cite{seiberg}.
Loosely speaking, the longitudinal part of the connection provides then the conjugate variable.
As is customary, all of this can be explicitly read from the boundary part of the first variation of the action \eqref{CSstart}, that is from the presymplectic\footnote{{\ Pre}symplectic means that gauge transformations have yet to be modded out and hence that the ensuing {pre}symplectic 2-form $\delta_{[1}\delta_{2]}CS|_{\partial}$ is degenerate. \label{ftnt1}} potential of the theory:
\be
\delta CS|_{\partial} = \frac{t}{4\pi}\int_{\partial M_3} \tr(\delta A_1 \wedge A_2) +  \frac{\bar t}{4\pi}\int_{\partial M_3} \tr(\delta \bar A_1 \wedge \bar A_2).
\label{deltaCS}
\ee
This leads us to define
\be
Z_{CS}\lt(S^3\setminus \G_5\,\big|\,A_1,\bA_1\rt)=\int_{A_1,\bA_1}\cd A\cd\bA\,e^{\frac{\rmi}{\hbar}\,CS\lt[S^3\setminus \G_5\,\big|\,A,\bA\rt]}\label{ZCS},
\ee
where $(A_1,\bA_1)$ set up the boundary conditions on $\Sig_6$, whereas the values of $(A_2,\bA_2)$ on the boundary are implicitly integrated over in the functional integral.

The prefactor to the action, $1/\hbar\in\R$, has to be viewed as a scaling parameter for the couplings $(t,\bar{t})$. In particular, the semiclassical limit $\hbar\to 0$ can be simply implemented by taking $(t,\bar{t})\to \infty$ uniformly. 

In the formula above, $Z_{CS}\lt(S^3\setminus \G_5\,\big|\,A_1,\bA_1\rt)$ is viewed as a ``wave-functional'' of Chern--Simons  theory, i.e. it is viewed as a (possibly distributional) state in the Hilbert space $\ch(\Sig_6)$ defined on the boundary $\Sig_6$. 
The Hilbert space $\ch(\Sig_6)$ is a quantization of $\cm_{\mathrm{flat}}(\Sig_6,\Slc)$, the moduli space of $\Slc$ flat connections on the closed genus-6 2-surface $\Sig_6$ \cite{DGV,Fock,FG2,gukov,Dimofte,knots}. 
In general, the moduli space of $\Slc$ flat connections on a closed genus-$g$ 2-surface $\Sig_g$, $\cm_{\mathrm{flat}}(\Sig_g,\Slc)$, is a hyper-K\"ahler variety of $\dim_\C=6g-6$, known as the Hitchin moduli space \cite{hitchin}. 

In order to study spaces of this type, it is convenient to decompose them into fundamental units. 
These are given by ``pair of pants'', or ``trinions'', which are nothing but 3-holed spheres.
A closed 2-surface $\Sig_g$ can be decomposed into pairs of pants by cutting through $3g-3$ closed meridian curves $\{c_m\}_{m=1}^{3g-3}$.
A flat connection on $\Sig_g$ hence defines, along the meridian cycles $\{c_m\}$, a set of $3g-3$ holonomies $\{H_m\}$ whose eigenvalues $\{x_m\}$ can be used as (a maximal commuting subset of the) canonical coordinates on $\cm_{\mathrm{flat}}(\Sig_6,\Slc)$. 
These, together with their canonically conjugate variables $\{y_m\}$, constitute the Fenchel--Nielsen (FN) coordinates on $\cm_{\mathrm{flat}}(\Sig_6,\Slc)$ \cite{FN}. They are commonly known as the length, $\{x_m\}$, and twist, $\{y_m\}$, FN coordinates. 

When written in terms of the FN coordinates, the Atiyah--Bott--Goldman symplectic 2-form on  $\cm_{\mathrm{flat}}(\Sig_g,\Slc)$, which can be obtained by symplectic reduction of the presymplectic form induced by the Chern--Simons action (see footnote \ref{ftnt1}), reads simply:
\be
\omega_{CS} = -\frac{t}{2\pi} \sum_{m=1}^{3g-3} \frac{\d y_m}{y_m} \wedge \frac{\d x_m}{x_m} + c.c.
\label{ABGsympl_xy}
\ee
We have used here the notation $\d y_m$ and $\d x_m$ to emphasize that these are coordinates on the moduli space of flat connections rather than on $\Sig_g$. This space is {\it finite} dimensional and so the symbol $\d$ does {\it not} indicate here any functional variation, just a standard differential on $\cm_{\mathrm{flat}}(\Sig_g,\Slc)$.

The construction of the FN coordinates and their relation to the 4-simplex geometry is detailed in the next section. The reader not interested in these details can safely skip it. For her, here is a very brief---heuristic---account of this construction: from equation \eqref{deltaCS}, one sees that the variable canonically conjugated to the meridian holonomy must be related to the longitudinal one; indeed, one can think of $y_m$ as being  given by the eigenvalue of the longitudinal holonomies $G_m$ transverse to $H_m$, once the source and target frames of $G_m$ have been appropriately fixed.

\subsection{Complex Fenchel--Nielsen Coordinates on $\cm_{\mathrm{flat}}(\Sig_6,\Slc)$}\label{FNCoords}

In this section, we review the construction of FN coordinates on $\cm_\text{flat}(\Sigma_6,\Slc)$. 
After choosing a pair of pants (or trinion) decomposition of $\Sigma$, we focus on two such pairs of pants ${\cal T}_a$ and ${\cal T}_b$, and the cylinder connecting them. 
At each pair of pants, we choose base points, $\fo_a$ and $\fo_b$. 
Starting and ending at these base points, we consider the holonomies $H_{ab}$ and $H_{ba}$, respectively, which encircle once the tube connecting ${\cal T}_a$ to ${\cal T}_b$.

The holonomies $H_{ab}$ and $H_{ba}$ can then be diagonalized by an appropriate choice of reference frame at $\fo_{a,b}$.
In formulas:
\be
H_{ab}=M_{ab}\begin{pmatrix}
  x_{ab} & 0 \\
  0 & x_{ab}^{-1}
 \end{pmatrix}M_{ab}^{-1},
 \qquad\text{where}\qquad 
 M_{ab}=\lt(\xi_{ab},J\xi_{ab}\rt),
 \label{2}
\ee
for some normalized spinors $\xi_{ab}\in\mathbb C^2$. 
A spinor $\xi$ is here said to be normalized, iff $\langle \xi , \xi \rangle = 1$, where
\be
\langle \xi, \eta \rangle = \bar\xi^1 \eta^1 + \bar \xi^2 \eta^2,
\ee
and the spinor $J\xi$ is orthogonal to $\xi$ and is defiened by the aaction of the antilinear map $J$:
\be
J\left(\begin{array}{c}
\xi^1 \\ \xi^2
\end{array} \right)
\equiv
\left(\begin{array}{c}
-\bar\xi^2 \\ \bar\xi^1
\end{array} \right).
\ee
For connections satisfying our boundary conditions, i.e. for $\Slc$ connections that reduce to $\Su$ connections in proximity of the graph vertices, the eigenvalue $x_{ab}$ must then satisfy $x_{ab} \in \mathrm{U}(1)$. 
Also, for the geometry to be non-degenerate, $x_{ab}\neq 1$. The latter condition will be assumed in the following.

Now, the parallel transport $G_{ab}$ is defined as the holonomy of the flat connection $A$ between $\fo_b$ to $\fo_a$ (holonomies compose from the right) along the tube connecting ${\cal T}_b$ to ${\cal T}_a$. 
Notice that there is no canonical choice of path along which to define $G_{ab}$. 

From the flatness of the connection $A$, it follows immediately that 
\be
H_{ab} = G_{ab} H_{ba} G_{ba},
\ee
and $G_{ba} \equiv G_{ab}^{-1}$.
From this equation, and equation \eqref{2}, one obtains 
\be
\begin{pmatrix}
  x_{ab} & 0 \\
  0 & x_{ab}^{-1}
 \end{pmatrix}M_{ab}^{-1}G_{ab}M_{ba}=M_{ab}^{-1}G_{ab}M_{ba}\begin{pmatrix}
  x_{ab} & 0 \\
  0 & x_{ab}^{-1}
 \end{pmatrix},
\ee
which in turn implies that $M_{ab}^{-1}G_{ab} M_{ba}$ is diagonal.\footnote{$M_{ab}$ is {\it not} the inverse of $M_{ba}$.} This is readily proved by expressing this matrix in the basis $\{1, \sigma_i\}$ and commuting it with $\frac12(x_{ab} + x^{-1}_{ab}) + \frac12(x_{ab} - x_{ab}^{-1})\sigma_3$.
Thus,
\be
M_{ab}^{-1}G_{ab} M_{ba} = \begin{pmatrix} 
\lambda_{ab} & 0 \\
0 & \lambda_{ab}^{-1}\\
\end{pmatrix},
\label{Gy}
\ee
and so,
\be
G_{ab} \xi_{ba} = \lambda_{ab} \xi_{ab}.
\ee

The Fenchel--Nielsen twist coordinate $y_{ab}$ is closely related to $\lambda_{ab}$, but the two are not precisely the same. 
They differ by a particular cross-ratio of inner products between the $\{\xi_{ab'},\xi_{a'b}\}$, where $b'$ ranges over the three links emanating from $\mathcal T_a$, and similarly for $a'$.

The cross ratios just guarantee scale invariance in the choice of a basis at $\fo_{a,b}$, while the inner products between the $\{\xi_{ab'},\xi_{a'b}\}$ are crucial to ``pick out the right components'' of $G_{ab}$ in the Poisson brackets. In other words, whereas the $H_{ab}$ start and end at the same point, and the trace is good enough to select the invariant part of these holonomies, the $G_{ab}$ must be sandwiched in between their starting and ending reference bases. 

The construction proceeds as follows. 
Start by observing that the canonical Atiyah--Bott--Goldman brackets (the Lie algebra basis is chosen to be $\tau_i \equiv \frac i 2 \sigma_i$)
\be
\{ A^i_\mu(x), A^j_\nu(y)\} = -\frac{8\pi}{t} \eps_{\mu\nu}\delta^{ij}\delta^{(2)}(x,y),
\ee
induce the following brackets between the parallel transports $H_{ab}$ and $G_{ba}$:
\be
\Big\{ G_{ba} \stackrel{\otimes}{,} H_{ab} \Big\} = \frac{2\pi}{t} \Big( G[\fo_b,p]\sigma_j G[p,\fo_a] \Big) \otimes \Big( H[\fo_a,p]\sigma_j H[p,\fo_a]\Big)
\label{GHpoissonbracket}
\ee
where $p$ is the (by construction) unique intersection point between the paths defining $G_{ba}$ and $H_{ab}$.\footnotemark~ 
The point  $p$, also splits these paths in two, and hence provides the decompositions $G_{ba}= G[\fo_b,p]G[p,\fo_a] $ and $H_{ab} = H[\fo_a,p]H[p,\fo_a]$.%
\footnotetext{Of course, according to our construction the paths defining $G_{ba}$ and $H_{ab}$ intersect at the point $\fo_a$, as well. However, it is not too hard to convince oneself that the construction is invariant under homotopic deformations of the paths. This means that one can regularize the paths for the $H_{ab}$ by choosing them to first follow the path defining $G_{ba}$ out of $\fo_a$ and up to the point $p$, and then to abruptly depart transversally. The only contributions to the Atiyah--Bott--Goldman brackets then come from the loop leaving and arriving at $p$, while the tail to $\fo_a$ plays no role. This trick can be used to regularize our expressions. Alternatively, one can consider a homotopically equivalent loop to define $H_{ab}$, which does not starts at $\fo_a$ and only intersects the path defining $G_{ba}$ at $p$. Since the result of the construction only involves the eigenvalues of $H_{ab}$, one can confidently extend this result to paths whose base point is arbitrarily close to $\fo_a$.\label{footnotepaths}} 

Defining the $\Slc$ invariant bi-linear inner product 
\be
\langle \xi \wedge \eta \rangle \equiv \langle J\xi, \eta\rangle = \eps_{\alpha\beta}\xi^\alpha \eta^\beta,
\ee
 we have, after a few lines of algebra,
\begin{align}
\Big\{ \langle\xi_{ba'} \wedge G_{ba}\xi_{ab}\rangle , \tr(H) \Big\} 
& =\frac{2\pi}{t} \langle G_{ab}\xi_{ba'} \wedge \sigma_j \xi_{ab} \rangle\; \tr(\sigma_j  H_{ab}),
\end{align}
where we used the identity $M\sigma_j M^{-1} \otimes \sigma_j = \sigma_j \otimes M^{-1}\sigma_j M$, and the fact that the holonomy $H_{ab}(p)$, representing $H_{ab}$ with base point parallel transported to $p$, can be expressed in two equivalent ways (thanks to the flatness of the connection $A$):
\be
H[p,\fo_a]H[\fo_a,p]=H_{ab}(p) = G[p,\fo_a]H_{ab}G[p,\fo_a]^{-1}.
\ee
Now, at $\fo_a$, $H_{ab}(\fo_a)\equiv H_{ab}$, is diagonal precisely in the basis $\{\xi_{ab},J\xi_{ab}\}$, thus using equation \eqref{2} and the identity $M_{ab}\sigma_j M_{ab}^{-1} \otimes \sigma_j = \sigma_j \otimes M_{ab}^{-1}\sigma_j M_{ab}$, we find 
\begin{align}
\Big\{ \langle\xi_{ba'} \wedge G_{ba}\xi_{ab}\rangle , \tr(H) \Big\} 
& = (x_{ab} - x_{ab}^{-1})\;\langle \xi_{ba'} \wedge  G_{ba}\xi_{ab} \rangle.
\end{align}  
Or, equivalently,
\be
\Big\{ \ln  \langle\xi_{ba'} \wedge G_{ab}\xi_{ab}\rangle  , \ln x_{ab} \Big\} = -\frac{2\pi}{t}.
\ee

Thus, we see that the contraction above extracts the non-trivial part of the brackets \eqref{GHpoissonbracket}.

To define the canonically conjugate FN coordinates we need to ({\it i}) symmetrize between the two choices of $a'$ in the above formula, i.e. between the two punctures at ${\cal T}_b$ not touched by the parallel transport along the tube connecting ${\cal T}_b$ to ${\cal T}_a$, and ({\it ii}) find a combination of the inner products that is invariant under meaningless rescaling of the basis vectors.
These two requirements are readily satisfied by the following definition of the FN twist coordinate:
\be
\tau_{ab} = -\frac{\langle \xi_{ba'} \wedge \xi_{ba''} \rangle  }{\langle \xi_{ba'} \wedge G_{ba}\xi_{ab} \rangle \; \langle \xi_{ba''} \wedge G_{ba}\xi_{ab} \rangle  } \frac{\langle \xi_{ab'} \wedge \xi_{ab} \rangle \; \langle \xi_{ab''} \wedge \xi_{ab} \rangle  }{\langle \xi_{ab'} \wedge \xi_{ab''} \rangle  },
\ee
where $a',\,a''\neq a$ (respectively, $b'\, b'' \neq b$) label the two other punctures at ${\cal T}_b$ (respectively, ${\cal T}_a$).
From this expression it is clear that only the two terms in the first denominator contribute to the Poisson bracket $\{\tau_{ab},x_{ab}\}$, while the other factors just ensure the requirements above are satisfied. 

Also, from the $\Slc$ invariance of the inner product $\langle \cdot \wedge \cdot \rangle$, it readily follows that the various factors entering $\tau_{ab}$ can be computed at any point of the surface, provided one defines the parallel transported sections $s_{ab}$, by $(d - A)s_{ab}=0$ and $s_{ab}(p_0) = s^0_{ab}$, for $s^0_{ab}$ an eigenvector (determined up to complex rescaling) of $H_{ab}(p_0)$ based at some point $p_0$ on the tube connecting ${\cal T}_a$ and ${\cal T}_b$. Similar definitions are understood for all the other choices of indices. We emphasize that both the complex normalization of $s^0_{ab}$ and the point at which it is defined are completely irrelevant at this point (cf. footnote \ref{footnotepaths}). This shows that the the FN coordinates can be ultimately defined in a completely geometrical way, without reference to any basis. For our purposes, however, it is easiest to work in the basis provided above, since it allows for a direct translation to the underlying simplicial geometry.

One last technical consideration is in order: being ``quadratic'' in $G_{ab}$, the complex FN twist variable $\tau_{ab}$ is actually a coordinate on the moduli space of $\mathrm{PSL}(2,\mathbb{C})$ flat connections, rather than on the moduli space of $\Slc$ flact connections. A lift to the coordinate $y_{ab}$, such that 
\be
y^2_{ab}=\tau_{ab},
\ee
is then needed to complete the construction of the $\Slc$ FN twist coordinate. 

To summarize, given the eigenvalues of $x_{ab}$ and $\lambda_{ab}$ of $H_{ab}$ and $G_{ab}$, respectively (the latter as expressed in the proper frame at each $\fo_a$, defined by $\{\xi_{ab}, J\xi_{ab}\}$ as above), the $\Slc$ FN coordinates and their Poisson brackets are 
\be
x_{ab}\quad\text{and}\quad y_{ab} = \lambda_{ab}\sqrt{\chi_{ab}(\xi)} 
\qquad\text{with}\qquad
\Big\{ \ln y_{ab} , \ln x_{cd} \Big\} = \frac{2\pi}{t} \delta_{(ab),(cd)},
\ee
where a branch of the square root has been arbitrarily chosen and $\chi_{ab}(\xi)$ stands for the cross ratio
\be
\chi_{ab}(\xi) = -\frac{\langle \xi_{ba'} \wedge \xi_{ba''} \rangle  }{\langle \xi_{ba'} \wedge \xi_{ba} \rangle \; \langle \xi_{ba''} \wedge \xi_{ba} \rangle  } \frac{\langle \xi_{ab'} \wedge \xi_{ab} \rangle \; \langle \xi_{ab''} \wedge \xi_{ab} \rangle  }{\langle \xi_{ab'} \wedge \xi_{ab''} \rangle  }.
\ee

\subsection{Holomorphic 3d Blocks and Quantum Flatness}\label{reviewCS}

The previous analysis suggests the following definitions for Darboux coordinates $(u_m,v_m)$: 
\be
x_m = e^{u_m}
\qquad\text{and}\qquad
y_m = e^{-\tfrac{2\pi}{t} v_m},
\ee
where an arbitrary branch of the logarithm has been chosen. Analogous equations are understood to define the complex conjugate variables $(\bar u_m, \bar v_m)$.
Hence, the Atiyah--Bott--Goldman symplecitc form on $\cm_{\mathrm{flat}}(\Sig_g,\Slc)$, see equation \eqref{ABGsympl_xy}, reads
\be
\omega_{CS} = \sum_{m=1}^{3g-3} \delta v_m \wedge \delta u_m + c.c. ,
\ee
or, in terms of Poisson brackets,
\be
\{u_m,v_n\} = \delta_{mn}  = \{ \bar u_m, \bar v_n\}.
\ee

The above relations lead one to introduce the quantum operators $(\hat{u}_m,\hat{v}_m)$ and $(\hat{\bar{u}}_m,\hat{\bar{v}}_m)$ with canonical commutation relations 
\be
\Big[\hat{u}_m,\hat{v}_n\Big]=\rmi \hbar\delta_{mn},\ \quad \text{and} \quad \ \lt[\hat{\bar{u}}_m,\hat{\bar{v}}_m\rt]=\rmi \hbar\delta_{mn}.
\ee
Equivalently, in terms of the operators $(\hat{x}_m, \hat{y}_m)$, one finds
\be
\hat{x}_m\hat{y}_m=e^{-\frac{2\pi i\hbar}{t}}\hat{y}_m\hat{x}_m,\ \ \ \ \text{and}\ \ \ \ \hat{x}_n\hat{y}_m=\hat{y}_m\hat{x}_n,\ \ \text{when}\ \ n\neq m,
\ee
with similar equations for $(\hat{\bar{x}}_m, \hat{\bar{y}}_m)$.

The Hilbert space $\ch(\Sig_g)$ of ``quantum flat-connections'' can hence be constructed as the (Schr\"odinger)  $L^2$-type representation of the above canonical commutation relations: a state in $\ch(\Sig_g)$ is a wave function of $(u,\bar{u})$, on which $(\hat u, \hat{\bar u})$ act by multiplication and $(\hat v,\hat{\bar v})$ by derivation, e.g. $\hat v = -i \hbar \partial_u$.
In particular, the path integral of equation \eqref{ZCS} should be written as $Z_{CS}\lt(M_3|u,\bar{u}\rt)$. 

The classical solutions to the Chern--Simons equations of motion $F(A)=0=F(\bA)$ on $M_3$ define  a holomorphic Lagrangian subvariety $\mathcal L_{\mathbf A}(M_3)$ in $\cm_\text{flat}(\Sigma_g=\partial M_3,\Slc)$ \cite{DGV,DGG}.
At least locally in $\cm_\text{flat}(\Sigma_g=\partial M_3, \Slc)$, this Lagrangian subvariety is described by a set of (Laurent) polynomial equations,
\be
\mathbf A_m (x,y) = 0 \quad m=1,\dots, 3g-3.
\ee
In quantum Chern--Simons theory, the holomorphic part of $\cl_{\mathbf A}(M_3)$ can then be quantized via the introduction of an operator version of the above equations,
\be
\hat{\mathbf A}_m (\hat x, \hat y; \hbar) \Psi(u) = 0.
\ee
Here $\hat{\mathbf A}_m (\hat x, \hat y; \hbar)$ is the quantization of ${\mathbf A}_m (x, y)$ defined by a specific operator ordering \cite{ABQ}, and consequently, $\Psi \in \ch(\partial M_3)$ is the physical wave function of { the holomorphic part of} $\Slc$ Chern-Simons theory associated with $M_3$. It is a holomorphic function of $u$ as a consequence of the holomorphicity of $\cl_{\mathbf{A}}$ and ${\mathbf{A}}_m(x,y)=0$.

The functional integral $Z_{CS}(M_3 | u, \bar u)$ of equation \eqref{CSstart} must satisfy at the same time the above operator constraint and its complex conjugate,
\be
\hat{\mathbf{A}}_m(\hat{x},\hat{y},\hbar)\,Z_{CS}\lt(M_3\,\big|\,u,\bar{u}\rt)=0=\hat{\mathbf{A}}_m(\hat{\bar{x}},\hat{\bar{y}},\hbar)\,Z_{CS}\lt(M_3\,\big|\,u,\bar{u}\rt).
\ee 
It was shown in \cite{analcs,DGLZ} that $Z_{CS}\lt(M_3\,\big|\,u,\bar{u}\rt)$ can be in fact written as a sum over branches of factorized wave functions
\be
Z_{CS}\lt(M_3\,\big|\,u,\bar{u}\rt)=\sum_{\a,\bar{\a}}\fn_{\a,\bar{\a}}\,Z_{CS}^{(\a)}\lt(M_3\,\big|\,u\rt)\,Z_{CS}^{(\bar{\a})}\lt(M_3\,\big|\,\bar{u}\rt) .\label{WKB}
\ee
This expression introduces the ``holomorphic 3d blocks'' $Z_{CS}^{(\a)}\lt(M_3\,\big|\,u\rt)$ which satisfy the holomorphic operator constraints:
\be
\hat{\mathbf{A}}_m\lt(\hat{x},\hat{y},\hbar\rt)Z_{CS}^{(\a)}\lt(M_3|\,u\rt)=0 \ \ \ \forall\ \a,\ \ \ \ \label{AZ=0}
\ee
and similarly for the antiholomorphic part
\be
\hat{\mathbf{A}}_m(\hat{\bar{x}},\hat{\bar{y}},\hbar)\,Z_{CS}^{(\bar{\a})}\lt(M_3\,|\,\bar{u}\rt)=0\ \ \ \forall\ \bar{\a}.
\ee

These are the central objects to be studied below. 
Each block, $Z_{CS}^{(\a)}\lt(M_3|\,u\rt)$, can be understood using Morse theory as a path integral of the holomorphic $\Slc$ Chern-Simons theory, heuristically identified with the holomorphic part of equation \eqref{CSstart}, as defined along a certain integration cycle which encloses a single saddle point $\alpha$, i.e. a classical solution in the form of a flat connection on $M_3$ \cite{analcs}. Each of the integration cycles defining a $Z_{CS}^{(\a)}\lt(M_3|\,u\rt)$ is known as a ``Lefschetz thimble'' of the Chern-Simons path integral.

Most interestingly for the purposes of this paper, the holomorphic 3d block, seen as an asymptotic expansion in $\hbar$, can also be understood from the viewpoint of a systematic WKB analysis of the above operator constraint equations.
At lowest order in $\hbar$, this reads\footnote{
In this equation, ``$\cdots$'' contains the subleading terms of $\log \hbar$ and $\sum_{n=0}^\infty S_n^{(\a)}(u) \hbar^n$. If $M_3$ is a knot complement \cite{borot}, known techniques  related to topological recursion allow one to recursively  \cite{DGLZ} compute all the quantum corrections $S_n(u)^{(\a)}$. 
The computation of coefficients $\fn_{\a\bar{\a}}$ in equation \eqref{WKB} is described in \cite{analcs}. Hence, the above discussion provides a perturbative definition of the holomorphic 3d block $Z_{CS}^{(\a)}\lt(M_3\big|\,u\rt)$. A nonperturbative definition has also been proposed in terms of a ``state-integral model'' \cite{DGLZ,knots,andersen}. 
}
\be
Z_{CS}^{(\a)}\lt(M_3\big|\,u\rt)=\exp\lt[\frac{i}{\hbar}\int\limits^{\lt(u,v{(\a)}\rt)}_{\substack{(u_0,v_0)\\ \Fc\subset\cl_{\mathbf{A}}}}\,\vth+\cdots\rt]
\qquad\text{and}\qquad 
Z_{CS}^{(\bar{\a})}\lt(M_3\big|\,\bar{u}\rt)=\exp\lt[\frac{i}{\hbar}\int\limits^{\lt(\bar{u},\bar{v}^{(\bar{\a})}\rt)}_{\substack{\lt(\bar{u}_0,\bar{v}_0\rt)\\ \Fc\subset\cl_{\mathbf{A}}}}\,\bar{\vth}+\cdots\rt].\label{block}
\ee
where $\vth$ and $\bar{\vth}$ are the holomorphic and anti-holomorphic parts of the Liouville 1-form (symplectic potential) on $\cm_\text{flat}(\partial M_3,\Slc)$.
As discussed in the previous section, these can be written locally in terms of the Fenchel--Nielsen coordinates $(x_m,y_m)$ and $(\bar x_m, \bar y_m)$ as
\be
\vth=\lt(-\frac{t}{2\pi}\rt)\sum_{m=1}^{3g-3}\ln y_m\frac{\d x_m}{x_m}
\qquad\text{and}\qquad 
\bar{\vth}=\lt(-\frac{\bar{t}}{2\pi}\rt)\sum_{m=1}^{3g-3}\ln \bar{y}_m\frac{\d \bar{x}_m}{\bar{x}_m}.
\ee
In these formulas, $\a$ labels the branches of the Lagrangian subvariety $\cl_{\mathbf{A}}$ that arise from solving ${\mathbf{A}}_m(x,y)=0$ and on which the $v_m{(\a)}$ are single-valued functions of $u_m$. 

The integral in equation \eqref{block} is performed along a contour $\Fc$ within the Lagrangian subvariety $\cl_{\mathbf{A}}$ connecting the flat connection $\lt(u,v{(\a)}\rt)$ in the branch $\a$ to a reference flat connection $\lt(u_0,v_0\rt)$. In our context, both flat connections at the end points of $\Fc$ are covered by a single FN coordinate chart. 
The last two equations provide the starting point of our semiclassical analysis leading to 4d simplicial quantum gravity in the next section.\\

For now, we conclude this discussion with a series of more technical remarks, which can be skipped on a first reading. 

\paragraph*{\it Overall phase} The freedom in fixing the overall phase of the wave function $Z_{CS}\lt(M_3\,\big|\,u,\bar{u}\rt)$ is, of course, related to the choice of a reference flat-connection $(u_0,v_0),\ (\bar{u}_0,\bar{v}_0)$. Let $(u,\bar{u})$ be the boundary values defining the path integral $Z_{CS}\lt(M_3,\big|\,u,\bar{u}\rt)$; we can choose the reference flat connection to be a pair of solutions $\lt(u,v{(\a_0)}\rt),\ \lt(\bar{u},\bar{v}{(\bar{\a}_0)}\rt)$ of ${\mathbf{A}}_m(u,v)=0$, such that $(\a_0,\bar{\a}_0)$ denote reference branches. Then, the phase difference between another pair of flat connections $\lt(u,v{(\a)}\rt),\ \lt(\bar{u},\bar{v}{(\bar{\a})}\rt)$ in the branches $\a,\bar{\a}$ and the reference pair $\lt(u,v{(\a_0)}\rt),\ \lt(\bar{u},\bar{v}{(\bar{\a}_0)}\rt)$ will be given by 
\be
Z_{CS}^{(\a)}\lt(M_3\big|\,u\rt)=\exp\lt[\frac{i}{\hbar}\int\limits^{\lt(u,v{(\a)}\rt)}_{\substack{(u,v{(\a_0)})\\ \Fc\subset\cl_{\mathbf{A}}}}\,\vth+\cdots\rt],\ \ \ \ 
Z_{CS}^{(\overline{\a})}\lt(M_3\big|\,\bar{u}\rt)=\exp\lt[\frac{i}{\hbar}\int\limits^{\lt(\bar{u},\bar{v}{(\overline{\a})}\rt)}_{\substack{\lt(\bar{u},\bar{v}{(\overline{\a}_0)}\rt)\\ \Fc\subset\cl_{\mathbf{A}}}}\,\bar{\vth}+\cdots\rt].
\label{ZZ0}
\ee
~\\

\paragraph*{\it Integer $k$} When $\mathrm{Re}(t)=k\in\Z$ and $\hbar^{-1}\in\Z$ (once again, $\hbar^{-1}$ is here understood solely as a scaling parameter for the couplings $(t,\bar{t})$), the Lagrangian subvariety $\cl_{\mathbf{A}}$ becomes quantizable, which means that the integrals of equation \eqref{block} do not depend on the choice of the contour since $\oint \vartheta \in 2\pi  \hbar \Z $ on $\cl_{\mathbf{A}}$. 
This fact has a beautiful algebraic $K$-theoretical interpretation: indeed, $\cl_{\mathbf{A}}$ is Lagrangian in a stronger sense, i.e. it is a $K_2$-\emph{Lagrangian subvariety} \cite{DGG,ABQ,DV}. A very brief explanation of this fact is given in Appendix \ref{Ktheory}. { In the case of knot-complement 3-manifolds, the fact that $\cl_{\mathbf{A}}$ is quantizable was understood very early on by \cite{NeumannZagier,Yoshida,Hodgson,Dunfield}. }

\paragraph*{\it Logarithmic variables} Although the Lagrangian subvariety $\cl_\mathbf{A}$ is defined by $\mathbf{A}_m(x,y)=0$ in terms of the variables $(x_m,y_m)$, the holomorphic 3d block $Z_{CS}^{(\a)}(M_3|u)$ is rather a function of the logarithmic coordinates $u_m$, satisfying equation \eqref{AZ=0}. 
This means that $Z_{CS}^{(\a)}(M_3|u)$ need not, in general, be a periodic function of $u$ under $u\to u+2\pi i$. 
Therefore, $Z_{CS}^{(\a)}(M_3|u)$ and $Z_{CS}^{(\a')}(M_3|u)$ have to be considered two {\it different} 3d holomorphic blocks even when $v{(\a)}$ and $v{(\a')}$ give the same $y_m=e^{-\frac{2\pi}{t} v_m}$. 
The reason is essentially that $Z_{CS}^{(\a)}(M_3|u)$ is defined by the path integral of an analytic continuation of Chern--Simons theory with $t$ extended to an arbitrary complex number (see \cite{analcs}, as well as the second reference in \cite{knots}):
by relaxing the requirement that $k\in\Z$, one defines $Z_{CS}^{(\a)}(M_3|u)$ as a path integral on the {\it covering space} of gauge equivalent classes of connections, which means that configurations related by {\it large} gauge transformations should not be identified. 

{The integration contour $\Fc$ appearing in the formulas above strictly speaking lies in the cover space of $\cl_{\mathbf{A}}$. 
In the analytic continued Chern--Simons theory, we have $\oint v\cdot\d u=0$ on the cover space of $\cl_{\mathbf{A}}$, which is explained in Appendix \ref{Ktheory}. 
Hence, fix $u$, and consider $(u,v{(\a)})$ and $(u,v{(\a')})$ two different solutions corresponding to the same flat connection $(x,y)$ on $M_3$, where $v{(\a)},v{(\a')}$ are different lifts of $y=e^{-\frac{2\pi}{t} v}$ to the cover space. Then, $v{(\a)}$ and $v{(\a')}$ differ by an integer multiple of $ i t$. Thus the classical terms in equation \eqref{block}, which are $\int^{(u,v{(\a)})}\vth$ and $\int^{(u,v{(\a')})}\vth$, must differ by an integer multiple of $i t\, u$. This can be used to show that there is no difference in the quantum corrections between $S^{(\a)}_n(u)$ and $S^{(\a')}_n(u)$ \cite{DGLZ}. }

\subsection{Asymptotics of holomorphic 3d Block and Simplicial Quantum Gravity}\label{3dblock}

After this general review on the quantization of flat connections on a three manifold, we turn our attention back to those connections satisfying the ``geometricity'' boundary conditions that we introduced in the first part of this paper.
These boundary conditions allow a one-to-one mapping between flat connections and homogeneously curved 4d simplicial geometries. 

The content of the needed boundary conditions is the following: in the vicinity of each vertex of $\G_5\in \mathbb S^3$, the $\Slc$ flat connection reduces to an $\Su$ flat connection. 
To express this conditions in terms of the FN coordinates, we introduce an adapted pair of pants decomposition of $\Sigma_6$, the tubular neighborhood of $\G_5$.
Since $\Sigma_6$ is already naturally decomposed into five 4-holed spheres $\mathcal S_a$, $a=1,\dots,5$, we just need to split each of these into two pairs of pants $(\mathcal T_a, \mathcal T'_a)$. As a result, we obtain the following two sets of FN coordinates: $\{x_{ab},y_{ab}\}_{a>b}$  which are attached to the tubes connecting two 4-holed spheres $\mathcal S_a$ and $\mathcal S_b$, and $\{ x_a, y_a\}_a$, which resolve the internal structure of each 4-holed sphere $\mathcal S_a$. 

As a consequence of the boundary conditions, the FN lengths coordinates must have unit norm, i.e.
\be
x_{ab}, \, x_a \in \mathrm{U}(1),
\ee 
and, in addition, the pairs $\{x_a,y_a\}$ have to parametrize an $\Su$ flat connection on $\mathcal S_a$ with given conjugacy classes $\{x_{ab}\}_{b,b\neq a}$ associated to its holes.

Thanks to the geometric correspondence explained in Section \ref{3/4}, a holomorphic 3d block $Z^{(\a)}(M_3|u)$ that solves the A-polynomial equation \eqref{AZ=0}, and moreover, satisfies the above boundary conditions can be readily interpreted as a quantum state of a 4d simplicial geometry peaked around a particular classical geometry.
This peakedness cannot be arbitrarily sharp, due to the Heisenberg relations between $x_a$ and $y_a$.

In this section, we analyze the asymptotic behavior of such a $Z^{(\a)}(M_3|u)$ as $\hbar\to 0$, and find evidence that it corresponds to a {\it physical} state of simplicial 4d Quantum Gravity. \\

Consider a set of boundary data $\lt[x_{ab}; x_a, y_a\rt]$ satisfying the geometricity (and non-degeneracy) conditions. Theorem \ref{parity} states that there are then exactly two connections $A$ and $\tilde A $ in $\cm^\text{BC}_\text{flat}(M_3=S^3\setminus \G_5,\Slc)$ that are consistent with the boundary data, and which correspond precisely to the two orientations of a geometric homogeneously curved 4-simplex. 

These two bulk connections $A$ and $\tilde A$ in $M_3=\mathbb S_3\setminus \G_5$, induce on the boundary $\partial M_3=\Sigma_6$ two different flat connections, which we call $\Fa$ and  $\tilde{\Fa}$, respectively.
The two connections, $\Fa=\lt[{x}_{ab},y_{ab};{x}_a,{y}_a\rt]$ and $\tilde{\Fa}=\lt[{x}_{ab},\tilde{y}_{ab};{x}_a,{y}_a\rt]$, are covered by a single FN coordinate chart. The 10 twist variables $y_{ab}$ differ from $\tilde{y}_{ab}$ by a parity transformation in the simplicial geometry, and must therefore be related by a simple transformation.
As the reconstruction theorem of Section \ref{3/4} suggests $x_{ab} = e^{u_{ab}}$, $y_{ab}= e^{-\frac{2\pi}{t}v_{ab}}$ and $\tilde y_{ab}= e^{-\frac{2\pi}{t}\tilde v_{ab}}$ are related to the areas and (hyper)dihedral angles of the 4-simplex in the following manner:
\be
u_{ab}&=&-i\nu\frac{\L}{6}\mathbf{a}_{ab}+i\pi\fs_{ab} + 2\pi i M_{ab},\\
v_{ab}{(\a)}&=&\frac{t}{4\pi}\nu\,\Theta_{ab}+\frac{it}{2\pi}\nu\,\theta_{ab}-\frac{t\ln\chi_{ab}(\xi)}{4\pi}-{it}\nu N'_{ab},\label{vandTheta}\\
\tilde{v}_{ab}{(\tilde{\a})}&=&-\frac{t}{4\pi}\nu\,\Theta_{ab}+\frac{it}{2\pi}\nu\,{\theta}_{ab}-\frac{t\ln\chi_{ab}(\xi)}{4\pi}-{it}\nu\tilde{N}'_{ab}\label{uvATheta},
\ee
where recall $v\in\{\pm1\}$ is a global sign, $\fs_{ab}\in\{0,1\}$, and $M_{ab}, \,N_{ab}, \, \tilde N_{ab} \in \mathbb Z$ are arbitrary integers related to the (necessitated) lift to logarithmic FN variables $(u,v)$. The relation between $u$ and the areas and between $v$ and the boost hyperdihedral angles are made more plausible by Eqs. \Ref{tildeHAr} and \Ref{GdihedralRel}, for further details and a proof see \cite{HHR}. 

A similar lift is presupposed to be chosen for the variables $\lt[x_a,y_a\rt]$ which parametrize the shape of the five tetrahedra of fixed areas $\{\mathbf{a}_{ab}\}$.\footnote{A tetrahedron is completely fixed by its 6 edges, therefore to the 4 areas two more parameters have to be added. See \cite{HHKR} for a detailed analysis of this fact in the homogeneously curved case.}
In the following, an important role will be played by the difference between $v_{ab}{(\a)}$ and $\tilde{v}_{ab}{(\tilde{\a})}$. This  is given by
\be
v_{ab}{(\a)}-\tilde{v}_{ab}{(\tilde{\a})}=\frac{t}{2\pi}\nu\Big(\Theta_{ab}+2\pi i{N}_{ab}\Big),
\ee 
where $ N_{ab}=\tilde{N}'_{ab}-N'_{ab}\in\Z$ (note that the logarithm branches of $\a$ and $\tilde \a$ need not be related).

As we have already discussed, there is an overall phase ambiguity in the 3d holomorphic blocks. 
Of course, this ambiguity cannot be removed, since it is intrinsic to the quantum formalism. 
However, what really matters in the WKB scheme discussed above, is the {\it phase difference} between the various contributions.
This quantity has an absolute meaning, and it is exactly what we are going to evaluate.
A convenient way of doing this is to fix the reference connection in the integrals of equation \eqref{block} (or \eqref{ZZ0}) to be e.g. $(u,v{(\tilde \a)})$. In this way, the phase we are interested in calculating is the leading order of 
\be
Z^{(\a)}\lt(M_3\big|\,u\rt)=\exp\lt[ \frac{i}{\hbar} I^\a_{\tilde \a}\Big(u,v(\a),v(\tilde \a) \Big) +\cdots\rt] \qquad \text{with} \qquad
I^\a_{\tilde \a}\Big(u,v(\a),v(\tilde\a) \Big)  = \int\limits^{\lt(u,v{(\a)}\rt)}_{\substack{\lt(u,v{(\tilde{\a})}\rt)\\ \Fc\subset\cl_{\mathbf{A}}}}\vth.
\ee
For completeness, we recall here that the Liouville 1-form $\vth$ is given by
\be
\vth=\sum_{a<b} v_{ab}{\d u_{ab}}+\sum_{a=1}^5v_{a}\d u_{a},
\ee
where $\d$ is a finite dimensional differential in the $(u,v)$ space (we adopted this notation to avoid confusion with the differential on $M_3$ or $\Sigma_6$).

\begin{figure}[h]
\begin{center}
\includegraphics[width=7cm]{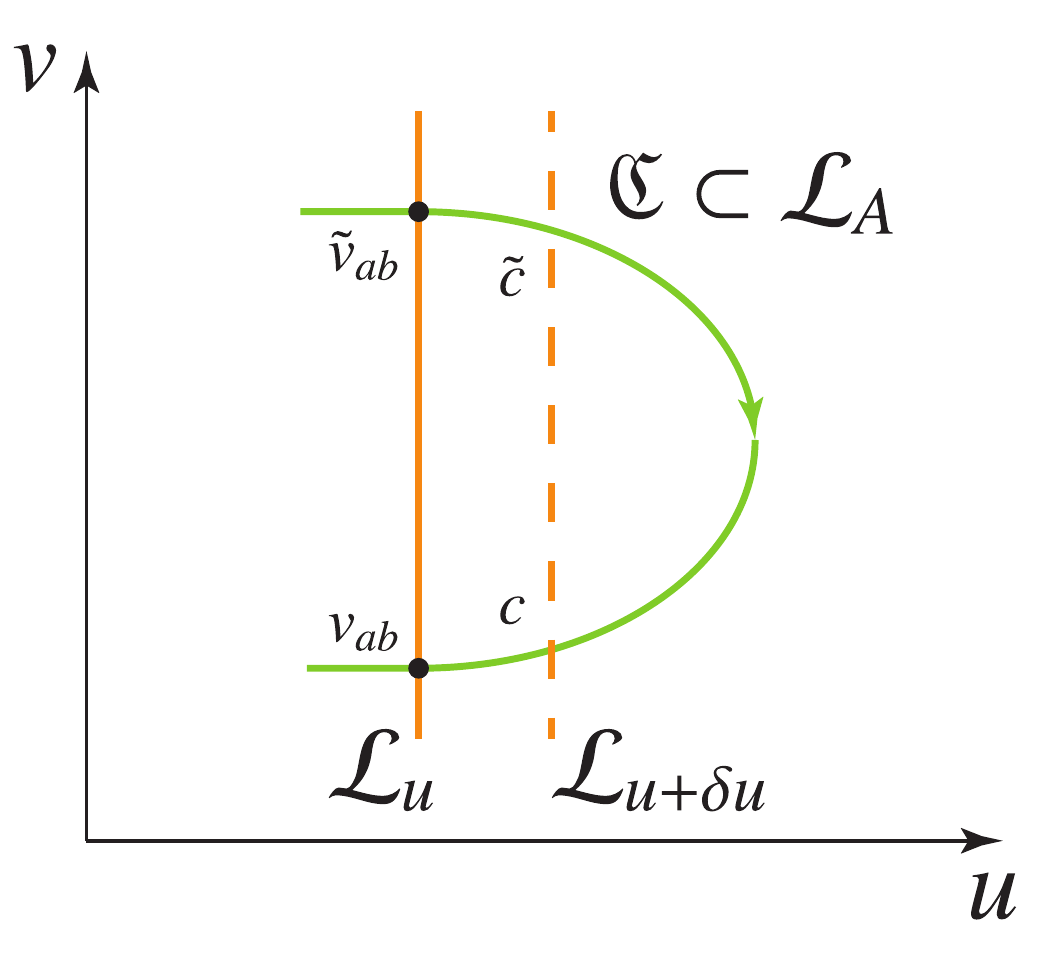}
\caption{The Lagrangian subvariety $\cl_{\mathbf{A}}$ and the plane $\cp_{\lt[{x}_{ab};{x}_a,{y}_a\rt]}$ intersect at 2 different points. The bent (green) curve is the integration contour $\Fc$ lying in $\cl_{\mathbf{A}}$, and connects the pair of intersection points. The (orange) vertical line represents the plane $\cp_{\lt[{x}_{ab};{x}_a,{y}_a\rt]}$ and intersects $\Fc$ at 2 points. The dashed (orange) vertical line represents the variation $\cp_{\lt[{x}_{ab};{x}_a,{y}_a\rt]}(\eta)$ from $\cp_{\lt[{x}_{ab};{x}_a,{y}_a\rt]}$. The second plane $\cp_{\lt[{x}_{ab};{x}_a,{y}_a\rt]}(\eta)$ intersects $\cl_{\mathbf{A}}$ at a different pair of points, which are also connected by the extended integration contour $\Fc(\eta)$. The 2 (green) segments in between the (orange) line and dashed (orange) line are the curve extensions $\delta\Fc=c\cup\tilde{c}$. In this figure we suppress the coordinates $x_a,y_a$. 
}
\label{intersection}
\end{center}
\end{figure}

To evaluate the integral above, it is useful to have a more geometric picture in mind. 
Recall that the set of solutions to the operator constraint of equation \eqref{AZ=0} defines a Lagrangian subvariety $\cl_{\mathbf A}=\cm_\text{flat}(M_3, \Slc)$ within $\cm_\text{flat}(\partial M_3, \Slc)$. 
Theorem \ref{parity} can be rephrased as stating that the plane $\cp_{\lt[{x}_{ab};{x}_a,{y}_a\rt]}$ of constant $\lt[{x}_{ab};{x}_a,{y}_a\rt]$ intersects $\cl_{\mathbf A}$ in precisely two points, $\lt[{x}_{ab}, y_{ab};{x}_a,{y}_a\rt]$ and $\lt[{x}_{ab}, \tilde y_{ab};{x}_a,{y}_a\rt]$, corresponding to two 4-simplices differing only be their parities.
This is schematically represented in Figure \ref{intersection}.
The idea is that, instead of directly attempting the calculation of the integral $I^\a_{\tilde\a}(u,v(\a),v(\tilde \a))$, we first evaluate its variation under a slight change of the planes $\cp_{\lt[{x}_{ab};{x}_a,{x}_a\rt]}$---or more precisely of their lifts $\cp_{\lt[{u}_{ab};{u}_a,{v}_a\rt]}$---and then integrate this variation.

To do this, we introduce a one-parameter family of boundary data $\lt[{x}_{ab}(\eta);{x}_a(\eta),{y}_a(\eta)\rt]$, with $\eta\in[0;1]$,  all compatible with some 4-simplex geometry.
This family can be readily lifted to $\lt[{u}_{ab}(\eta);{u}_a(\eta),{v}_a(\eta)\rt]$. The variations involved in this family being smooth, they do not allow for changes in the lifts nor in the branches $\a$ and $\tilde\a$ in which the intersections $v_{ab}^{(\a)}(\eta)$ and $v_{a}^{(\tilde\a)}(\eta)$ live. 
Hence, we define the variation
\be
\delta_\eta I^\a_{\tilde \a}(\eta) = I^\a_{\tilde{\a}}(\eta+\delta\eta)-I^\a_{\tilde{\a}}(\eta)
\ee
where $I^\a_{\tilde{\a}}(\eta)$ is a shorthand notation for $I^\a_{\tilde{\a}}\Big(u(\eta),v{(\a)}(\eta),v^{(\tilde \a)}(\eta)\Big)$.
Because $\mathcal{L}_A$ is Lagrangian there is a freedom in the contour of integration $\frak C$ entering the definition of $I^\a_{\tilde \a}$, we use this freedom to deform the contour so that it contains the path parametrized by $\eta$, that is ${\frak C}\supset C_\eta$ where 
\be
C_\eta = \bigcup_\eta \Big(\cl_{\mathbf A}\cap \cp_{[u_{ab}(\eta);u_a(\eta),v_a(\eta)]}\Big).
\ee

Now, $C_\eta$  is composed by two portions, $C_\eta = c \cup \tilde c$, contained in the branches $\a$ and $\tilde \a$, respectively (see Figure \ref{intersection}).
As a result, $\delta_\eta I^\a_{\tilde \a}$ can be expressed as the sum of two line integrals contained in $c$ and $\tilde c$. 
Developing these integrals at first order in $\delta \eta$, we find
\be
\delta_\eta I^\a_{\tilde \a} = \lt(\;\int\limits_{\substack{0\\ c\subset \cl_{\mathbf A}}}^{\delta\eta} - \int\limits_{\substack{0\\ \tilde c\subset \cl_{\mathbf A}}}^{\delta\eta} \rt) \lt( \sum_{a<b} v_{ab}\d u_{ab} + \sum_{a} v_a \d u_a \rt) = \sum_{a<b} \lt( v{(\a)}_{ab} - v{(\tilde a)}_{ab}\rt) \d_\eta u_{ab} + \cdots.
\ee
Here the dots stand for second order corrections in $\delta \eta^2$, while $\delta_\eta u_{ab}$ is the first order development of $u_{ab}(\eta + \delta \eta) - u_{ab}(\eta)$. Furthermore, we used the fact that the contributions coming from the two branches to the integral of $\sum_{a} v_a \d u_a$ cancel each other (exactly). This is because the integrations along each section have opposite orientation and $v_a$ and $u_a$ are the same on each portion because they are fixed by the boundary conditions. 

Now, using the geometric correspondence of equation \eqref{uvATheta}, one finds that at first order
\be
\delta_\eta I^\a_{\tilde \a} = \lt( \frac{\Lambda t}{12 \pi i} \rt) \sum_{a<b}   \Theta_{ab} \delta_\eta \mathbf{a}_{ab}+ \lt( \frac{\Lambda t}{6} \rt) \sum_{a<b}  N_{ab} \delta_\eta \mathbf{a}_{ab}.
\ee
Since $\eta$ was supposed to parametrize a continuous family of actual 4-simplex geometries, the above variation can be integrated thanks to the Schl\"afli identities, which state that for a continuous family of homogeneously curved 4-simplices,
\be
 \delta_\eta \text{Vol}_4^\Lambda =\Lambda^{-1} \sum_{a<b}\mathbf{a}_{ab} \delta_\eta \Theta_{ab} , 
\ee
where $\text{Vol}_4^\L$ is the 4-volume of the homogeneously curved 4-simplex of curvature $\Lambda$. Hence, 
\be
I^\a_{\tilde{\a}}=\lt(\frac{\L t}{12\pi i}\rt)\lt(\sum_{a<b} \mathbf{a}_{ab}\Theta_{ab}-\L \mathrm{Vol}^\L_4+C^\a_{\tilde{\a}}\rt)+\lt(\frac{\L t}{6}\rt)\sum_{a<b} {N}_{ab}{\mathbf{a}_{ab}}.
\ee
Notice that in this expression the dependence of branches $\a,\tilde{\a}$ is contained in the integration constant $C^\a_{\tilde{\a}}$, as well as in the terms $\lt(\frac{\L t}{6}\rt)\sum_{a<b}{N}_{ab}{\mathbf{a}_{ab}}$. We will comment more about them in the later paragraphs.
The proof of Schl\"afli identity can be found in e.g. \cite{curvedRegge,eva}, see also \cite{HHKL} for a symplectic and semiclassical perspective.

From the previous equations, we deduce the following leading order expression for the holomorphic 3d block with our boundary condition imposed, $Z_{BC}^{(\a)}(M_3| u)$:
\be
Z_{BC}^{(\a)}(M_3| u) = \exp\lt[ \frac{i}{\hbar} \lt(\frac{\L t}{12\pi i}\rt)\lt(\sum_{a<b} \mathbf{a}_{ab}\Theta_{ab}-\L\mathrm{Vol}^\L_4\rt)+\frac{i}{\hbar}\lt(\frac{\L t}{12 \pi i}\rt)C^\a_{\tilde{\a}} +\frac{i}{\hbar}\lt(\frac{\L t}{6}\rt)\sum_{a<b}  {N}_{ab}{\mathbf{a}_{ab}} + \cdots\rt].
\ee

To go further with our analysis, it is important to recall that---although the holomorphic 3d block studied above is the fundamental unit of the CS quantum state---the full quantum state is given by a sum of products of holomorphic and antiholomorphic blocks, as in equation \eqref{WKB}.
In particular, the product of the holomorphic block above and its antiholomorphic counterpart gives
\be
&&Z_{BC}^{(\a)}\lt(M_3\big|\,u\rt)Z_{BC}^{(\bar {\a})}\lt(M_3\big|\,\bar{u}\rt)=\nonumber\\
&&\qquad \exp\lt[\frac{i}{\hbar}2\mathrm{Re}\lt(\frac{\L t}{12\pi i}\rt)\lt(\sum_{a<b} \mathbf{a}_{ab}\Theta_{ab}-\L \mathrm{Vol}^\L_4\rt)+\frac{i}{\hbar}2\mathrm{Re}\lt(\frac{\L t}{12\pi i}C^\a_{\tilde{\a}}\rt)+\frac{i}{\hbar}2\mathrm{Re}\lt(\frac{\L t}{6}\rt)\sum_{a<b}  {N}_{ab}\mathbf{a}_{ab}+\cdots\rt]. \nonumber\\\label{ZZasymp}
\ee
The anti-holomorphic block is defined by complex conjugation of $(u, v{(\alpha)})$ with the reference being the complex conjugate of $(u, \tilde{v}{(\tilde{\alpha})})$.

{
Given our phase convention, we find that at leading order in $\hbar$ the phase of the above product vanishes for the branch given by $\a \to \tilde \a$.
Therefore, assuming that the coefficient $\fn_{\a,\bar\a}$ in equation \eqref{WKB} is the same for $\a$ and $\tilde \a$ (which is more than reasonable given the symmetry which relates the two branches), we find that the total wave function---up to an irrelevant global phase---is
\be
Z_{BC}(M_3| u,\bar u) \sim \cos\lt[\frac{\L \mathrm{Im}(t)}{12\pi  \hbar}\lt(\sum_{a<b} \mathbf{a}_{ab}\Theta_{ab}-\L \mathrm{Vol}^\L_4\rt)+\mathrm{Im}\lt(\frac{\L t}{12\pi \hbar}C^\a_{\tilde{\a}}\rt)+\mathrm{Re}\lt(\frac{\L t}{6\hbar}\rt)\sum_{a<b}  {N}_{ab}\mathbf{a}_{ab}+\cdots\rt]. \notag\\
\ee
This is our main result. A few comments are in order.
First of all let us explain the notation:
$\sim$ highlights the fact that an irrelevant overall phase has been neglected (and, conversely, the presence of a cosine highlights the relevant interference between the two branches), whereas the ellipsis $\cdots$ indicates that, as usual, only the leading order in $\hbar$ has been taken into account. }
Second, and most importantly, in the expression above we recognize the appearance of the Hamilton--Jacobi functional for General Relativity on a homogeneously curved 4-simplex. This is the on-shell Regge action for such a 4-simplex \cite{Taylor:2005,curvedRegge}:
\be
S_\text{Regge} = \frac{1}{8\pi \rm G_N}\lt(\sum_{a<b} \mathbf{a}_{ab}\Theta_{ab}-\L \mathrm{Vol}^\L_4\rt).
\ee
This observation allows us to identify the (inverse) Newton constant with the imaginary part of the CS coupling $t$:
\be
{\rm G_N} = \lt| \frac{3}{2 \L {\rm Im}(t)} \rt|.
\label{Gnewton}
\ee
Or, in terms of the (squared) Planck length $\ell_\text{Pl}^2 = 8 \pi \hbar {\rm G_N}$,
\be
\lt|\frac{{\rm Im}(t)}{\hbar} \rt| = \lt|\frac{12\pi}{\L\ell_\text{Pl}^2} \rt|,
\ee
where dimensionless quantities are now being compared.
This relation with the on-shell Regge action is what allows us to claim a relation between the quantization of $\Slc$ flat connections on $S^3\setminus \G_5$ with appropriate boundary conditions and simplicial 4d Quantum Gravity with cosmological constant.

The Regge action, though, is not the only term appearing in the leading order expression of $Z_{BC}(M_3|u,\bar u)$. 
This takes us back to the integration constant $C^\a_{\tilde \a}$, which---as such---must be independent of the geometry. {This constant is actually expected to depend on the behavior of $\cl_{\mathbf A}$ at the singularity $y_{ab}=\tilde y_{ab}$ where the two branches meet, and the geometry degenerates. These kinds of contributions have been studied extensively in the literature on WKB and semiclassical approximations, where they are known as Maslov indices \cite{Berry:1972, Esterlis:2014}. This term is similar to the phase offset which appears in the asymptotics of the 3d Ponzano--Regge model with respect to the standard 3d Regge action \cite{PR, Roberts:1999,Aquilanti:2012}. }
Finally, we are left with the ambiguity associated with the logarithmic lifts, which is given by
\be
\mathrm{Re}\lt(\frac{\L t}{6\hbar}\rt)\sum_{a<b} {N}_{ab}\mathbf{a}_{ab}.
\ee
To start with, let us notice that this ambiguity does not affect the asymptotics if ${\rm Re}(t)=0$. 
However, requiring that this ambiguity is not present in the generic case is equivalent to asking that the areas $\mathbf{a}_{ab}$ of the 4-simplex triangles be quantized:
\be
\mathbf{a}_{ab} \in \frac{12\pi \hbar}{\L {\rm Re}(t)} \Z.
\label{Qcondition}
\ee
This corresponds to an equidistant spacing in the area spectrum given by
\be
\Delta \mathbf{a} = \frac{{\rm Im}(t)}{{\rm Re}(t) } \ell_\text{Pl}^2.
\ee
{This condition will be analyzed again in the next section, where it will acquire a special meaning in relation to an explicit imposition of the desired boundary conditions inspired by Loop Quantum Gravity.}

\section{Wilson Graph Operator and Boundary Conditions}\label{wilson}

In the previous sections we studied complex Chern--Simons theory on the $\G_5$ graph complement 3-manifold $M_3$, with certain boundary conditions.
We also saw how our boundary conditions correspond to quantum states of the Chern--Simons theory on $\partial M_3 = \Sigma_6$ encoding quantum 4-simplicial geometries with a semiclassical amplitude given by a discretized form of the Einstein--Hilbert functional. 

We now show how these precise boundary conditions can be imposed by introducing a specific Wilson graph operator supported on $\G_5$ within the complex Chern--Simons theory. In this section, we restrict our analysis to the case where $\mathrm{Re}(t)=k\in\mathbb Z$ and $\hbar^{-1}\in \mathbb Z$.

The idea is the following: a general feature of topological quantum field theory is that imposing specific boundary conditions on the path integral (on $M_3$) corresponds to evaluating the amplitude of a specific quantum state in $\ch(\Sigma_6)$, associated to the boundary $\partial M_3=\Sigma_6$ (see e.g. \cite{knots}). In formulas 
\be
\A [x_{ab};x_a,y_a]:= \lag\,\Psi^{\G_5}_{[x_{ab};x_a,y_a]}\,\Big | \, Z_{CS}(M_3)\,\rag_{\ch(\Sig_6)}, \label{inner}
\ee
where $\Psi^{\G_5}_{[x_{ab};x_a,y_a]}\in \ch(\Sig_6)$ imposes the boundary conditions $[x_{ab};x_a,y_a]$.
More explicitly, the state $\Psi^{\G_5}_{[x_{ab};x_a,y_a]}$ can be defined via a path integral through the insertion of a Wilson graph operator at the center of the tubular neighborhood of $\G_5$, $N(\G_5)$. Indeed the boundary of $N(\G_5)$ is $\bar \Sigma_6$, i.e. identical to $\partial M_3$, but with opposite orientation---a fact that ensures the above contraction is natural.

Specifically, we consider $\Slc$ Chern--Simons theory on $N(\G_5)$ and define a knotted Wilson graph operator $\G_5^{[j_{ab},\xi_{ab}]}[A,\bar A]$ located at the core of $N(\G_5)$ \cite{HHR}, such that $\Psi^{\G_5}_{[x_{ab};x_a,y_a]}$ can be written as 
\be
\Psi^{\G_5}_{[x_{ab};x_a,y_a]}\lt(A_1,\bA_1\rt):=\int_{A_1,\bA_1}\cd A\cd \bA\ e^{\frac{i}{\hbar}\,CS\lt[N(\G_5)\,|\,A,\bA\rt]}\ \G_5^{[j_{ab},\xi_{ab}]}[A,\bar A].\label{psigamma5}
\ee 
The relation between the operator labels $[j_{ab},\xi_{ab}]$ and the state or boundary condition labels $[x_{ab}; x_a,y_a]$ will be spelled out soon.
With the above definitions, the properties of the inner product in $\ch(\Sigma_6)$ ensure that
\be
\A [x_{ab};x_a,y_a]=\int\cd A\cd \bA\ e^{\frac{i}{\hbar}\,CS\lt[S^3\,|\,A,\bA\rt]}\ \G_5^{[j_{ab},\xi_{ab}]}[A,\bA].\label{API}
\ee

We come now to the definition of the knotted Wilson graph operator $\G_5^{[j_{ab},\xi_{ab}]}[A,\bA]$ (see also \cite{HHR}), which is conveniently presented as a list:

\begin{itemize}

\item Each edge $\ell_{ab}$ connecting two 4-valent vertices of the graph $\G_5$ is labeled by a unitary irreducible representaiton (``irrep'') of $\Slc$ in the principal series (such irreps are necessarily infinite dimensional as a consequence of the non-compactness of $\Slc$). 
These representations are required to be of a specific form.
Before specifying this form, let us recall that the unitary irreps of $\Slc$ in the principle series depend on two parameters $(j,\rho)$, with $j\in \frac12 \mathbb Z^+ $ and $\rho\in\R$ \cite{ruhl}.
Moreover, these irreps can be decomposed as an infinite tower of SU(2) irreps, i.e. their Hilbert spaces decompose as $V^{j,\rho}=\oplus_{k\geq j}V_k$, where $V_k$ is the SU(2) irrep with spin $k\in\frac12\mathbb N$. 
Using this decomposition, a basis of $V^{j,\rho}$ is given by $|(j,\rho);k,m\rangle$. 
Coming back to our own Wilson graph operator $\G_5^{[j_{ab},\xi_{ab}]}$, we require that the specific irreps attached to the edges $\ell_{ab}$  have the form $(j_{ab},\rho_{ab})=(j_{ab}, \gamma j_{ab})$, for some fixed  $\gamma \in \mathbb R$.\footnote{In the next section we will see that $\g$ corresponds to the Barbero--Immirzi parameter of Loop Quantum Gravity.}

\item Each of the two end points of an edge $\ell_{ab}$ in $\G_5$ is equipped with an $\Su$ Perelemov coherent state, $|j_{ab},\xi_{ab}\rangle\in V_{j_{ab}}$, and  $|j_{ba},\xi_{ba}\rangle\in V_{j_{ab}}$, respectively. The state $|j,\xi\rangle$ is defined via an $\Su$ action on the highest weight vector $|j_{ab},j_{ab}\rangle$ \cite{coherentstate}. Specifically, denoting the Wigner matrix of $g$ in the $\Su$ irrep $V_j$ as $D^j(g):V_j\to V_j$, we have
\be
|j,\xi\rangle:=D^j(g_\xi)|j,j\rangle
\quad\text{where}\quad
 g_\xi\equiv \begin{pmatrix}
  \xi^1 & -\bar{\xi}^2  \\
  \xi^2 & \ \ \ \bar{\xi}^1
 \end{pmatrix}\in\Su,
\ee 
for some normalized 2-spinor $\xi$, $\langle \xi, \xi\rangle = \bar{\xi}^{1}\xi^1+\bar{\xi}^2\xi^2=1$. 
The coherent states $|j,\xi\rangle$ form an over-complete basis in $V_j$ and provide the resolution of the identify
\be
\mathbf{1}_j=(2j+1)\int_{S^2} \rmd\mu( \xi)\ |j,\xi\rangle\langle j,\xi |.
\ee 
Here, the integration domain is the coset space $S^2=\mathrm{SU(2)}/\mathrm{U(1)}$, since $|j,\xi\rangle\mapsto e^{i\phi}|j,\xi\rangle=|j,e^{i\phi}\xi\rangle$ leaves the integrand invariant. 
The phase convention for $\xi$ needs to be fixed by convention in order to define the coherent state basis. 
Once  the phase information is fixed, we can think of $|j,\xi\rangle$ as labeled by a unit 3-vector $\hat{n}$, rather than a spinor.
Indeed, in its spin 1 representation, $g_\xi\in\Su$ rotates the 3-vector $\hat{z}=(0,0,1)^t$ to the unit vector $\hat{n}_\xi=\lag\xi,\vec{\sig}\xi\rag$, where $\vec{\sig}$ is the vector of Pauli matrices. 

Since the edges of $\G_5^{[j_{ab},\xi_{ab}]}$ are labeled  by irreps of $\Slc$, and thus naturally carry $\Slc$ group elements, we need to produce states in $\ch^{j,\gamma j}$ from the $\Su$ coherent states just described.
This is achieved using the injection map 
\be
{Y}:V_j\hookrightarrow V^{j,\g j},\quad |j,\xi\rangle \mapsto Y|j,\xi\rangle :=|(j,\g j);j,\xi\rangle,
\ee
which identifies the $\Su$ irrep $V_j$ with the lowest subspace in the tower $V^{j,\g j}=\oplus_{k\geq j} V_k$. 
At the end of this construction, the two end points of the edges $\ell_{ab}$ carry the two states $|(j_{ab},\g j_{ab});j_{ab},\xi_{ab}\rangle$ and $|(j_{ab},\g j_{ab});j_{ab},\xi_{ba}\rangle$ in $V^{j_{ab},\g j_{ab}}$.

\item Finally, the Wilson graph operator $\G_5^{[j_{ab},\xi_{ab}]}[A,\bA]$ is defined by a product over all edges $\ell_{ab}$ of inner products in each $V^{j_{ab},\g j_{ab}}$:\footnote{This knotted Wilson graph operator is strictly related to the projected spin-network functions of $\Slc$ \cite{EPRL,DL}.}
\begin{align}
&\G_5^{[j_{ab},\xi_{ab}]}[A,\bA]:=\lt[\prod_{a=1}^5\int_{\Slc}\rmd g_a\rt]\prod_{a<b}\lag (j_{ab},\g j_{ab});j_{ab},\xi_{ab} \Big|\,g_a^{-1}G_{ab}g_b\,\Big|(j_{ab},\g j_{ab});j_{ab},\xi_{ba}\rag, \label{G_5operator}
\end{align}
where
\be
G_{ab}=\cp \exp\int_{\ell_{ab}} A
\ee
is the holonomy of $A$ along $\ell_{ab}$ oriented from $b$ to $a$.
Note that $\G_5^{[j_{ab},\xi_{ab}]}[A,\bA]$ is gauge invariant thanks to the Haar integrals $\prod_{a=1}^5\int_{\Slc}\rmd g_a$ (in fact, one of these integrals is completely redundant for this purpose, and has to be dropped to avoid meaningless divergences). 
Importantly, these inner products are {\it not} holomorphic functions on the complex group $\Slc$, since they come from unitary irreps.

\item In formulas \Ref{psigamma5} and \Ref{API},  it is convenient to make a partial gauge fixing. Making use of the invariance of $e^{\frac{i}{\hbar}\, CS}$ under large gauge transformation (when $k\in\Z$), we are allowed to fix the $g_a=1$ for all $a$, while at the same time dropping all the associated integrals. 
In the following, we will abuse notation and, despite fixing $g_a\equiv 1$, still denote the Wilson graph operator \Ref{G_5operator} by $\G_5^{[j_{ab},\xi_{ab}]}[A,\bA]$.
\end{itemize}

The knotted Wilson graph operator $\G_5^{[j_{ab},\xi_{ab}]}[A,\bA]$ can be split into contributions from the edges $\{\ell_{ab}\}$ and of those from the vertices $\{a\}$.
To this purpose, we rewrite the inner products in (the gauge fixed) $\G_5^{[j_{ab},\xi_{ab}]}[A,\bA]$ as 
\begin{equation}
\int_{(\mathbb{CP}^1)^{\times 2}}\rmd z_{ab}\rmd z_{ba}\lag (j_{ab},\g j_{ab});j_{ab},\xi_{ab} \big|\,G_{a}^{(\ell_{ab})-1}\big| z_{ab}\rag\lag z_{ab}\big|G'_{ab}\big|z_{ba}\rag\lag z_{ba}\big|G_b^{(\ell_{ab})}\big|(j_{ab},\g j_{ab});j_{ab},\xi_{ba}\rag,
\label{innerProds}
\end{equation}
where the edge $\ell_{ab}$ has been split into three pieces, and the holonomy $G_{ab}$ is, accordingly, written as the product $G_a^{(\ell_{ab})-1} G_{ab}'G_b^{(\ell_{ab})}$. 

Let us further explain the notation used in Eq. \Ref{innerProds}. Here, we denote the representation of the vector $| f\rangle \in V^{j,\rho}$ by a homogeneous function of two complex variables $\lag z | f\rag =: f(z)$, i.e. $f(z)\equiv f(z^1,z^2, \bar{z}^1,\bar{z}^2)$, such that for any $\a\in\mathbb C$,
\be
V^{j,\rho}\ni f(\alpha z) = \a^{-1+i\rho+j}\bar \a^{-1 + i\rho -j} f(z) .
\ee
The inner product in $V^{j,\rho}$ is $L^2$, i.e. $\lag f|f'\rag=\int_{\mathbb{CP}^1}\rmd z\, \bar{f}(z)\,f'(z)$ with $\rmd z= \frac{i}{2}(z^1\rmd z^2-z^2\rmd z^1)\wedge(\bar{z}^1\rmd \bar{z}^2-\bar{z}^2\rmd \bar{z}^1)$ an homogeneous measure on $\mathbb C^2$.
For details, see e.g. \cite{ruhl} (or also \cite{hanPI}, for a brief summary).  These equations spell out the meaning of the integrals and of the first and last term in the product.  We are left with the factor $\lag z_{ab}\big|G'_{ab}\big|z_{ba}\rag$.

This can be written as a path integral that implements the $\Slc$ coadjoint orbit quantization.
The idea is that the unitary irreps of a Lie group $G$ can be obtained by geometric quantization of its coadjoint orbits. 
For reviews see \cite{orbit}, or \cite{Dimofte} for a nice summary, or the succinct account in Appendix \ref{coadj}.

To be more explicit, let us recall that the $\Slc$ coadjoint orbit associated to a generic element $\lambda\in\mathfrak{sl}_2\mathbb C$ is the 4-dimensional manifold $\Omega_\lambda\:=\Slc/\mathrm U(1)_\mathbb{C} = \mathrm T^\ast S^2$.
The base space $S^2$ can be identified with $\mathbb{CP}^1=\Slc/B$, with $B$ the Borel subgroup of invertible upper-triangular matrices, and the $\mathbb{CP}^1$ here is the same one that appeared above. Hence, the variable $z\in\mathbb{CP}^1$ is precisely the position variable of a Schr\"odinger representation of $\Omega_\lambda=\mathrm T^\ast S^2$.
This correspondence should clarify the meaning of $\lag z_{ab}\big|G'_{ab}\big|z_{ba}\rag$, as well as its path integral representation 
\be
\lag z_{ab}\big|G'_{ab}\big|z_{ba}\rag\equiv\lag z_{ab}\big|\mathcal P e^{\int_{\ell_{ab}'} A }\big|z_{ba}\rag=\int_{z_{ba}}^{z_{ab}} \cd \fg_{ab}\cd\bar{\fg}_{ab}\,e^{iS_{ab}'[\fg_{ab},\bar{\fg}_{ab};A,\bA]}\label{zGz},
\ee
where the first-order action functional (here $A$ and $\bA$ should be understood as external sources) is 
\begin{equation}
S_{ab}'[\fg_{ab},\bar{\fg}_{ab};A,\bA]=-\half\int_{\ell'_{ab}}\tr \lt[(\nu+\kappa)\fg_{ab}^{-1}(\rmd+A^T)\fg_{ab}+(\nu-\kappa)\bar{\fg}_{ab}^{-1}(\rmd+\bA^T)\bar{\fg}_{ab}\rt]\label{zGzS},
\end{equation}
and the choice of weight $\lambda$ is encoded in the matrices
\be
\nu=-\g j_{ab}\begin{pmatrix}
  1 & 0  \\
  0 & -1
 \end{pmatrix}
 \quad\text{and}\quad
  \kappa={i j_{ab}}\begin{pmatrix}
  1 & 0  \\
  0 & -1
 \end{pmatrix}.
\ee 

On a first encounter the bounds of integration of Eq. \Ref{zGz} may be obscure. To clarify these bounds notice that although the path integral is carried out over maps $\fg_{ab}:\ell_{ab}'\to \Slc$, a gauge symmetry is present that effectively reduces the integration space to maps with range the coadjoint orbit, i.e. $\ell'_{ab} \to \Omega_\lambda=\Slc/\mathrm U(1)_\mathbb{C}=\mathrm T^\ast S^2$. 
In this sense, the above path integral can be consistently viewed as a quantum particle moving on its ``position space'' $\mathbb{CP}^1\ni z$, with boundary conditions at the two end points of $\ell'_{ab}$ given by $z_{ba}$ and $z_{ab}$.

This rewriting of $\lag z_{ab}\big|G'_{ab}\big|z_{ba}\rag$ allows detailed study of the path integral defining the state $\Psi^{\G_5}_{[x_{ab};x_a,y_a]}$ in a tubular neighborhood $N(\ell_{ab}')\subset N(\G_5)$ of $\ell_{ab}'$. Topologically, $N(\ell'_{ab}) \cong [0,1]\times D^2$, where $D^2$ is a 2-disk.
We parametrize this space with $(t,x^1,x^2)$, where $t\in[0,1]$ and  $(x^1,x^2)\in D^2$ so that $(x^1,x^2)=(0,0)$ is the location of the Wilson line.
Accordingly, the Chern--Simons connection in $N(\ell'_{ab})$ can be decomposed into a time component $A_t$ along $\ell_{ab}'$ and a spatial component $A_\perp$. 
With this decomposition, and after an integration by parts, the contribution of $N(\ell'_{ab})$ to the action $CS[N(\G_5)|A,\bA]$ becomes%
\be
CS\lt[N(\ell'_{ab})\big|A,\bA\rt]=\frac{t}{8\pi}\int_{N(\ell'_{ab})}\tr\lt(A_\perp\wedge \rmd A_{\perp}\rt)+ 2 \tr\lt(F_\perp\wedge A_{t}\rt)+\mathrm{c.c.},\label{CSN}
\ee
where $F_\perp=\rmd A_\perp+A_\perp\wedge A_\perp$ is the curvature of $A_\perp$.%
\footnote{On the boundary $\partial N(\ell'_{ab})\cong [0,1]\times S^1$, the two components of the connection $(A_1, A_2)$ are the pullbacks of $(A_\perp,A_t)$, respectively.} Here, the boundary term coming from the integration by parts cancels exactly the boundary term present in the Chern--Simons action, i.e. $\frac{t}{8\pi}\int_{\partial N(\ell'_{ab})}\tr\lt(A_1\wedge A_2\rt) $.  

In the definition of $\Psi^{\G_5}_{[x_{ab};x_a,y_a]}$, the Chern--Simons theory on $N(\ell_{ab}')$ appears to be coupled to the coadjoint orbit path integral of equation \Ref{zGz}. The total action is linear in $A_t$ and $\bA_t$. 
Thus, integrating these out we obtain two functional delta functions on the space of $(A_\perp,\bA_\perp)$, which constrain $F_\perp$ and $\bar{F}_\perp$ to be given by
\be
\frac{t}{4\pi\hbar}F^T_\perp=\frac{1}{2}\fg\lt(\nu+\kappa\rt)\fg^{-1}\delta^{(2)}(x)\rmd x^1\wedge \rmd x^2, \quad \text{and} \nonumber\\
\frac{\bar{t}}{4\pi\hbar}\bar{F}^T_\perp=\frac{1}{2}\bar{\fg}\lt(\nu-\kappa\rt)\bar{\fg}^{-1}\delta^{(2)}(x)\rmd x^1\wedge \rmd x^2, \quad \phantom{and}
\label{equationFF}
\ee 
where $\delta^{(2)}(x)$ is a delta function on $D^2$ such that for any 1-form $f$, $\int_{N(\ell_{ab}')}\delta^{(2)}(x)\rmd x_1\wedge \rmd x_2\wedge f=\int_{\ell_{ab}'}f$. These constraints fix the conjugacy class of the meridian holonomies $H_{ab}$ and $\bar H_{ab}$, i.e.
\be
H_{ab}\sim \begin{pmatrix}
  \mathrm{q}^{j_{ab}} & 0  \\
  0 &  \mathrm{q}^{-j_{ab}}
 \end{pmatrix} \ \ \ \ \text{with}\ \ \ \ \mathrm{q}=e^{\frac{2\pi i\hbar}{t}(1+i\g)},
\ee
and similarly for $\bar H_{ab}$.
Note that when the parameters $t$ and $\g$ satisfy 
\be
\frac{2\pi\hbar}{t}(1+i\g)\in\R,
\label{equationreal}
\ee
the eigenvalues of $H_{ab}$ satisfy the boundary conditions of Section \ref{b.c.}:
\be
x_{ab}=\mathrm{q}^{j_{ab}}\in \mathrm{U(1)}.
\label{xqjequation}
\ee
Reinserting the constrained value of $F_\perp$ in the first term of $CS\lt[N(\ell'_{ab})\big|A,\bA\rt]$ in equation \Ref{CSN} and using the identity $\tr\lt(A_\perp\wedge \rmd A_{\perp}\rt)=\tr\lt(A_\perp\wedge F_\perp \rt)$, one finds that this term vanishes identically since $F_\perp$ is constrained to be proportional to $\rmd x^1\wedge \rmd x^2$. 

As a result, the contribution coming from $N(\ell_{ab}')$   to the wave function $\Psi^{\G_5}_{[x_{ab};x_a,y_a]}$ gives a product of delta functions:
\be
\prod_{a<b} \delta\lt(x_{ab}\, , \,\mathrm{q}^{j_{ab}}\rt)\delta\lt(\bar{x}_{ab}\, ,\,\bar{\mathrm{q}}^{j_{ab}}\rt).\label{deltafunc}
\ee
Therefore, we see that the boundary data $x_{ab}=\mathrm{q}^{j_{ab}}$ is imposed strongly by the Wilson graph operator. 

In the previous section we studied the semiclassical behavior of the Chern--Simons path integral with the geometric boundary conditions imposed. This was achieved in that context by simply sending $\hbar \to 0$.
Here, the boundary conditions are imposed through the insertion of a Wilson graph operator, and as a consequence the relation between the operator's labels and the boundary conditions is mediated by terms containing $\hbar$, as in Eqs. \Ref{equationFF}--\Ref{xqjequation}.
Therefore, in order to reproduce the semiclassical behavior obtained in the previous section in this context, together with $\hbar$ being sent to zero, the representation labels $j_{ab}$ must be sent uniformly to infinity in such a way that the boundary data $x_{ab}=\mathrm{q}^{j_{ab}}=\exp\lt({2\pi i\hbar{(1+i\g)}j_{ab}/t}\rt)$ stay fixed.
Specifically, we see that the right semiclassial limit is now the double-scaling limit
\be
\hbar\to0
\quad\text{and}\quad
 j_{ab}\to\infty,
 \quad\text{while keeping} \quad
\hbar j_{ab} = \text{const}.
\ee

We studied precisely this double scaling limit via stationary phase techniques in \cite{HHR}. Here we quickly review that analysis.
Using the $\Slc$ irreps described above, the full (gauge-fixed) $\G_5$ Wilson graph operator can be written in the following integral form: 
\be
\G_5^{[j_{ab},\xi_{ab}]}[A,\bA]=\int_{\mathbb{CP}^1}\prod_{a<b}\rmd \mu( z_{ab})\ e^{I_{\G_5}},
\ee
where the measure is  $\rmd \mu(z) = {\rmd z}/{\langle z,z\rangle^2}$, and the ``Wilson graph action'' $I_{\G_5}$ is 
\be
I_{\G_5}=\frac1\hbar\sum_{a<b}\hbar j_{ab}\ln\frac{\lag G_{ab}^\dagger z_{ab},\xi_{ba}\rag^2\lag\xi_{ab},z_{ab}\rag^2}{\lag G_{ab}^\dagger z_{ab},G_{ab}^\dagger z_{ab}\rag \lag z_{ab},z_{ab}\rag}+i\g \hbar j_{ab}\ln \frac{\lag G_{ab}^\dagger z_{ab},G_{ab}^\dagger z_{ab}\rag}{\lag z_{ab},z_{ab}\rag}
\ee
(by construction the choice of a branch for the logarithm is irrelevant).
Using the Cauchy--Schwarz inequality, it is immediate to see that $\mathrm{Re}(I_{\G_5})\leq 0$.

This leads to study of the stationary points of $I_{\G_5}$ coupled to Chern-Simons theory on $S^3$ in the double scaling limit, as in Eq. \Ref{API}.
Doing so, one finds the following stationarity equations

\begin{description}

\item[Parallel Transport:] From $\delta_{z_{ab}} I_{\G_5}=0$ and $\mathrm{Re}(I_{\G_5})=\mathrm{max\, Re}(I_{\G_5})=0$, one obtains the following parallel transport relations for the coherent state labels $\xi_{ab}$:
\be
\xi_{ab}=\frac{||z_{ab}||}{||G_{ab}^\dagger z_{ab}||} e^{i\theta_{ab}}G_{ab} \xi_{ba},
\quad\text{and}\quad
J\xi_{ab}=\frac{||G_{ab}^\dagger z_{ab}||}{||z_{ab}||} e^{-i\theta_{ab}}G_{ab} J\xi_{ba},
\label{gluing}
\ee
which relate the 2-spinors $\xi_{ab}$ and $\xi_{ba}$ at the two end-points of the edge $\ell_{ab}$.

\item[Monodromies:] Variation with respect to the Chern-Simons connections $A$ \& $\bar{A}$ gives the distributional curvature on $S^3$, 
\be
\eps^{\mu\rho\sig}F_{\rho\sig}^i(x)=\frac{8\pi\hbar(1+i\g)}{t}\sum_{a<b}j_{ab}\lag  G_{sb}^\dagger\sig_i(G_{sb}^\dagger)^{-1}\xi_{ba},\,\xi_{ba}\rag \delta^{(2)\ \mu}_{\ell_{ab}}(x).\label{epsF}
\ee
As expected, $\bar{F}$ satisfies the complex conjugate equation. 
Here, again $\sig_i$ are the Pauli matrices, and $\delta_{\ell}^{(2)\;\mu}(x):=  \int_0^1 \delta^{(3)}(x-\ell(s)) \frac{\D \ell^\mu}{\D s} \D s $.
With a slight abuse of notation, we use the parameter $s\in[0,1]$ to label intermediate points on the edge $\ell_{ab}$, so that 
\be
G_{sb} = \mathcal P \exp{\int_0^s A_\mu(\ell(s')) \frac{\rmd \ell_{ab}^\mu}{\rmd s'} \d s'},
\ee
with the reasonable requirements $\ell_{ab}(s=0)=b$ and $\ell_{ab}(s=1)=a$.
As expected, the curvature is only supported distributionally on the graph $\G_5$, while $F=\bar{F}=0$ on the graph complement $S^3\setminus\G_5$.

Integrating equation \Ref{epsF} over a disk using the non-Abelian Stokes theorem, one obtains nontrivial holonomies along the non-contractible cycles $c_{ab}(s)$ transverse to $\ell_{ab}$ in the vicinity of the point $\ell_{ab}(s)$:
\be
H_{ab}(s)=\exp\lt[\frac{4\pi\hbar(1+\rmi\g)}{ t} j_{ab}\lag  G_{sb}^\dagger{\sig_j}(G^\dagger_{sb})^{-1}\xi_{ba},\xi_{ba}\rag\frac{i\sig_j}{2}\rt],\ \ \ (a<b).\label{monodromy}
\ee
These holonomies should be thought of as being based at the vertex $b$. Notice that the parallel transport equations for the $\xi_{ab}$ guarantee consistency if one were to choose instead vertex $a$ as the base point.
Notice also that the conjugacy class of $H_{ab}(s)$ is consistent with the delta function equation \Ref{deltafunc}.

\end{description}

So far, our analysis has focused mostly on the edges $\ell_{ab}$.
Let us now focus on the neighborhood of a vertex $a$.
Start by considering a 2-sphere with radius $s$ enclosing the vertex $a$, and denote $H_l(s)\:=H_{ab_l}(s)$ $(l=1,\cdots,4)$. 
As a consequence of the flatness on the graph complement $S^3\setminus \G_5$, we obtain 
\be
\fg_4(s)H_4(s)\fg_4(s)^{-1}\fg_3(s)H_3(s)\fg_3(s)^{-1}\fg_2(s)H_2(s)\fg_2(s)^{-1}\fg_1(s)H_1(s)\fg_1(s)^{-1}=1,\label{4gHg}
\ee
where $\fg_{l}\in\Slc$ stands for the holonomy connecting the base point of each $H_l(s)$ to a common base point on the sphere (for details on a convenient choice of paths and their relation to the framing of $\G_5$, see \cite{HHR} ). 
Again because of the flatness in $S^3\setminus \G_5$, one finds
\be
\fg_{l}(s)^{-1}\fg_{l-1}(s)=G_{as_l}^{-1}G_{as_{l-1}}. 
\ee
On the other hand, using Eqs. \Ref{gluing} and \Ref{monodromy}, each $H_l(s)$ can be brought to an element of $\Su$ using the adjoint action of $G_{as_l}^{-1}$. Of course, this holds under the condition that the parameters $t$ and $\g$ satisfy equation \Ref{equationreal},  $\frac{2\pi\hbar}{t}(1+i\g)\in\R$, i.e. 
\be
G_{as}H_{ab}(s)G_{as}^{-1}=H_b(a)=\exp\lt[\frac{4\pi\hbar(1+i\g)}{ t} j_{ab}\widehat{n_\xi}_{ab}\frac{i\sig_j}{2}\rt],
\ee
where $\widehat{n_\xi}=\lag\xi, \vec{\sig}\xi\rag$ is the $\R^3$ unit vector encoded in the spinor $\xi$. 
Then, equation \Ref{4gHg} reduces to a product of four $\Su$ matrices 
\be
\overleftarrow{\prod_{b:b\neq a}} H_{b}(a)=
\overleftarrow{\prod_{b:b\neq a}}\exp\lt[\frac{4\pi\hbar(1+i\g)}{ t} j_{ab}\hat{n}_{ab}\frac{i\sig_j}{2}\rt]=1.
\label{4jn}
\ee
Recall from Section \ref{b.c.}  that this equation is the starting point of the tetrahedral reconstruction.

Moreover, this equation shows that, after removing the intersection points with the graph $\G_5$ (as well as a tubular neighborhood thereof), the pull-back of the connection to the resulting 4-holed sphere is essentially an $\Su$ flat connection. These 4-holed spheres are essentially the $\{\cs_{a=1,\cdots,5}\}=\Sig_6\setminus\{c_{ab}\}_{a<b}$. Thus, we see that the full set of geometricity boundary conditions in Section \ref{b.c.} derive naturally from the insertion of the $\G_5$ Wilson graph operator, albeit part of it only in the semiclassical limit (i.e. in the double-scaling limit). 
This fact was already expected: while the $\{x_{ab}\}$ are strongly fixed to be in $\mathrm U(1)$ (see Eq. \Ref{deltafunc}), due to the Heisenberg uncertainty principle, the pairs of conjugated variables $\{x_a, y_a\}$ cannot be rigidly restricted at the same time. The latter restriction is the one ensuring that on each $\mathcal S_a$ the $\Slc$ flat connection effectively restricts to an $\Su$ one. This restriction emerges strictly speaking only in the double scaling limit.
In this sense, the state $\Psi^{\G_5}_{[x_{ab};x_a,y_a]}$ can be viewed as the ``semiclassical'' state that is on the one hand sharply peaked on the configuration variables $x_{ab}=\mathrm{q}^{j_{ab}}\in\mathrm U(1)$, and on the other ``coherently'' peaked at some phase space point $(x_a,y_a)$ fully determined by the graph data $[j_{ab},\xi_{ab}]$.

Summarizing, the stationary point equations deduced in the semiclassical (i.e. double scaling) limit are found to define an $\Slc$ connection on the graph complement $M_3=S^3\setminus \G_5$, which satisfies---in the limit---the geometricity boundary conditions. 
According to Theorem \ref{parity} there are exactly two such connections $A$ and $\tilde{A}$, which correspond to a parity related pair of convex, Lorentzian, constant curvature 4-simplices.
In particular, the network of relations between $j_{ab}$ and $x_{ab}=e^{u_{ab}}$, and between $u_{ab}$ and the triangle areas $\mathbf{a}_{ab}$, implies that 
\be
\nu\frac{\L}{6}\mathbf{a}_{ab} = 
-\frac{2\pi \hbar}{ t} \lt(\frac{1}{\g}+i\rt)\g j_{ab}+\pi\fs_{ab}\ \ \text{mod}\ \ 2\pi\,\Z .\label{jAr}
\ee
Although this relation seems to give a non-unique value for $\mathbf{a}_{ab}$, the theorem ensures that there is only one geometrically viable choice. Also, as we have shown in the last section, the ambiguities above play no role in the evaluation of the semiclassical action provided a specific quantization condition for the areas is introduced (and $k=\mathrm{Re}(t)\in\Z$).  
It is, however, straightforward to check that this quantization condition, Eq. \Ref{Qcondition}, is automatically satisfied when the boundary conditions are imposed by the Wilson graph insertion studied in this section:
\be
2\pi \hbar\, \mathbb Z \ni 2\mathrm{Re}\lt( \frac{\Lambda t}{6} \rt) \Ar_{ab} = 2\pi \hbar \lt( 2 j_{ab} + 2\pi \hbar^{-1} k \fs_{ab}  + 4\pi \hbar^{-1} k\, \mathbb Z\right).
\ee

All these results, together with those established in the previous section, imply that inserting the solution of the equations of motion (i.e. a flat connection corresponding to a geometrical 4-simplex) back into the total action $I_{\G_5}+\frac{i}{\hbar}CS$, we find that the leading behavior of $\sa[x_{ab};x_a,y_a]$ in the semiclassical limit is the same as that of the 3d block of equation \Ref{ZZasymp}, which in turn reproduces the 4-dimensional Regge action of the constant curvature 4-simplex with a cosmological constant term.
In this way, we see that, while the Wilson graph operator imposes the geometricity boundary condition, the asymptotic behavior of $\A[x_{ab};x_a,y_a]$ is basically determined by that of the Chern-Simons 3d block. 
A heuristic reason why Chern--Simons theory on $S^3\setminus\G_5$ should ``know'' about 4-dimensional geometry is given in \cite[Sect. 3]{HHR}.

In \cite{HHR}, the following result is also shown: Under the double-scaling limit $\hbar\to 0,\ j_{ab}\to\infty$ with $j_{ab}\hbar$ fixed, the Chern-Simons expectation value $\A [x_{ab};x_a,y_a]$ of $\G_5$ graph operator in Eq. \Ref{API} has the following asymptotic behavior 
\be
\A [x_{ab};x_a,y_a]\sim e^{\frac{i}{\ell_P^2}S_{Regge}^\L+\cdots}+e^{-\frac{i}{\ell_P^2}S_{Regge}^\L+\cdots}\label{Aasymptotics}
\ee
up to an overall phase factor. The two exponentials come from the two solutions $A$ and $\tilde{A}$ respectively. The ellipsis $\cdots $ stand for quantum corrections. The constant curvature Regge action of simplicial gravity $S_{Regge}^\L$ reads
\be
S_{Regge}^\L=\sum_{a<b} \mathbf{a}_{ab}\Theta_{ab}-\L \mathrm{Vol}^\L_4
\ee
and $\ell_P^{-2}=\mathrm{Re}\lt(\frac{\L t}{12\pi i\hbar}\rt)$. We have assumed here the Chern-Simons couplings $t=k+is$ and $\bar{t}=k-is$ satisfy $k\in\Z$ and $s\in\R$. The parameter $\hbar^{-1}$ is an integer and just scales the parameters $t$ and $\bar{t}$.

In the semiclassical limit, $\A [x_{ab};x_a,y_a]$ has the same asymptotic behavior as the sum of a pair of Chern-Simons 3d blocks (up to an overall phase):
\be
\A [x_{ab};x_a,y_a]\sim  Z_{CS}^{(\a)}(u)Z_{CS}^{(\overline{\a})}(\bar{u})+ Z_{CS}^{(\tilde{\a})}(u)Z_{CS}^{(\overline{\tilde{\a}})}(\bar{u})
\ee
here $Z_{CS}^{\a}(u)$ and $Z_{CS}^{\tilde{\a}}(u)$ correspond to the pair of flat connection $A$ and $\tilde{A}\in \cm_{\mathrm{flat}}(S^3\setminus\G_5,\Slc)$ with 2 arbitrary lifts $\a$ and $\tilde{\a}$, respectively. Note that the analysis in Section \ref{3dblock} has only a single exponential because we computed the phase difference (or ratio) between two 3d blocks $Z_{CS}^{(\a)}(u)Z_{CS}^{(\overline{\a})}(\bar{u})$ and $Z^{(\tilde{\a})}_{CS}(u)Z_{CS}^{(\overline{\tilde{\a}})}(\bar{u})$.

In addition, it is interesting that the cosmological constant term in Eq. \Ref{Aasymptotics} comes from the evaluation of the Chern-Simons functional on $S^3$ at the connection $A$ that satisfies the critical equations. This connection is now viewed as a distributional connection on $S^3$ (with a distributional curvature supported on the graph) instead of being a flat connection on $S^3\setminus\G_5$. The following difference between the evaluations at $A$ and $\tilde{A}$ gives the constant curvature 4-volume of the 4-simplex:
\be
CS\big[S^3\big|A,\overline{A}\big]-CS\big[S^3\big|\tilde{A},\overline{\tilde{A}}\big]=\frac{2\L}{\ell_P^2}\mathrm{Vol}_4^\L+ 2\pi i\Z.
\ee

\section{Relation with Loop Quantum Gravity}\label{LQG}

If we take the asymptotic ``decoupling limit'' by turning off the Chern-Simons coupling in $\A [u_{ab};u_a,v_a]$ via $t,\bar{t}\to\infty $ while keeping $j_{ab}$ fixed, the path integral Eq. \Ref{API} is localized on the solution of Chern-Simons equations of motion $F=\bar{F}=0$ on $S^3$; this gives a trivial connection on $S^3$. The Wilson graph $\G_5\lt[j_{ab},\xi_{ab}\big|A,\bA\rt]$ evaluated at trivial connection gives the Engle-Pereira-Rovelli-Livine (EPRL) spinfoam 4-simplex amplitude $\A_{EPRL}[j_{ab},\xi_{ab}]$ in LQG. This relation was the original motivation for the definition of the $\G_5$ Wilson graph operator. 

The relations among Chern-Simons theory, 4-dimensional LQG and 4-dimensional simplicial gravity can be summarized in the following diagram:
\be
\A [u_{ab};u_a,v_a]\ &\ \ \  \ \stackrel{\hbar\to 0,\ j\to\infty,\ j\hbar\ \mathrm{fixed}}{\longrightarrow}\ \ \ \ &\ e^{\frac{i}{\ell_P^2}S_{Regge}^\L}+e^{-\frac{i}{\ell_P^2}S_{Regge}^\L}\nonumber\\
&& \nonumber\\
\Big\downarrow \stackrel{t\to\infty}\ \ \ \ &&\ \ \ \ \ \Big\downarrow \stackrel{\L\to 0}\nonumber\\
&& \nonumber\\
\A_{EPRL}[j_{ab},\xi_{ab}] \ &\ \ \ \ \stackrel{j\to\infty}{\longrightarrow}\ \ \ \ &\ e^{\frac{i}{\ell_P^2}S_{Regge}}+e^{-\frac{i}{\ell_P^2}S_{Regge}}
\ee
where $\A [u_{ab};u_a,v_a]$, defined in Eq. \Ref{API}, is the $\Slc$ Chern-Simons evaluation of the $\G_5$ Wilson graph operator. The relation along the lower line states that the large-$j$ asymptotics of the EPRL spinfoam amplitude reproduces the flat simplicial geometry and Regge action without cosmological constant $\L$ and was proved in \cite{semiclassical,HZ}. This diagram suggests that the Chern-Simons expectation value $\A [u_{ab};u_a,v_a]$ can be viewed as a deformation of the EPRL spinfoam amplitude, which includes a cosmological constant into the framework of LQG.

The 4-dimensional spinfoam amplitude of LQG, which defines a quantum 4d geometry, describes the quantum transition between boundary states for quantum 3d geometries. The boundary states of a 4-dimensional spinfoam amplitude are SU(2) spin-network states. The latter states form the kinematical framework of LQG (see \cite{book,rev}) and describe quantum 3d geometries. A spin-network state is a triple $(\G,\vec{j},\vec{i})$ consisting of: an oriented graph $\G$; a map $\vec{j}=\{j_\ell\}_{\ell\in E(\G)}$ from the set of graph edges $E(\G)$ to the space of unitary irreps of SU(2) labeled by $j_\ell$; and $\vec{i}=\{i_v\}_{v\in V(\G)}$, a map from the set of graph vertices $V(\G)$ to the invariant tensors (intertwiners) $v\mapsto i_v\in \mathrm{Inv}_{\Su}(V_{j_1}\otimes\cdots\otimes V_{j_n})$, where $j_1,\cdots,j_n$ are the spin labels on the edges incident to $v$. The spin-network states are a basis for the LQG Hilbert space and diagonalize the geometrical operators, e.g. quantum area and volume operators. The discrete spectrum of the area operator is parametrized by the spins $j_\ell$ (and is linear in $j_\ell$ when $j_\ell\gg1$), and the discrete spectrum of volume is parametrized by both the spins $j_\ell$ and the invariant tensors $i_v$ \cite{RS,Hal}. The invariant tensor $i_v$ carries even more information, it parametrizes the space of quantum (zero curvature) polyhedra with face areas being proportional to the incident $j_\ell$ \cite{CF,shape}. 

The spin-network data $(\G,\vec{j},\vec{i})$ is well adapted to the framework in the present paper and can be identified with the boundary data of the flat connections we have been discussing. The identification of the spin-network graph with $\G_5$ is immediate since it appears in the definition of the Wilson graph operator and its Chern-Simons evaluation $\A [u_{ab};u_a,v_a]$. The spin $j_\ell$ is mapped by $Y$ to an $\Slc$ principle series irrep $(j_\ell,\g j_\ell)$ for each edge, where $\g$ is the Barbero-Immirzi parameter of LQG. At each vertex, we employ the SU(2) coherent state basis and consider $i_v$ to be a coherent intertwiner, which is mapped by $Y$ to an $\Slc$ invariant tensor in the Wilson graph operator.

Given a graph, e.g. $\G_5$ in our context, and its tubular neighborhood $N(\G_5)$, let us consider the quantization of SU(2) flat connections on the closed 2-surface $\Sig_6=\partial N(\G_5)$. By specifying the meridian closed curves $c_{ab}$ as in Section \ref{b.c.}, we arrive at a set of local symplectic coordinates for $\cm_{\mathrm{flat}}(\Sig_6,\Su)$: $x_{ab}=e^{u_{ab}},y_{ab}=e^{-\frac{\pi}{k} v_{ab}}\in \mathrm{U}(1)$ with $\{u_{ab},v_{ab}\}=1$. Quantizing these coordinates, as well as the flat connections $\cm_{\mathrm{flat}}(\cs_a,\Su)$ of the 4-holed spheres with fixed conjugacy class $x_{ab}$ at each hole will provide a quantization for the full space $\cm_{\mathrm{flat}}(\Sig_6,\Su)$. The quantization of $x_{ab},y_{ab}$ is a quantization of $S^1\times S^1$. The prequantum line bundle over $S^1\times S^1$ has a curvature $\o=-\frac{k}{\pi} \rmd \ln x_{ab}\wedge\rmd\ln y_{ab}$. Weyl's integrality criterion then implies that $k\in\Z$. We choose the polarization such that the wave function is written as $f(u_{ab})$ and satisfies both periodicity and Weyl invariance $f(u_{ab})=f(-u_{ab})=f(u_{ab}+2\pi i)$. Periodicity in both $u_{ab}$ and $v_{ab}$ directions implies that $u_{ab}$ can only take $k+1$ discrete values $u_{ab}=0, \frac{i\pi}{k},\frac{2i\pi}{k},\cdots,{i\pi}$, i.e.
\be
x_{ab}= e^{\frac{2\pi i}{ k} j_{ab}},\quad \text{with} \quad \ j_{ab}=0,\half,\cdots,\frac{k}{2}.
\ee
The quantization of the flat connections $\cm_{\mathrm{flat}}(\cs_a,\Su)$ with fixed conjugacy classes $x_{ab}$ results in the Hilbert space $\ch(\cs_a)$ spanned by Wess-Zumino-Witten (WZW) conformal blocks $\cf(i_a)$ of level $k\in\Z$ on a 4-holed sphere \cite{seiberg}. Each conformal block $\cf(i_a)$ is associated with a 4-valent SU(2) intertwiner $i_a$ with the above spins $j_{ab}$. A restricted subclass of SU(2) intertwiners is allowed because of the restrictions on the ranges of the spins $j_{ab}$ and the spin in the recoupling channel.  The dimension of the intertwiner space, $\ch(\cs_a)$, consequently is given by the famous Verlinde formula \cite{verlinde}. As a result, we obtain the Hilbert space for the full quantization of $\cm_{\mathrm{flat}}(\Sig_6,\Su)$; it is spanned by the basis 
\be
\psi_{(\G_5,\vec{j},\vec{i})}=\prod_{a<b}\delta(x_{ab},e^{\frac{2\pi i}{ k} j_{ab} }) \prod_{a=1}^5\cf(i_a).
\ee
 The above discussion can be straightforwardly generalized to arbitrary graphs $\G$. Now we see that the quantization of SU(2) flat connections on $\Sig_g=\partial N(\G)$ for any graph $\G$ naturally gives the spin-network data $(\G,\vec{j},\vec{i})$ with $j\leq k/2$ and a restricted subclass of intertwiners. The restricted class of spin-network data is likely to be the right subclass for LQG when a cosmological constant is included.

By the analysis in Section \ref{b.c.}, the SU(2) flat connections on a 4-holed sphere with fixed conjugacy classes $x_{ab}$ correspond to  constant curvature tetrahedral geometries with fixed face areas. Therefore the Hilbert space $\ch(\cs_a)$ of conformal blocks is the space of ``quantum constant curvature tetrahedra'' with ``quantum areas'' proportional to $j_{ab}$. We may consider an overcomplete coherent state basis $\psi_{x_a,y_a}^k$ peaked at the phase space point with conjugate coordinates $(x_a,y_a)$. For these coherent states and $\G=\G_5$ the spin-network data $(\G,\vec{j},\vec{i})$  can be mapped to the $\Slc$ flat connection data $(x_{ab};x_a,y_a)$  on $\Sig_6=\partial S^3\setminus\G_5$ subject to the restriction of spins and intertwiners just discussed.  

In order to be the boundary data of an $\Slc$ Chern-Simons theory, we make the following identification:
\be
x_{ab}= e^{\frac{2\pi i}{ k} j_{ab}}=e^{\frac{2\pi i}{ t} \lt({1}+i\g\rt) j_{ab}} \label{xandj}
\ee
where Eq. \Ref{deltafunc} has been used (and we have set $\hbar=1$). Here $k\in\Z$ has been identified with $\mathrm{Re}(t)$, and both $\g$ and $\frac{1}{ t} \lt(1+i\g\rt)$ have been assumed to be real numbers, so that $\g=s/k$. Given that $(x_{ab};x_a,y_a)$ comes from spin-network data, the boundary condition in Section \ref{b.c.}  and the quantization condition Eq. \Ref{Qcondition} are satisfied following the same argument as given in Section \ref{wilson}. It is interesting to notice that when $t$ is purely imaginary ($k=0$ or $\g\to\infty$), the spectrum of $x_{ab}$ is not discrete anymore, while the quantization condition Eq. \Ref{Qcondition} is satisfied trivially. This possibility is beyond the regime of spin-network data, but still well-controlled by the 3d blocks of Chern-Simons theory discussed in Section \ref{3dblock}.\footnote{$\Slc$ Chern-Simons theory with purely imaginary $t$ relates to the quantum Lorentz group with real $q$ \cite{roche}. The 3d blocks of Chern-Simons theory with our boundary conditions implemented may relate to the spinfoam model defined in \cite{QSF}.}

The discussion above provides a map from spin-network data to the boundary data $(x_{ab};x_a,y_a)$ of $\Slc$ Chern-Simons theory satisfying the boundary condition in Section \ref{b.c.}. When there exists an $\Slc$ flat connection $A$ on $S^3\setminus\G_5$ whose boundary value is consistent with the boundary data $(x_{ab};x_a,y_a)$, we may use these data to construct a Chern-Simons 3d block $Z_{CS}^{(\a)}(u)Z_{CS}^{(\overline{\a})}(\bar{u})$. The Chern-Simons 3d block $Z_{CS}^{(\a)}(u)Z_{CS}^{(\overline{\a})}(\bar{u})$ studied in Section \ref{3dblock} may play an interesting role in LQG as could the amplitude $\sa[u_{ab};u_a,v_a]$ (compare $\A_{EPRL}[j_{ab},\xi_{ab}]$). 

As we have seen in Section \ref{wilson}, the Regge-action asymptotic behavior of $\sa[u_{ab};u_a,v_a]$ crucially depends on the peakedness of the Chern-Simons state created by the Wilson graph operator. However, different Wilson graph operators may produce the same peakedness in the boundary data, and thus lead to the same asymptotics of $\sa[u_{ab};u_a,v_a]$. The close relationship with the EPRL 4-simplex amplitude has led us to study the particular type of Wilson graph operators $\G_5[j_{ab},\xi_{ab}|A,\bA]$. Independent of the choice of Wilson graphs, the Chern-Simons 3d block $Z_{CS}^{(\a)}(u)Z_{CS}^{(\overline{\a})}(\bar{u})$ on $S^3\setminus\G_5$ with the right boundary condition imposed is the essential ingredient behind the Regge-action asymptotics of $\sa[u_{ab};u_a,v_a]$. Although we have defined $Z_{CS}^{(\a)}(u)$ perturbatively on the cover space parametrized by the logarithmic data $u$ instead of $x$, it can be defined non-perturbatively, as in \cite{3dblock,3d3dduality}. These references show that the non-perturbative $Z_{CS}^{(\a)}(u)$ manifestly depends on $x=\exp(u)$. Therefore $Z_{CS}^{(\a)}(u)Z_{CS}^{(\overline{\a})}(\bar{u})$ depends on the boundary or spin-network data in the desired manner.

When we generalize our framework from a 4-simplex to a generic simplicial manifold, the class of 3d blocks $Z_{CS}^{(\a)}(u)Z_{CS}^{(\overline{\a})}(\bar{u})$ that asymptotically reproduce classical gravity may ultimately span the physical Hilbert space $\ch_{Phys}$ in LQG. The operator constraint equation that quantizes the Lagrangian subvariety $\cl_{\mathbf{A}}$,
\be
\hat{\mathbf{A}}_m(\hat{x},\hat{y},\hbar)Z_{CS}^{(\a)}(u)=0,
\ee
may relate to the quantization of the Hamiltonian constraint equation in LQG \cite{QSD,links,valentin}, provided the proper boundary conditions are implemented. 

There is a perspective that we would like to point out before we conclude this section. In \cite{lowE}, it is suggested that the simplicial 4d geometries correspond to the dynamical vacua of LQG, namely, to solutions of the critical equations of the spinfoam amplitude. In the present work and in \cite{HHR}, we have made the correspondence between simplicial 4d geometry and $\Slc$ flat connections on the graph complement 3-manifold $S^3\setminus\G_5$ explicit, and shown that the solutions of the critical point equations arising from $\A[u_{ab};u_a,v_a]$ give the $\Slc$ flat connections on $S^3\setminus\G_5$. Therefore we suggest that the moduli space of LQG dynamical vacua can be embedded into the moduli space $\cm_{\mathrm{flat}}(S^3\setminus\G_5,\Slc)$, where the image of the embedding map is specified by the boundary condition in Section \ref{b.c.}. We expect that the dynamical properties of the LQG vacua, including the perturbative behavior of LQG, should be largely controlled by $\Slc$ Chern-Simons theory.

\section{Beyond A Single 4-Simplex }\label{beyond}

The above analysis is primarily about the geometry of a single 4-simplex and its correspondence with flat connections on $S^3\setminus\G_5$. This analysis can be generalized to an arbitrary simplicial decomposition of a 4-dimensional manifold into an arbitrary number of simplices. In this section we give the idea of the construction and results, more details appear in \cite{3d3dduality}. 

\begin{figure}[h]
\begin{center}
\includegraphics[width=15cm]{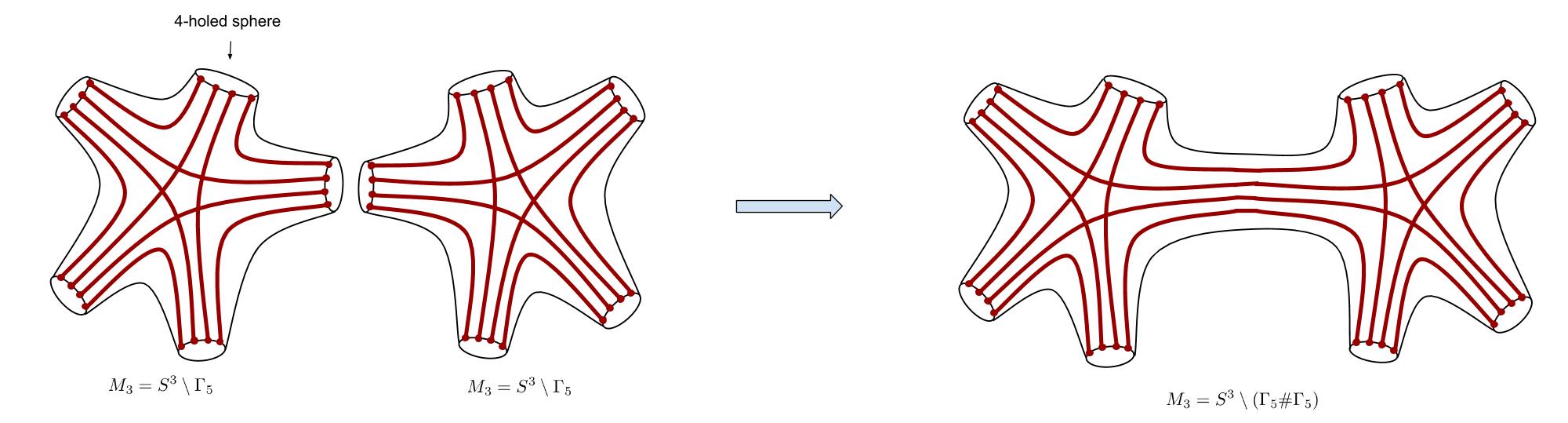}
\caption{Left: Two copies of the graph complement 3-manifold $S^3\setminus\G_5$ viewed from 4 dimensions. Each $S^3\setminus\G_5$ is drawn by suppressing 1 dimension. The 3-manifold $S^3\setminus\G_5$ has five ``big boundary'' components, which are 4-holed spheres and correspond to the five vertices of $\G_5$. The manifold $S^3\setminus\G_5$ also has ten ``small boundary'' components, which are ten cylinders and correspond to the ten edges of $\G_5$. Removing the tubular neighborhoods of the 10 edges results in the red `tunnel' curves that connect the holes in the big boundary components. The tunnels give the 10 small boundary components. The union of big and small boundary components gives the closed 2-surface $\Sig_6=\partial (S^3\setminus\G_5)$. Right: The graph complement 3-manifolds can be glued through a pair of big boundary components, i.e. a pair of 4-holed spheres, via an identification of the holes. After gluing, some of the tunnels are continued from one $S^3\setminus\G_5$ to the other. Note that in this figure, the properties of crossings are not shown.}
\label{gluing3-fold}
\end{center}
\end{figure}

A 4-dimensional simplicial complex $\ck_4$ is built by gluing 4-simplices $\sig$. The simplicial geometry of $\ck_4$ is made up of the constant curvature geometry of each 4-simplex together with the distributional curvature located at the 2d hinges at the 4-simplex-gluing interfaces. The simplicial geometries on $\ck_4$ again correspond to a class of $\Slc$ flat connections on a 3-manifold $\mathscr{M}_3$. The 3-manifold $\mathscr{M}_3$ is obtained by gluing $N$ copies of $S^3\setminus\G_5$, where $N$ is the number of 4-simplices in $\ck_4$, as in Fig. \ref{gluing3-fold}. 

The 3-manifold $\mathscr{M}_3$ can be constructed in the following way (see Fig. \ref{gluing3-fold}): Corresponding to gluing a pair of 4-simplices in 4d through a pair of tetrahedra, a 3-manifold is constructed by gluing a pair of $S^3\setminus\G_5$ through a pair of 4-holed spheres. The boundary $\Sig_6$ of $S^3\setminus\G_5$ can be decomposed into 2 types of components: the ``big boundaries'', which here consist of five 4-holed spheres that resulted upon removing the neighborhood of the five vertices in $\G_5$, and the ``small boundaries'', which here consist of the 10 cylinders that resulted upon removing the tubular neighborhood of the 10 edges of $\G_5$.\footnote{The two types of boundary components are also called ``geodesic boundaries'' and ``generalized cusps''.} When a pair of $S^3\setminus\G_5$ are glued through a pair of 4-holed spheres via a certain identification of holes, the resulting 3-manifold is a graph complement $S^3\setminus(\G_5\#\G_5)$ of a bigger graph. The graph sum $\G_5\#\G_5$ is obtained by removing a vertex in each $\G_5$, and connecting the resulting 4 pairs of open edges. Using this procedure repeatedly, we can obtain 
\be
\mathscr{M}_3=\underbrace{(S^3\setminus\G_5)\cup\cdots\cup(S^3\setminus\G_5)}_{N\ \mathrm{copies}}=\Fx_3\setminus\G_5^{\# N}.
\ee
Here $N$ is the number of 4-simplices in the 4-manifold and $\Fx_3$ is, in general, a more complicated closed 3-manifold than $S^3$. For example, $\pi_1(\Fx_3)$ may be nontrivial, as can be seen when we glue 2 pairs of 4-holed spheres between 2 copies of $S^3\setminus\G_5$.

We impose the boundary conditions of Section \ref{b.c.} to specify the $\Slc$ flat connection on $\sm_3$, i.e. the flat connections restricted to the big boundary components of $\sm_3$ become SU(2) flat connections on 4-holed spheres. However, in addition to the boundary condition, we have to require that on the 4-holed sphere that serves as the interface for the gluing of the $(S^3\setminus\G_5)$'s, the $\Slc$ flat connection has to reduce to SU(2) as well. This is required so that the flat connection in each copy of $(S^3\setminus\G_5)$ determines a constant curvature 4-simplex geometry. Given an $\Slc$ flat connection on $\sm_3$ satisfying the above requirements, it determines a convex constant curvature 4-simplex geometry for each copy of $S^3\setminus\G_5$ by Theorem \ref{3d/4d}.

The fundamental group of $\sm_3$ is obtained simply by forming the product of $\pi_1(S^3\setminus\G_5)$'s and identifying the generators corresponding to the 4-holed spheres that serve as the gluing interface. In terms of holonomies, there may be need for a parallel transport between the base points of loops $\fl_{ab}$ in different copies of $S^3\setminus\G_5$. Given a pair of glued $S^3\setminus\G_5$, the uniqueness Lemma \ref{topounique} guarantees that the isomorphisms Eq. \Ref{Xmap} gives identifications between the loops in the two copies of $S^3\setminus\G_5$ and the simple paths in the two 4-simplices. These isomorphisms induce 2 isomorphisms $S_1$ and $S_2$, as in Eq. \Ref{Smap}, between the identified loops and the simple paths in the two tetrahedra from the two 4-simplices. Since the loops are identified, the composed map $S_1\circ S_2^{-1}$ identifies the simple paths in the two tetrahedra. This $\Slc$ flat connection in $\sm_3$ gives an SU(2) flat connection on the interface 4-holed sphere, which determines uniquely a convex constant curvature tetrahedron by Theorem \ref{2d/3d}. This tetrahedron is shared by the 2 geometrical 4-simplices, since the simple paths of the tetrahedra seen from each side of the gluing have been identified. Therefore the geometrical 4-simplices determined by each copy of $S^3\setminus\G_5$ glue geometrically and form a large simplicial geometry. All 4-simplices and tetrahedra have the same constant curvature $\L$.

Note that the large simplicial geometry on the simplicial complex is not necessarily constant curvature. It can approximate arbitrary Lorentzian geometry on a 4-dimensional manifold because of the hinging at 4-simplex interfaces; this is just as in Regge calculus \cite{Regge}. 

In a single copy of $S^3\setminus\G_5$, an $\Slc$ flat connection $A$ corresponding to 4-simplex geometry is always accompanied by its parity partner $\tilde{A}$, which determines the same geometry but with different 4d orientation by Theorem \ref{parity}. The pair $A$ \& $\tilde{A}$ are related by complex conjugation with respective to the complex structure of $\Slc$ and therefore $A$ \& $\tilde{A}$ give the same SU(2) flat connection on 4-holed spheres. On an $\sm_3$ formed by gluing $N$ copies of $S^3\setminus\G_5$ there are $2^N$ parity-related flat connections, which determine the same geometry on the simplicial complex. Each of the $2^N$ flat connections associates with a choice of 2 possible orientations in each individual 4-simplex.\footnote{The same phenomena happens in the asymptotics of LQG spinfoam models \cite{HZ}.} All of the parity-related flat connections give the same set of SU(2) flat connections on all 4-holed spheres, including the big boundary components and gluing interfaces. Among the $2^N$ parity-related flat connections, there are only 2 flat connections associated with the 2 possible uniform 4d orientations on the entire simplicial complex, which we call the \emph{global parity pair} and denote again by $A$ \& $\tilde{A}$.

In terms of complex FN coordinate on $\cm_{\mathrm{flat}}(\partial\sm_3,\Slc)$, the global parity pair $A,\tilde{A}\in\cm_{\mathrm{flat}}(\sm_3,\Slc)$ can be written as 
\be
A=[x_\ell,y_{\ell};x_B,y_B], \quad \text{and}  \quad \ \ \tilde{A}=[x_\ell,\tilde{y}_{\ell};x_B,y_B],
\ee
where $x_\ell$ and $y_{\ell}$ are the complex length and twist variables of a small boundary component $\ell$.\footnote{When the small boundary component $\ell$ is a torus cusp, $x_\ell,y_{\ell}$ are simply the eignvalues of meridian and longitude loop holonomies.} Here $x_B$ and $y_B$ are the canonical coordinates of $\cm_{\mathrm{flat}}(\text{4-holed sphere},\Su)$ at a big boundary component and the variables $[x_\ell;x_B,y_B]$ are treated as the boundary data. 

A small boundary component $\ell$ corresponds to a unique triangle $\Delta_\ell$ in the simplicial complex $\ck_4$.\footnote{The triangle $\Delta_\ell$ is an internal triangle when $\ell$ is a torus cusp and a boundary triangle when $\ell$ is a cylinder connecting 2 big boundary components. If the 4-manifold is closed and the simplicial complex does not have boundary, the corresponding $\sm_3$ has only torus cusps.} The coordinate $x_{\ell}$ relates to the triangle area $\Ar_\ell$ of $\Delta_\ell$ in the same way as before, e.g. in Eq. \Ref{uvATheta}. The relation between $y_\ell$ and the hyperdihedral angles is given by a sum over all the 4-simplices sharing $\Delta_\ell$, i.e.
\be
\ln y_{\ell}&=&-\frac{1}{2}\nu\,\sgn(V_4)\sum_{\sig,\Delta_\ell\subset\sig}\Theta_\ell(\sig)-i\nu\,\theta_{\ell}+\frac{\ln\chi_\ell(\xi)}{2} \qquad \text{mod}\ \ 2\pi i N_{\ell},\ \ \ \ N_{\ell}\in\Z,
\ee
where $\Theta_\ell(\sig)$ is the hyperdihedral (boost) angle in the 4-simplex $\sig$ hinged by $\Delta_\ell$. The sign $\sgn(V_4)$ is a global sign determined by the uniform 4d orientation and $y_\ell$ and $\tilde{y}_{\ell}$ relate to two different signs, $\sgn(V_4)=\pm1$ respectively.

We define the logarithmic variables $u,v$ in the same way as before, and choose a canonical lift to the cover space for the boundary data $[x_\ell;x_B,y_B]\mapsto [u_\ell;u_B,v_B]$. We also choose two arbitrary lifts $\a,\tilde{\a}$ for $y_\ell\mapsto v^\a_\ell$ and $\tilde{y}_\ell\mapsto \tilde{v}^{\tilde{\a}}_\ell$. The holomorphic 3d block $Z^{(\a)}_{CS}\lt(\sm_3\big|u\rt)$ of $\Slc$ Chern-Simons theory on $\sm_3$ can be constructed in the same way as Eq.\Ref{block}, for $(A,\a)$ with the reference $(\tilde{A},\tilde{\a})$. The Liouville 1-form is now given by
\be
\vth=\sum_\ell v_{\ell}\rmd u_{\ell}+\sum_B v_B\rmd u_B.
\ee
The integration contour of $\int_{\Fc}\vth$ is in $\cl_{\mathbf{A}}\simeq\cm_{\mathrm{flat}}(\sm_3,\Slc)$, which is a holomorphic Lagrangian subvariety in $\cm_{\mathrm{flat}}(\partial\sm_3,\Slc)$. 

The semiclassical asymptotic behavior of $Z^{(\a)}_{CS}\lt(\sm_3\big|u\rt)$ can be analyzed in the same way as in Section \ref{3dblock}, which leads to the following generalization of Eq.\Ref{ZZasymp}
\be
Z_{CS}^{(\a)}\lt(\sm_3\big|\,u\rt)Z_{CS}^{(\overline{\a})}\lt(\sm_3\big|\,\bar{u}\rt)
&=&\exp\lt[\frac{i}{\hbar}2\mathrm{Re}\lt(\frac{\L t}{12\pi i}\rt)\lt(\sum_{\ell} \mathbf{a}_{\ell}\sum_{\sig,\Delta_\ell\subset\sig}\Theta_\ell(\sig)-\L \sum_\sig \mathrm{Vol}^\L_4(\sig)\rt)\rt]\nonumber\\
&&\times\ \exp\lt[\frac{i}{\hbar}2\mathrm{Re}\lt(\frac{\L t}{12\pi i}C^\a_{\tilde{\a}}\rt)+\frac{i}{\hbar}2\mathrm{Re}\lt(\frac{\L t}{6}\rt)\sum_{\ell}\Delta N_{\ell}\mathbf{a}_{\ell}+\cdots\rt],
\ee
where the lift-independent term
\be
S^\L_{Regge}=\sum_{\ell} \mathbf{a}_{\ell}\sum_{\sig,\Delta_\ell\subset\sig}\Theta_\ell(\sig)-\L \sum_\sig \mathrm{Vol}^\L_4(\sig)
\ee
is the Lorentzian Regge action of Einstein gravity on the simplicial complex $\ck_4$ \cite{foxon,Regge,FFLR}. The sum $\sum_{\sig,\Delta_\ell\subset\sig}\Theta_\ell(\sig)$ is the Lorentzian deficit angle when $\Delta\ell$ is an internal triangle in $\ck_4$, while it is a hyperdihedral boost angle when $\Delta_\ell$ is a boundary triangle of $\ck_4$. The gravitational constant $G_{\mathrm{N}}$ is given by Eq. \Ref{Gnewton} and $C^\a_{\tilde{\a}}$ is again an integration constant. The term $\frac{i}{\hbar}2\mathrm{Re}\lt(\frac{\L t}{6}\rt)\sum_{\ell}\Delta N_\ell\mathbf{a}_{\ell}$ is lift-dependent and takes discrete values. This term disappears when the quantization condition $2\mathrm{Re}\lt(\frac{\L t}{6}\rt)\sum_{\ell}\Delta N_\ell\mathbf{a}_{\ell}\in 2\pi \hbar \Z$ or $t\in i\R$ is satisfied.

\section*{Acknowledgements}

HMH gratefully acknowledges sabbatical support from the Perimeter Institute for Theoretical Physics. MH would like to thank V. V. Fock, J. Mour\~ao, R. Van der Veen, Z. Sun for several fruitful discussions. MH would also like to thank F. Vidotto for an invitation and for hospitality while visiting Radboud Universiteit Nijmegen, and for her interesting comments on this work.  MH acknowledges funding received from the People Programme (Marie Curie Actions) of the European Union's Seventh Framework Programme (FP7/2007-2013) under REA grant agreement No. 298786, from the Alexander von Humboldt Foundation, from the US National Science Foundation through grant PHY-1602867, and a Start-up Grant at Florida Atlantic University, USA. This work is  supported  by  Perimeter  Institute  for  Theoretical  Physics.   Research  at  Perimeter  Institute is supported by the Government of Canada through Industry Canada and by the Province of Ontario through the Ministry of Research and Innovation.

\appendix

\section{$K_2$-Lagrangian Subvariety}\label{Ktheory}

In this appendix, we provide a very brief introduction to the notion of $K_2$-Lagrangian subvariety, and explain its relation to quantizability. The discussion here follows \cite{Fock}. See also \cite{DGG,DV} for more detailed discussions.

Let $\C^*=\C\setminus\{0\}$, we define the Abelian group $\C^*\wedge\C^*=\wedge^2\C$ generated by $a\wedge b$, with $a,b\in\C^*$ and the relations
\be
a\wedge b=-b\wedge a,\ \ \ \ \ (ab)\wedge c=a\wedge c+b\wedge c.
\ee 

Let $\cm$ be a complex variety, and denote the set of holomorphic functions $U_\a\to\C^*$ on the chart $U_\a$ as $\C^*(U_\a)$ . A $K_2$-\emph{symplectic structure} on $\cm$ is an element $\o^K_\a\in \C^*(U_\a)\wedge\C^*(U_\a)$ on every coordinate chart $U_\a$, such that on $U_\a\cap U_\b$, $\o^K_\a-\o_\b^K=\sum_{I}z_I\wedge (1-z_I)$ for some $z_I\in \C^*(U_\a\cap U_\b)$. In other words, A $K_2$-\emph{symplectic structure} on $\cm$ belongs to the group $K_2(\C)$, which is the quotient of $\C^*\wedge\C^*$ by the subgroup generated by $z\wedge(1-z)$. 

We define a map $\rmd\ln\wedge\rmd\ln$ from $\C^*\wedge \C^* $ to the space of holomorphic 2-forms $ \O^2(\C)$ by 
\be
\rmd\ln\wedge\rmd\ln:\ x\wedge y\mapsto \rmd\ln x\wedge \rmd\ln y.
\ee
It is easy to see that $\rmd\ln\wedge\rmd\ln$ is essentially a map from $K_2(\C)\to\O^2(\C)$, since $\rmd\ln z\wedge\rmd \ln(1-z)=0$. Moreover, given a $K_2$-symplectic structure $\o^K=\sum_m x_m\wedge y_m$, the map $\rmd\ln\wedge\rmd\ln:\ \sum_m x_m\wedge y_m\mapsto \sum_m \rmd\ln x_m\wedge \rmd\ln y_m$ is a closed 2-form (pre-symplectic form) on the complex variety $\cm$. 

Let $\cm$ be a complex variety with a $K_2$-symplectic form $\o^K\in K_2(\C)$ such that $\lt[\rmd\ln\wedge\rmd\ln\rt](\o^K) =\o$ is a symplectic structure. A $K_2$-Lagrangian subvariety $\cl^K\in\cm$ is a subvariety with $\dim\cl^K=\half\dim\cm$ and 
\be
\o^K|_{\cl^K}=\sum_{I}z_I\wedge (1-z_I), 
\ee 
for some holomorphic functions $z_I$  on $\cm$. References \cite{DGG,DV,FG2} show that, at least on the generic part of $\cm_{\mathrm{flat}}(\Sig_g,\Slc)$ that is of interest, the symplectic structure $\o=\sum_m\frac{\rmd x_m}{x_m}\wedge \frac{\rmd y_m}{y_m}$ has a $K_2$-avatar $\o^K\in K_2(\C)$ such that $\lt[\rmd\ln\wedge\rmd\ln\rt](\o^K) =\o$. The moduli space $\cm_{\mathrm{flat}}(M_3,\Slc)=\cl_{\mathbf{A}}$ with $\partial M_3=\Sig_g$ is a $K_2$-Lagrangian subvariety in $\cm_{\mathrm{flat}}(\Sig_g,\Slc)$, i.e. $\o^K|_{\cl_{\mathbf{A}}}=\sum_{I}z_I\wedge (1-z_I)$ for some holomorphic functions $z_I$.

Define two homomorphisms $\vth_k$ and $\vth_\sig$ from $K_2(\C(\cl_{\mathbf{A}}))$ to $H^1(\cl_{\mathbf{A}},\R)$ (up to a $4\pi^2\Z$ covering for $\vth_k$) by
\be
\vth_\sig: x\wedge y\mapsto \vth_\sig(x\wedge y)&:=& \ln|y|\,\rmd (\arg x)-\ln|x|\,\rmd (\arg y), \quad \text{and}\nonumber\\
\vth_k: x\wedge y\mapsto \vth_k(x\wedge y)&:=& \ln|y|\,\rmd (\ln |x|)+\arg x\, \rmd (\arg y).
\ee
Having chosen a polarization, the Lagrangian subvariety $\cl_{\mathbf{A}}$ is quantizable when the following conditions are satisfied for all closed path $\a\in\cl_{\mathbf{A}}$ (when the real part of Chern-Simons coupling $\mathrm{Re}(t)=k\in\Z$) \cite{gukov}:
\be
\oint_\a\vth_\sig(\o^K|_{\cl_{\mathbf{A}}})=0, \ \  \text{and} \ \ \oint_\a\vth_k(\o^K|_{\cl_{\mathbf{A}}})\in 4\pi^2\mathbb{Q}.
\ee
Since $\cl_{\mathbf{A}}$ is a $K_2$-Lagrangian subvariety with respect to $\o^K$, then $\vth_\sig(\o^K)$ is given by
\be
\vth_\sig(\o^K|_{\cl_{\mathbf{A}}})= \sum_I \ln|1-z_I|\,\rmd (\arg z_I)-\ln|z_I|\,\rmd (\arg (1-z_I))=-\sum_I\rmd D(z_I),
\ee
where $D(z_I)$ is the Bloch-Wigner dilogarithm function
\be
D(z)=\ln|z|\arg(1-z)+\mathrm{Im}\lt(\mathrm{Li}_2(z)\rt).
\ee
Then $\oint_\a\vth_\sig(\o^K|_{\cl_{\mathbf{A}}})=0$ is satisfied since $D(z)$ is a continuous function on $\C$. Similarly,
\be
\vth_k(\o^K|_{\cl_{\mathbf{A}}})&=& \sum_I\ln|1-z_I|\,\rmd (\ln |z_I|)+\arg z_I\, \rmd (\arg (1-z_I))\nonumber\\
&=&-\sum_I\rmd \lt[\mathrm{Re}\lt(\mathrm{Li}_2(z_I)\rt)-\arg z_I\,\arg (1-z_I)\rt].
\ee
The real part of the dilogarithm $\mathrm{Re}\lt(\mathrm{Li}_2(z_I)\rt)$ is also a continuous function on $\C$, while $\sum_I\oint_\a\rmd\lt[\arg z_I\,\arg (1-z_I)\rt]\in 4\pi^2\Z$. Thus $\oint_\a\vth_k(\o^K|_{\cl_{\mathbf{A}}})\in 4\pi^2\mathbb{Q}$ indeed holds. We conclude that $\cl_{\mathbf{A}}$ being $K_2$-Lagrangian implies that $\cl_{\mathbf{A}}$ is quantizable. In addition, the fact that the 1-forms $\vth_k(z\wedge(1-z))$ and $\vth_\sig(z\wedge(1-z))$ are exact up to $4\pi^2\Z$ shows that they are indeed homomorphisms from $K_2(\C(\cl_{\mathbf{A}}))$ to $H^1(\cl_{\mathbf{A}},\R)$ up to a $4\pi^2\Z$ covering for $\vth_k$.

When we consider the analytic continuation of Chern-Simons theory with generic non-integer $k$, the Lagrangian subvariety $\cl_{\mathbf{A}}$ has to be replaced by its cover space $\overline{\cl}_{\mathbf{A}}$, on which $\ln z_I$ is single-valued. This is because we do not  quotient out large gauge transformation for the analytically continued Chern-Simons theory. So, the 1-forms $\vth_k(z\wedge(1-z))$ and $\vth_\sig(z\wedge(1-z))$ are indeed exact on the cover space $\overline{\cl}_{\mathbf{A}}$, i.e. $\oint_\a\vth_\sig=\oint_\a\vth_k=0$ on $\overline{\cl}_{\mathbf{A}}$.

\section{Quantization of Coadjoint Orbit, Unitary Representations of $\Slc$, and a Path Integral Formula for the Wilson Line}\label{coadj}

In this appendix, we give a brief review of the geometric quantization of the coadjoint orbits of $\Slc$, which gives the unitary irreducible representations of $\Slc$. We also give a quick review of the path integral formula for unitary Wilson line, which is a consequence of the coadjoint orbit quantization. More extensive reviews of these topics can be found in \cite{orbit} (see also \cite{Dimofte} for a nice summary).  

As a complex Lie algebra, $\slc$ is generated by the basis
\be
H=\begin{pmatrix}
  1 & 0  \\
  0 & -1
 \end{pmatrix},\ \ \ \ 
E=\begin{pmatrix}
  0 & 1  \\
  0 & 0
 \end{pmatrix},\ \ \text{and}  \ \ 
F=\begin{pmatrix}
  0 & 0  \\
  1 & 0
 \end{pmatrix}.
\ee
If $\slc$ is viewed as a real Lie algebra, it is generated by $\{E,\ F,\ H, \tilde{E}=iE,\ \tilde{F}=iF,\ \tilde{H}=iH\}$. Given $\slc$ (viewed as a real Lie algebra) and its complexification $(\slc)_{\C}\simeq \slc\times \slc$, a nondegenerate trace form $\lag\ ,\ \rag: (\slc)_{\C}\times\slc\to\C$ is given by 
\be
\lag(X_L,X_R),Y\rag=\half\tr(X_L Y)+\half\tr(X_R \bar{Y}),
\ee
where $X_{L},X_{R},$ and $Y$ are $2\times 2$ matrices. The trace form is a complexification of the invariant bilinear form of $\slc$. Using the trace form, a weight $\underline{\l}\in(\slc^*)_{\C}$ can be identified as a pair of $2\times 2$ matrices $(\l_L,\l_R)$ in $(\slc)_{\C}$. The coadjoint orbit is defined by 
\be
(\O_\l)_\C:=\{g(\l_L,\l_R) g^{-1}\}_{g\in\Slc_\C}\simeq \Slc/\Fh^L_{\l}\times\Slc/\Fh^R_{\l}
\ee 
where $\Fh^{L,\,R}_\l$ is the stabilizer $\Fh^{L,\,R}_\l=\{\,h\in\Slc\,|\, h\l_{L,\,R} h^{-1}=\l_{L,\, R}\}$. Here the stabilizer is precisely the Cartan subgroup (or maximal torus) $\Fh_\l=\mathbb{T}_\C$, thus the coadjoint orbit is given by 
\be
(\O_\l)_\C=\Slc/\mathbb{T}_\C\times\Slc/\mathbb{T}_\C\simeq T^* S^2\times T^* S^2,\ \ \ \ \text{with}\ \ \ \ \O_\l=\Slc/\mathbb{T}_\C=T^* S^2.
\ee 
For our present purposes it is sufficient to consider the real form $\O_\l$ of the coadjoint orbit; this is achieved by viewing the second copy of $T^*S^2$ as the complex conjugate of the first copy.

Let $\underline{\nu},\underline{\kappa}\in(\slc^*)_\C$ be the linear functionals defined by $\underline{\nu}(H)=-i w,\ \underline{\kappa}(\tilde{H})=m$ ($w,n\in\C$),  and $\underline{\nu}(\tilde{H})=\underline{\kappa}({H})=0$, while both $\underline{\nu}$ and $\underline{\kappa}$ annihilate $E,F,\tilde{E},\tilde{F}$. The above trace form results in the identification $\underline{\nu}\longleftrightarrow(\nu,\nu)$ and $\underline{\kappa}\longleftrightarrow(\kappa,-\kappa)$ with $\nu$ and $\kappa$ the  $2\times 2$ matrices
\be
\nu=-\frac{iw}{2}\begin{pmatrix}
  1 & 0  \\
  0 & -1
 \end{pmatrix},\ \ \text{and} \ \ \kappa=-\frac{im}{2}\begin{pmatrix}
  1 & 0  \\
  0 & -1
 \end{pmatrix}.
\ee 
The weight $\underline{\l}$ satisfies $\underline{\l}=\underline{\nu}\oplus\underline{\kappa}\longleftrightarrow (\l_L,\l_R)=(\nu+\kappa,\nu-\kappa)$. The coadjoint orbit $\O_\l$ has a natural $\Slc$ invariant symplectic structure:
\be
\o_{\nu,\kappa}=\frac{1}{2}\tr\lt[(\nu+\kappa)g^{-1}\rmd g\wedge g^{-1}\rmd g\rt]+\frac{1}{2}\tr\lt[(\nu-\kappa)\overline{g^{-1}\rmd g\wedge g^{-1}\rmd g}\rt]. \label{gsymplectic}
\ee

To proceed with geometric quantization, a line-bundle $\Fl\to\O_\l$ must be defined over the phase space $\O_\l$, with $\o_{\nu,\kappa}$ the curvature of $\Fl$. Due to the compact cycle $S^2\subset\O_\l$, Weyl's integrality criterion requires $\o_{\nu,\kappa}$ to have $m\in\Z$ in order that $\Fl$ is prequantizable. Reality of the curvature $\o_{\nu,\kappa}$   implies $w\in i\R$. The prequantum line-bundle $\Fl$ can be obtained by taking the quotient of $\C\times \Slc$  by the representation $\Fh_\l=\mathbb{T}_\C$ acting on $\C$. The representation is given by $(f,x)\mapsto (\sig(h) f,xh)$, so that the quotient is given by the identification:
\be
(f,xh)=(\sig(h^{-1})f,x)\ \  \ \ \text{or}\ \ \ \ f(xh)=\sig(h^{-1})f(x),\ \ \ \ \text{with}\ \ f\in\C,\ x\in\Slc,\ h\in\mathbb{T}_\C.\label{quotient}
\ee 
The representation $\sig(h^{-1})$ is given by $e^{(i\underline{\nu}+\underline{\rho})\oplus i\underline{\kappa}}(h^{-1})$. Here $\underline{\rho}\in\slc^*$ is the restricted positive root $\underline{\rho}(H)=2,\ \underline{\rho}(\tilde{H})=0$ ($\underline{\rho}$ annihilates $E,F,\tilde{E},\tilde{F}$). The above quotient gives the prequantum line-bundle $\Fl\to\O_\l$ where $\Slc$ acts on the sections $f$ by
\be
g\act f(x)=f(g^T x).\label{gpaction}
\ee 

An element of $\Slc$ can be written as
\be
g=\begin{pmatrix}
  z^1 & -x^2  \\
  z^2 & x^1
 \end{pmatrix}\ \ \ \ \text{with}\ \ \ \ z^1x^1+z^2x^2=1.
\ee
In the coadjoint orbit $\Slc/\mathbb{T}_\C$ there is an equivalence $(z^1,z^2,x^1,x^2)\sim (\a z^1,\a z^2,\a^{-1}x^1,\a^{-1}x^2)$ for $\a\in \C^{*}$. We use a polarization such that the resulting sections of $\Fl$ depend only on the projective coordinate $z^1/z^2$. Because of the above quotient procedure Eq. \Ref{quotient}, the sections transform in the following way:\footnote{Let $h=e^{tH}$ with $t\in\C$, we have $\lt[(i\underline{\nu}+\underline{\rho})\oplus i\underline{\kappa}\rt](tH)=\half\tr\lt[(i\nu+i\kappa+\rho)tH\rt]+\half\tr\lt[(i\nu-i\kappa+\rho)\bar{t}H\rt]=\frac{t}{2}(w+k+2)+\frac{\bar{t}}{2}(w-k+2)$, where the $2\times2$ matrix $\rho$ equals $\nu$ when $w=2i$.}
\be
f(\a z^1,\a z^2,\bar{\a} \bar{z}^1,\bar{\a} \bar{z}^2)=\a^{-\half (w+m)-1}\bar{\a}^{-\half (w-m)-1}f( z^1,z^2,\bar{z}^1, \bar{z}^2),\ \ \ \ \a\in \C^*.\label{scaling}
\ee
This transformation is precisely the scaling property of the homogeneous function/section in the principle series representation when $w\in i\R$ and $m\in\Z$. In our analysis of knotted graph operators, the parameters $w,m$ are given by
\be
w=-2i\g j_{ab}, \ \ \ \ m=-2j_{ab},\ \ \text{and} \ \ j_{ab}\in\Z/2.
\ee
The group action of Eq. \Ref{gpaction} gives the representation:
\be
\begin{pmatrix}
  a & b  \\
  c & d
 \end{pmatrix}\act f(z,\bar{z})=\Big(bz+d\Big)^{-\half (w+m)-1}\lt(\overline{bz+d}\rt)^{-\half (w-m)-1}f\lt(\frac{az+c}{bz+d}\rt), \label{grep}
\ee
where $z={z_1}/{z_2}$ is a projective coordinate on $\C\mathbb{P}^1$. The space of these sections on $\mathbb{CP}^1$, completed using the $L^2$ inner product with measure $\rmd z=\frac{i}{2}(z^1\rmd z^2-z^2\rmd z^1)\wedge(\bar{z}^1\rmd \bar{z}^2-\bar{z}^2\rmd \bar{z}^1)$, carries the principle series unitary irreducible representation of $\Slc$ labeled by $(m,w)$.  The carrier space is denoted by $\ch^{m,w}$ or $\ch^{j,\rho}$ with $m=-2j$ and $w=-2i\rho$. There is an isomorphism between the representations with labels $(m,w)$ and $(-m,-w)$. 

In the above representation, expressed in terms of sections on $\mathbb{CP}^1$, the variables $z^1$ and $z^2$ are ``position variables'' and correspond to multiplication operators on (a dense domain of) $\ch^{m,w}$. The variable $x^1$ and $x^2$ are ``momentum variables'' and correspond to the derivative operators:
\be
x^1=\lt(\frac{2}{w+m}\rt)\frac{\partial}{\partial z^1},\ \ \ \ x^2=\lt(\frac{2}{w+m}\rt)\frac{\partial}{\partial z^2},\ \ \ \ 
\bar{x}^1=\lt(\frac{2}{w-m}\rt)\frac{\partial}{\partial \bar{z}^1},\ \ \ \ \bar{x}^2=\lt(\frac{2}{w-m}\rt)\frac{\partial}{\partial \bar{z}^2}.
\ee
The scaling property of Eq. \Ref{scaling} implies
\be
z^1\frac{\partial}{\partial z^1}+z^2\frac{\partial}{\partial z^2}=-\half(w+m)-1,\ \ \ \ \bar{z}^1\frac{\partial}{\partial \bar{z}^1}+\bar{z}^2\frac{\partial}{\partial \bar{z}^2}=-\half(w-m)-1.
\ee

Note that the unitary irrep constructed above is an \emph{induced representation} $\mathrm{ind}^{\Slc}_B(\sig)$ on the sections of a line-bundle over the coset $\Slc/B\simeq\mathbb{CP}^1$. Here $B$ is the Borel subgroup of upper-triangular matrices, whose Lie algebra is generated by $H,\tilde{H},E,\tilde{E}$. The sections are obtained from the functions $f$ on $\Slc$ that satisfy 
\be
f(xb)=\sig(b^{-1})f(x), 
\ee
where $b\in B,\ x\in\Slc$, and $\sig$ is given by $\sig=e^{(i\underline{\nu}+\underline{\rho})\oplus i\underline{\kappa}}$, viewed as a representation of $B$.

The Wilson line in the unitary irrep $(m,w)$ can be written as a path integral. When we consider its matrix element in the $z$-space representation ($z$ is the projective coordinate of $\mathbb{CP}^1$):
\be
\lag z \big|\,\cp e^{\int_\ell A}\,\big| z'\rag_{\ch^{m,w}}&=&\int_{z'}^z \cd g\cd\bar{g}\,e^{iS[g,\bar{g};A,\bA]},\label{holonomyPI}
\ee
where the action $S[g,\bar{g};A,\bA]$ is given by:
\be
S[g,\bar{g};A,\bA]&=&-\half\int_\ell\tr \lt[(\nu+\kappa)g^{-1}(\rmd+A^T)g+(\nu-\kappa)\bar{g}^{-1}(\rmd+\bA^T)\bar{g}\rt].
\ee
The path integral has a first-order Lagrangian depending on the $\Slc$-valued functions $g:\ell\to \Slc$. The boundary condition for the path integral is that the ``position variables'' $g$ at the source and target of $\ell$ are equal to $z'$ and $z$. The above path integral can be viewed as a quantum particle moving through the ``position space'' $\mathbb{CP}^1$.

However there is a gauge symmetry of the action, i.e. $S[g,\bar{g};A,\bA]$ is invariant under $g\mapsto gh$ with $h\in\Fh_\l=\mathbb{T}_\C$ when $h$ is trivial on the boundary. Therefore the path integral is essentially defined over the maps $g:\ell\to \Slc/\mathbb{T}_\C=\O_\l$, where $\O_\l$ is the coadjoint orbit, except for the integral at the boundary of $\ell$. If we consider the gauge transformation $g\mapsto gh$ with $h\in\Fh_\l=\mathbb{T}_\C$ non-trivial on the boundary, the path integral Eq. \Ref{holonomyPI} transforms non-trivially. Evaluation of the path integral defines a section in the line-bundle over $\mathbb{CP}^1\times\mathbb{CP}^1$. Indeed, let us consider an arbitrary gauge transformation $g\mapsto gh$ with $h=e^{\t H},\ \t\in\C $. The action $S$ transforms as
\be
S[g,\bar{g};A,\bA]\mapsto S[g,\bar{g};A,\bA]+\frac{i}{2}(w+m)\int_\ell\rmd\t+\frac{i}{2}(w-m)\int_\ell\rmd\bar{\t}.\label{Strans}
\ee 
Under the transformation $g\mapsto gh$ the coordinates $z^1,z^2,x^1,x^2$ for the quotient $\Slc/\mathbb{T}_\C$ scale as 
\be
\begin{pmatrix}
  z^1 & -x^2  \\
  z^2 & x^1
 \end{pmatrix}\begin{pmatrix}
  \a & 0  \\
  0 & \a^{-1}
 \end{pmatrix}=\begin{pmatrix}
  \a z^1 & -\a^{-1}x^2  \\
  \a z^2 & \a^{-1}x^1
 \end{pmatrix},\ \ \ \ \text{where}\ \ \ \ \a=e^{\t}.
\ee
Eq. \Ref{Strans} implies that the path integral transforms in the same way as Eq.\Ref{scaling}, which has to be the case in order that Eq.\Ref{holonomyPI} is correct and the $L^2$ inner product with $\rmd z$ is scale invariant, i.e. 
\be
\int_{z'}^{\l z} \cd g\cd\bar{g}\,e^{iS[g,\bar{g};A,\bA]}&=&\a^{-\frac{1}{2}(w+m)-1}\bar{\a}^{-\frac{1}{2}(w-m)-1}\int_{z'}^{z} \cd g\cd\bar{g}\,e^{iS[g,\bar{g};A,\bA]}\nonumber\\
\text{and} \quad \int_{\l z'}^{ z} \cd g\cd\bar{g}\,e^{iS[g,\bar{g};A,\bA]}&=&\a^{\frac{1}{2}(w+m)-1}\bar{\a}^{\frac{1}{2}(w-m)-1}\int_{z'}^{z} \cd g\cd\bar{g}\,e^{iS[g,\bar{g};A,\bA]}.
\ee
Note that a factor $\a^{-1}\bar{\a}^{-1}$  above comes from the path integral measure at the boundary.

Using the boundary conditions on $z$ and $z'$, the variational equations of motion can be derived from $S$:
\be
\lt[\nu+\kappa,\ g^{-1}(\rmd+A^T)g\rt]=0,\ \quad \text{and} \quad \ \lt[\nu-\kappa,\ \bar{g}^{-1}(\rmd+\bar{A}^T)\bar{g}\rt]=0,
\ee
which implies that $g$ is the gauge transformation diagonalizing the component of $A_t$ along the curve $\ell$, or,
\be
\frac{\rmd}{\rmd t}g+A^T_t g \propto_\C g H,\ \quad \text{and}  \quad \ \frac{\rmd}{\rmd t}\bar{g}+\bar{A}^T_t \bar{g} \propto_\C \bar{g} H.
\ee
Again expressing $g$ using the coordinates $z^1,z^2,x^1,x^2$, we find that $\frac{\rmd}{\rmd t}z+A^T_t z\propto_\C z$ (and similarly for $\bar{z}$) where $z=(z^1,z^2)^T$. Then the on-shell relation for the boundary data $z,z'$ of the path integral is:
\be
z\propto_\C \cp e^{-\int_\ell A^T} z'.
\ee

Hamiltonian analysis of $S[g]$ reproduces the symplectic structure $\o_{\nu,\kappa}$ of Eq. \Ref{gsymplectic}, and gives the Hamiltonian $\mathbf{H}=p\cdot\partial_t q -\mathbf{L}$:
\be
\mathbf{H}=\half\tr \lt[(\nu+\kappa)g^{-1}A_t^Tg+(\nu-\kappa)\bar{g}^{-1}\bA_t^T\bar{g}\rt],
\ee
where $A_t$ is the component of $A$ along the curve $\ell$. We replace the variables in $g$ by the corresponding operators in the $z$-space representation to define the Hamiltonian operator $\hat{\mathbf{H}}$. Here $A_t$ is treated as an external variable, so its components in an $\slc$ basis are treated as c-numbers. Consequently, it can be shown that, the resulting Hamiltonian operator $-i\hat{\mathbf{H}}$ is precisely the representation of $A_t: \ell\to\slc$ in the unitary irrep as an operator on $\ch^{m,w}$. In other words,  if we expand the $2\time 2$ matrix $A_t=a H+b E+c F$, then
\be
-i\hat{\mathbf{H}}=a \hat{H}+b \hat{E}+c \hat{F},
\ee
where $\hat{H},\hat{E}$ and $\hat{F}$ are the differential operators representing $H,E$, and $F\in\slc$ and generating infinitesimally the representation Eq. \Ref{grep}. As a result, the path integral of Eq. \Ref{holonomyPI} for Wilson line follows from the quantum mechanical relation:
\be
\lag z \big|\,T e^{-i\int \hat{\mathbf{H}}\,\rmd t }\,\big|z'\rag=\int_{z'}^z \cd g\cd\bar{g}\,e^{iS[g,\bar{g};A,\bA]},
\ee
where $T$ denotes the time-ordering corresponding to the path ordering $\cp$ of the Wilson line.

\end{document}